\documentclass[12pt,article]{report}

\usepackage{booktabs,tabularx,graphicx}
\usepackage{caption}
\usepackage{multirow} 
\usepackage{subcaption}
\usepackage{dirtytalk}
\usepackage{amsmath}
\usepackage{stackengine}
\usepackage{geometry}
\usepackage{fancyhdr}
\usepackage{afterpage}
\usepackage{graphicx}
\usepackage{amsmath,amssymb,amsbsy}
\usepackage{dcolumn,array}
\usepackage{tocloft}
\usepackage{setspace} 
\usepackage{xcolor,soul,framed} 
\usepackage{textcomp}
\usepackage{multirow}
\usepackage{tikz}
\usetikzlibrary{shapes.geometric, arrows}
\usetikzlibrary{arrows.meta, positioning, shadows}
\usepackage{mathptmx} 

\usepackage{etoolbox}
\usepackage{cite}
\usepackage{breqn}
\usepackage{blindtext}
\usepackage{parskip}
\usepackage{titlesec}
\usepackage{enumitem}
\usepackage{tocloft}
\usepackage{IEEEtrantools}
\usepackage{longtable}
\usepackage{algorithm} 
\usepackage{algpseudocode} 
\usepackage{appendix}

\usepackage[acronym,toc]{glossaries}

\usepackage{lineno,hyperref}
\modulolinenumbers[5]

 \makeatletter
\def\@makechapterhead#1{%
  {\parindent \z@ \raggedright \normalfont
    \ifnum \c@secnumdepth >\m@ne
        \huge\bfseries \@chapapp\space \thechapter
        \par\nobreak
        \vskip 1\p@
    \fi
    \interlinepenalty\@M
    \Huge \bfseries #1\par\nobreak
    \vskip 40\p@
  }}
\def\@makeschapterhead#1{%
  {\parindent \z@ \raggedright
    \normalfont
    \interlinepenalty\@M
    \Huge \bfseries  #1\par\nobreak
    \vskip 40\p@
  }}
\makeatother

\makeglossaries

\newacronym{PAR}{PAR}{Phase Angle Regulator}
\newacronym{ISPAR}{ISPAR}{ Indirect Symmetric Phase Angle Regulator}
\newacronym{CT}{CT}{Current Transformer}
\newacronym{ML}{ML}{Machine Learning}
\newacronym{ANN}{ANN}{Artificial Neural Network}
\newacronym{FFNN}{FFNN}{Feed Forward Neural Network}
\newacronym{PNN}{PNN}{Probabilistic Neural Network }
\newacronym{SVM}{SVM}{Support Vector Machine} 
\newacronym{DT}{DT}{Decision Tree }
\newacronym{RFC}{RFC}{Random Forest Classifier }
\newacronym{ANFIS}{ANFIS}{ Adaptive neuro-fuzzy  inference  system }
\newacronym{WT}{WT}{Wavelet Transform}
\newacronym{WF}{WF}{Wind Farm}
\newacronym{WTG}{WTG}{Wind Turbine Generator}
\newacronym{GBC}{GBC}{Gradient Boosting Classifier}

\newacronym{kNN}{kNN}{k-Nearest Neighbour}
\newacronym{NN}{NN}{Neural Network}
\newacronym{RF}{RF}{Random Forest}
\newacronym{PSCAD}{PSCAD}{Power System Computer Aided Design}
\newacronym{EMTDC}{EMTDC}{Electromagnetic Transients including DC} 

\newacronym{t-t}{t-t}{turn-to-turn}
\newacronym{w-w}{w-w}{winding-to-winding}
\newacronym{ST}{ST}{switching time}
\newacronym{ph}{PH}{phase}
\newacronym{kv}{kV}{Kilovolt}
\newacronym{PTS}{PTS}{percentage of turns shorted}

\newacronym{FR}{FR}{fault resistance}
\newacronym{FL}{FL}{fault location}
\newacronym{FIT}{FIT}{fault inception time}
\newacronym{FU}{FU}{faulty unit}
\newacronym{FT}{FT}{fault type}
\newacronym{PS}{PS}{phase shift}
\newacronym{DIF}{DIF}{Detection of internal faults}
\newacronym{LFU}{LFU}{Localization of faulty unit}
\newacronym{CTD}{CTD}{Classification of transient disturbances}
\newacronym{ED}{ED}{Event Detector}
\newacronym{WE}{WE}{Wavelet Energy}
\newacronym{WC}{WC}{Wavelet Coefficient}
\newacronym{TD}{TD}{Time Domain}
\newacronym{CRB}{CRB}{conventional relay block}
\newacronym{mRMR}{mRMR}{Maximum Relevance Minimum Redundancy}
\newacronym{DWT}{DWT}{Discrete Wavelet Transform}
\newacronym{Xgb}{Xgb}{Extreme Gradient Boost}
\newacronym{NB}{NB}{Naive Bayes}
\newacronym{GP}{GP}{Gaussian Process}
\newacronym{TP}{TP}{True Positive}
\newacronym{TN}{TN}{True Negative}
\newacronym{FN}{FN}{False Negative}
\newacronym{FP}{FP}{False Positive}
\newacronym{SMOTE}{SMOTE}{Synthetic Minority Over-sampling Technique}
\newacronym{SNR}{SNR}{Signal-to-Noise-ratio}
\newacronym{dB}{dB}{decibels}

\newacronym{LTC}{LTC}{Load Tap Changer}
\newacronym{GC}{GC}{Grading Capacitance}
\newacronym{nom8}{$Q$}{Reactive Power}
\newacronym{FFT}{FFT}{fast Fourier Transform}
\newacronym{CDF}{CDF}{change detection filter}
\newacronym{nom11}{p.u.}{Per Unit}
\newacronym{PQ}{PQ}{Power Quality}
\newacronym{CBR}{CBR}{Capacitor bank rating}
\newacronym{RFD}{RFD}{Residual flux density}
\newacronym{FA}{FA}{Firing Angle}
\newacronym{FRT}{FRT}{Fault Ride Through}
\newacronym{DG}{DG}{distributed generations}
\newacronym{DFIG}{DFIG}{doubly-fed induction generator}
\newacronym{AR}{AR}{Auto-regressive}
\newacronym{PST}{PST}{Phase Shift Transformer}

\newacronym{OST}{OST}{out of step tripping}
\newacronym{PSB}{PSB}{power swing blocking}
\newacronym{WSCC}{WSCC}{Western Systems Coordinating Council}

\newacronym{nom13}{$N_a$}{Moving Average Window Size}
\newacronym{nom14}{$\Delta t$}{Time Interval}

\newacronym{nom16}{$R$}{Resistance}
\newacronym{nom17}{$X$}{Reactance}
\newacronym{nom18}{$B$}{Susceptance}

\newacronym{nom19}{$v$}{Wind Speed}

\newacronym{sd}{SD}{Standard Deviation}

\begin{document}

\pagenumbering{roman} 
\pagestyle{empty} 

\begin{center}
    \LARGE{\textbf{Data-driven Protection of Transformers, Phase Angle Regulators, and Transmission Lines in Interconnected Power Systems}}
\end{center} \vspace{10mm}

\begin{center}
    By\\\vspace{8mm}
   \textbf{\large{Pallav Kumar Bera}}\\
    B.Tech., Haldia Institute of Technology, 2011\\
    M.Tech., Indian Institute of Technology, 2014\\ \vspace{3.0cm}
    {\Large{\textbf{Dissertation}}}\\\vspace{2mm}
    Submitted in partial fulfillment of the requirements for the degree\\
    of Doctor of Philosophy in Electrical \& Computer Engineering\\ 
    \vspace{3cm}
    Syracuse University\\
    August 2021
\end{center}
\newpage
\begin{center}
   Copyright~\copyright Pallav Kumar Bera 2021\\
    All rights reserved
\end{center}

\newpage
\begin{center}
    \Large{\textbf{DEDICATION}}\\
\end{center}
This dissertation is dedicated to my mother Sumita Bera, my father Nitai Chand Bera, my wife Samita Rani Pani, my sister Laboni Bera, and my adorable son \textit{Aariv}.

\pagestyle{plain} 
\newpage
\chapter*{Acknowledgements}
\addcontentsline{toc}{chapter}{Acknowledgements}
\label{aknow}
\vspace{-1cm}
This dissertation would not have been possible without the support of many people. First and foremost, I'd like to express my gratitude to Prof. Can Isik, my Ph.D. advisor. Can has been an extraordinary mentor as well as an administrator. I appreciate his openness and support throughout this Ph.D. I thank him for the freedom he gave me, whether it was in selecting a research topic or a professional path.

\hspace{8mm}I am also thankful to Prof. Tomislav Bujanovic, Prof. Sara Eftekharnejad, Prof. Prasanta K Ghosh, Prof. Makan Fardad, and Prof. Jamie L Winders for being on my dissertation committee. I appreciate your responsiveness in all communications and your insightful suggestions and helpful comments on my thesis work.

\hspace{8mm}I'd like to express my gratitude to Rajesh Kumar and Vajendra Kumar, my wonderful co-authors. Working with you has been a fantastic experience. They provided me with invaluable advice on research methods, paper writing, and presentation. Aside from my co-authors, I'm grateful to Naveed Tahir and Jack Vining, my friendly, helpful, and supportive lab mates. I'd also like to express my gratitude to Prof. Jae C Oh, our Department Chair, for his support and frequent visits to our lab. 

\hspace{8mm} My parents, wife, sister, extended family members, and in-laws have all made greater sacrifices than me in order to obtain this degree. Without their love, inspiration, and support, I would not be where I am now.  I greatly appreciate my parents for backing my decisions and always letting me choose what I love and my wife who has been instrumental in my endeavor in the United States, while she stayed back in India. Finally, I'd want to express my gratitude to Syracuse University for its financial and logistical help.

\newpage
\tableofcontents
\printglossary

\newpage
\listoffigures
\addcontentsline{toc}{chapter}{List of Figures}

\newpage
\listoftables
\addcontentsline{toc}{chapter}{List of Tables}
\newpage
\begin{center}
    \LARGE{\textbf{Abstract}}
\end{center}

This dissertation highlights the growing interest in and adoption of machine learning approaches for fault detection in modern electric power grids. Once a fault has occurred, it must be identified quickly and a variety of preventative steps must be taken to remove or insulate it. As a result, detecting, locating, and classifying faults early and accurately can improve safety and dependability while reducing downtime and hardware damage. Machine learning-based solutions and tools to carry out effective data processing and analysis to aid power system operations and decision-making are becoming preeminent with better system condition awareness and data availability. 

\hspace{8mm}Power transformers, Phase Shift Transformers or Phase Angle Regulators, and transmission lines are critical components in power systems, and ensuring their safety is a primary issue. Differential relays are commonly employed to protect transformers, whereas distance relays are utilized to protect transmission lines. Magnetizing inrush, overexcitation, and current transformer saturation make transformer protection a challenge. Furthermore, non-standard phase shift, series core saturation, low turn-to-turn, and turn-to-ground fault currents are non-traditional problems associated with Phase Angle Regulators. Faults during symmetrical power swings and unstable power swings may cause mal-operation of distance relays, and unintentional and uncontrolled islanding. The distance relays also mal-operate for transmission lines connected to type-3 wind farms.

\hspace{8mm}The conventional protection techniques would no longer be 
adequate to address the above-mentioned challenges due to their limitations
in handling and analyzing the massive amount of data, limited
generalizability of conventional models, incapability to model non-linear
systems, etc. These limitations of conventional differential and distance
protection methods bring forward the motivation of using machine
learning techniques in addressing various protection challenges.

\hspace{8mm} The power transformers and Phase Angle Regulators are modeled to simulate and analyze the transients accurately. Appropriate time and frequency domain features are selected using different selection algorithms to train the machine learning algorithms. The boosting algorithms outperformed the other classifiers for detection of faults with balanced accuracies of above 99\% and computational time of about one and a half cycles. The case studies on transmission lines show that the developed methods distinguish power swings and faults, and determine the correct fault zone. The proposed data-driven protection algorithms can work together with conventional differential and distance relays and offer supervisory control over their operation and thus improve the dependability and security of protection systems.

\pagenumbering{arabic} 
\pagestyle{myheadings}\makeatletter
\fancyhf{}
\fancyheadoffset{0cm}
\renewcommand{\headrulewidth}{0pt} 
\renewcommand{\footrulewidth}{0pt}
\fancyhead[R]{\thepage}
\fancypagestyle{plain}{
   \fancyhf{}
   \fancyhead[R]{\thepage}
}


\newpage
\setlength{\parindent}{9mm}
\chapter{Introduction}
\section{Overview}
Machine  Learning, today, represents a field of intensive research in various applications of control systems, computer vision, pattern recognition, financial trading, healthcare, and load forecasting, fault and failure analysis, demand-side management,
non-intrusive load monitoring, cyberspace
security, electricity theft detection, and islanding detection in power systems, and elsewhere. They have also been used for the purpose of protection of Power Transformers, Phase Angle Regulators (\acrshort{PAR}), and Transmission lines, and several such instances are documented in publications since 1990s \cite{ch1990ann,ch11992ann1,ch11992ann2}.

Power Transformers are one of the essential elements in power systems and their protection is a top priority. Predominantly differential relays are used in that process,  which involves comparing the primary current and secondary currents. Magnetizing inrush, overexcitation, and Current Transformer (\acrshort{CT}) saturation make the transformer protection a challenge. Magnetizing inrush occurs during the energization of transformers which sometimes results in a high current in the order of 10 times the full load current resulting in mal-operation of the relay. Overexcitation occurs when magnetic flux in the transformer core increases above the typical design level, resulting in a higher magnetizing current. In order to avoid such mal-operations, differentiating the transient disturbances from the fault currents are necessary. The second and fifth harmonic restraint concept is used to distinguish faults from magnetizing inrush and overexcitation. However, in certain cases (CT saturation, presence of parallel capacitance, or distributed capacitances), internal fault currents have a considerable amount of second and fifth harmonics \cite{chtx3pro}. Further, the use of low-loss amorphous core materials in modern transformers produce lower harmonic contents in inrush currents \cite{chmodern_core}. The inefficient setting of commonly used dual-slope biased differential relays may also result in maloperations in cases of CT saturation during external faults \cite{chctsat1}.

PARs are a special class of transformers that are used to control real power flow in networked power systems and make certain that the ratings of transmission equipment are not exceeded during contingency conditions. The performance of PARs affects the continuous and stable operation in a power system.  Generally, differential protection is used for PARs and its operation highly depends on appropriate analysis of the different electromagnetic transient events. Like in the case of the Power Transformers, discriminating external faults with CT saturation, magnetizing inrush, and other transient disturbances from internal faults is a challenge for the protection systems of PARs. Further, methods used to compensate the phase for differential relays in regular transformers with a fixed phase shift are not applicable in PARs with variable phase shift \cite{chpstguide}.

Transmission lines facilitate the movement of electrical power from generating stations to the consumers. The security and dependability of the Transmission line protection system are tested during power swings. Distance relays are predominantly used for the protection of high-voltage networks because of their selective and dependable tripping for line faults and simple time coordination of relays across the system. The relays trip with a predefined time delay when the impedance enters one of the protective zones as seen during faults. However, in the case of power swings, the impedance trajectory may also encroach the zones and the distance relay mal-operates. Mal-operation in the distance relays is one of the primary reasons for cascaded outages\cite{chphadke}.

\section{Motivation}
Many researchers have proposed the use of intelligent techniques to protect Power Transformers, PARs, and Transmission lines. The authors detected high-impedance faults in power distribution networks using Artificial Neural Networks (\acrshort{ANN}) in \cite{ch1990ann}.  Higher-harmonic components were evaluated with ANN in a power system in \cite{ch1992ann}. The authors of \cite{channtline1993} suggested the use of neural networks to achieve adaptive relaying protection of Transmission lines considering effect of fault resistance.
A feed-forward neural network (\acrshort{FFNN}) was proposed as an alternative method to discriminate between transformer magnetizing inrush and fault currents in a digital relay implementation \cite{ch1994ann}. ANN-based digital protection was proposed in \cite{ch1997ann} where reliable operation of the protection system was established in cases of inrush, inrush with simultaneous or slightly delayed short circuit, faults with second harmonic component, and partial saturation of current transformers. Again, in \cite{chann1998} the inrush and internal fault were distinguished using ANN on experimental and simulated data. In\cite{chann1}, magnetizing inrush, fault with CT saturation, and internal faults were classified using spectral energies of wavelet components and ANN. 
Probabilistic Neural Network (\acrshort{PNN}) has been used to detect different conditions in Power Transformer operation in \cite{chtripathy}.
Support Vector Machines (\acrshort{SVM}) and Decision Tree (\acrshort{DT}) based transformer protection were proposed in \cite{chSVM1,chSVM2} and \cite{chdt1,chdt2,chdtwt}, respectively. Random Forest Classifier (\acrshort{RFC}) was proposed to discriminate internal faults and inrush in \cite{chShahrfc}.

Literature investigating differential protection using ML methods in PARs are limited. However, attempts were made in \cite{chtencon} where internal faults are distinguished from magnetizing inrush using Wavelet Transform (\acrshort{WT}) and then the internal faults are classified using ANN and in \cite{chisspit} where the internal faults in series and exciting transformers of the Indirect Symmetric Phase Angle Regulator (\acrshort{ISPAR}) are classified using RFC.

Publications reporting the applications of ML methods to differentiate power swings from faults also exist. In \cite{chseethalekshmi}, SVMs are used to distinguish faults during power swing and voltage instability and then classify power swing and voltage instability using real power, reactive power, current, voltage, and delta and their changes as input features. An adaptive neuro-fuzzy inference system (\acrshort{ANFIS}) with inputs change of positive sequence impedance, positive and negative sequence currents, and power swing center voltages were used in \cite{chanfis}.

The early articles were received with considerable skepticism and still, there are doubts about the feasibility of Machine Learning (\acrshort{ML}) in practical implementations. However, the growing interest in the field of {Tiny Machine Learning} \footnote{TinyML is a platform that combines embedded Machine Learning (ML) applications, algorithms, hardware, and software. TinyML is distinct from traditional machine learning and involves both software and embedded hardware expertise.} which focuses on more energy-efficient computing is changing the status quo. The pervasiveness of ultra-low-power embedded devices, coupled with embedded ML frameworks will enable widespread use of AI-powered devices.

With the increasing penetration of large-scale power electronics devices including renewable generations interfaced with converters, low loss transformers, and non-linear loads there is an anticipated impact on the performance of traditional power system protection. 
ML-based solutions can be used to provide a holistic answer to the challenge of changing power system dynamics, as well as to support the operation of existing protection where it is found vulnerable. The proposed ML-based protection and classification techniques do not depend on the equivalent circuit of power system element (Power Transformer, PAR, or Transmission line) and the harmonic contents in the differential and relay currents, rather they make decisions based on the current signature. Therefore, for the protection of modern transformers and Transmission lines with unpredictable and changing harmonic components in the line currents, the ML-based fault detection and classification of transients method would be more effective. The methods are simple but robust and with the advent of high speed and dedicated microprocessors, their practical implementation is not a distant dream.

\begin{figure}[ht]
\centerline{\includegraphics[width=5.5 in, height= 2 in]{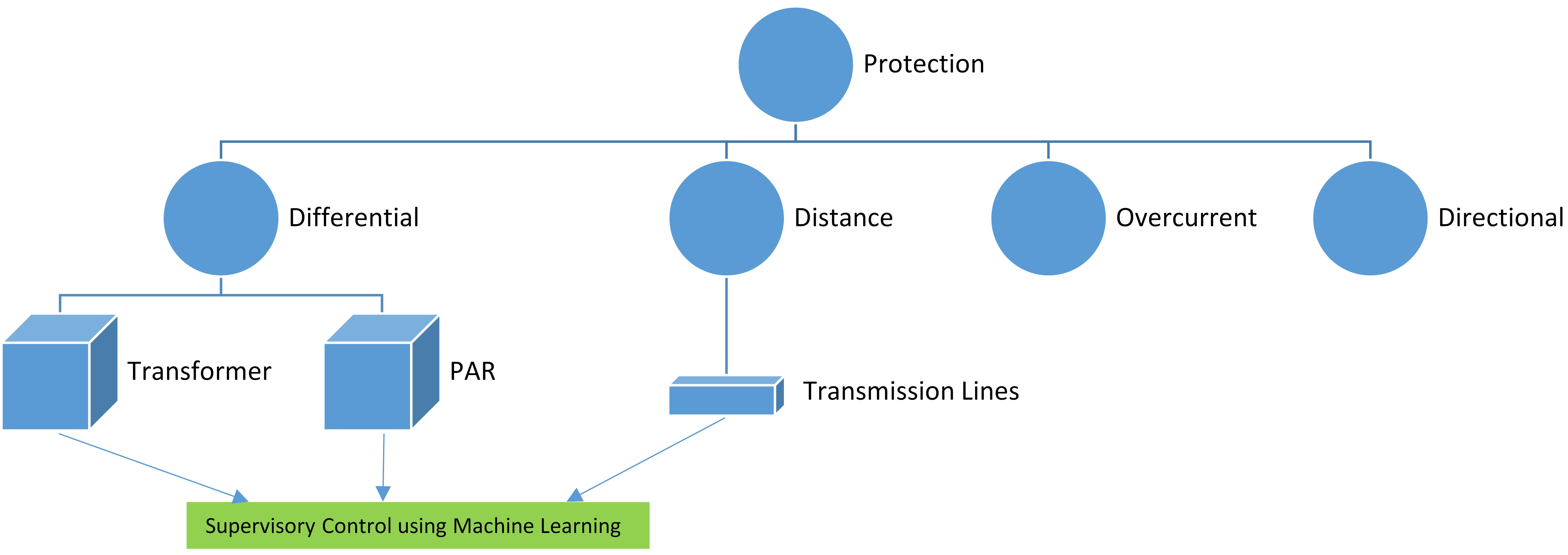}}
\caption{Types of protection and power system elements considered for supervisory control with ML algorithms }
\label{ch1chart}
\end{figure}

In the chapters that follow, the effectiveness and feasibility of different ML algorithms are evaluated for detecting and classifying different power system transients in Power Transformers, Phase Angle Regulators, and Transmission lines. Fig.\ref{ch1chart} shows the types of protection and power system elements considered for supervisory control with ML algorithms in the chapters to come.

\section{Main Contributions}
The research presents a comprehensive study of the various power system transients which includes faults, magnetizing inrush, sympathetic inrush, overexcitation, external faults with CT saturation, ferroresonance, non-linear load switching, capacitor switching, CT saturation, faults during inrush, series core saturation, and low current faults, among other things, while taking into account various traditional ML algorithms trained on an extensive list of 3-phase current features. The individual contributions of the research results presented in this dissertation are summarized as follows:
\subsection{Chapter 2}
\begin{itemize}
    \item Modeling of 2- and 3-winding transformers and modeling of ISPAR.
    \item Identification of relevant features to distinguish internal faults and other transients.
    \item Validation of proposed protection scheme under conditions such as fault during magnetizing inrush, saturation of series-winding, CT saturation, and presence of inverter interfaced wind turbine; and with different transformer ratings, tap positions, and noise levels.
\end{itemize}

\subsection{Chapter 3}
\begin{itemize}
    \item Modeling of 5-bus interconnected system with Phase Angle Regulators and Power Transformers and simulation of 101,088 transients cases.
    \item Identification of relevant features to distinguish internal faults and other transients.
    \item Validation of proposed protection scheme under CT saturation, different transformer ratings and connections, and noise levels.
\end{itemize}

\subsection{Chapter 4}
\begin{itemize}
    \item Study of uncertain operation of distance relay connected to the Wind Farm (\acrshort{WF}) during balanced faults.
    \item Development of protection scheme for Transmission lines connected to Type-3 Wind Turbine Generators (\acrshort{WTG}).
    \item Verification of the proposed protection scheme on different test systems and under various conditions.
\end{itemize}

\subsection{Chapter 5}
\begin{itemize}

\item Modeling and simulation of faults, faults during swing, and power swing cases in a 9-Bus Western Systems Coordinating Council (\acrshort{WSCC}) 3-machine system.  
\item Differentiate faults and faults during power swings from power swings and classify the power swings into stable and unstable swings. Thus, avoiding mal-operation of distance relays during faults during power swings and misoperation during unstable power swings ensuring the security and dependability of the protection system. 
\end{itemize}

\section{Organization of the Dissertation}
The dissertation has been organized into six chapters.
\subsection{Chapter 2: Differential Protection and Classification of Transients for Phase Angle Regulators}

Differential relays associated with Phase Angle Regulators mal-operate for several traditional and non-traditional transient conditions. This chapter explores the suitability of time and time-frequency feature-based estimators to distinguish internal faults from other transient conditions like overexcitation, external faults with CT saturation, and magnetizing inrush for ISPARs.  Two and three-winding transformer models are developed for creating the internal faults including inter-turn and inter-winding faults.
Subsequently, the faulty core unit (series or exciting) is located, and the transients are identified. Six well-known classifiers are trained on features extracted from one cycle of post transient 3-phase differential currents filtered by an event detector. Maximum Relevance Minimum Redundancy, Random Forest, and exhaustive search with Decision Trees are used to select the relevant wavelet energy, time-domain, and wavelet coefficient features respectively. The fault detection scheme trained on XGBoost classifier with hyperparameters obtained from Bayesian Optimization gives an accuracy of 99.8\%. The reliability of the proposed scheme is verified with varying tap positions, noise levels, and ratings; and under different conditions like CT saturation, fault during magnetizing inrush, series core saturation, low current faults, and integration of wind energy. As a potential application, the methodology can be deployed to supervise microprocessor-based differential relays to improve the security and dependability of the protection system.

\subsection{Chapter 3: Differential Protection of Power Transformers and Phase Angle Regulators}
This chapter solves the problem of accurate detection of internal faults and classification of transients in a 5-bus interconnected system for Phase Angle Regulators (PAR) and Power Transformers. The analysis prevents mal-operation of differential relays in case of transients other than faults which include magnetizing inrush, sympathetic inrush, external faults with Current Transformer (CT) saturation, capacitor switching, non-linear load switching, and ferroresonance.
A gradient boosting classifier (\acrshort{GBC}) is used to distinguish the internal faults from the transient disturbances based on 1.5 cycles of 3-phase differential currents registered by a change detector. After the detection of an internal fault, GBCs are used to locate the faulty unit (Power Transformer, PAR series, or exciting unit) and identify the type of fault. In case a transient disturbance is detected, another GBC classifies them into the six disturbances. Five most relevant frequency and time domain features obtained using Information Gain are used to train and test the classifiers. The proposed algorithm distinguishes the internal faults from the other transients with a balanced accuracy ($\bar{\eta}$) of 99.95\%. The faulty transformer unit is located with $\bar{\eta}$ of 99.5\%  and the different transient disturbances are identified with $\bar{\eta}$ of 99.3\%. {Moreover, the reliability of the scheme is verified for different ratings and connections of the transformers involved, CT saturation, and noise level in the signals.} These GBC classifiers can work together with a conventional differential relay and offer supervisory control over its operation.

\subsection{ Chapter 4: Distance Protection of Transmission Lines Connected to Wind Farms}
Distance relays mal-operate for Transmission lines connected to type-3 WFs. This chapter proposes a waveshape property-based protection of the intertie zone between WF and grid during 3-phase faults. It mitigates the challenges faced by the normally used distance relays and ensures the protection systems' security and dependability. The proposed scheme uses the autoregressive coefficients of the 3-phase currents obtained from the Current Transformer at one end to distinguish the faults fed by the type-3 WFs and the primary grid. The validity of the technique is verified on three test systems. The results obtained with different wind speeds, crowbar resistance, fault resistance, inception time, and fault locations are encouraging and suggest the possible utilization of feature-based algorithms to improve the power system distance relaying system.

\subsection{Chapter 5: Identification of Stable and Unstable Power Swings}
Faults during symmetrical power swings cause mal-operation of distance relay. The undesired operation also occurs during unstable power swings causing uncontrolled islanding. Faster detection of faults during power swings and classification of power swings can assist the protection system in making reliable decisions on blocking or unblocking a relay's operation. This chapter segregates the faults, faults during power swing from power swings in one cycle with an accuracy of 99.3\%. It then identifies the different power swings in 10 cycles that occur in a 9-bus WSCC system. Support Vector Machines (SVM), Decision Tree (DT), and k-Nearest Neighbor (kNN) classifiers are trained and tested on six features obtained from 3-phase(ph) relay voltage and current to test the validity of the detection and classification scheme. 

\subsection{Chapter 6: Conclusion and Future Work} This chapter summarizes the main contributions, findings, and results of this dissertation; and presents the concluding remarks. Directions and ideas for future research related to the protection of micro-grids are also presented.


    
\newpage

\chapter{Differential Protection and Classification of Transients for Phase Angle Regulators}


\section{Introduction}\label{ch2sec:introduction}
Phase Shift Transformers or Phase Shifters or Phase Angle Regulators (PARs) control the steady-state power flow in parallel transmission lines and sometimes connect two independent grids. They ensure that contingency conditions do not exceed the ratings of transmission equipment. 
Their performance affects the continuous and stable operation of the power system. With a lower successful operating rate than the transmission lines, transformer protection systems are challenged under various conditions. Internal faults are electrically detected in a transformer mainly with differential, overcurrent, and ground fault relays. Differential relays detect and clear faults faster and locate them accurately. In general, electromechanical, solid-state, analog, and microprocessor-based relays are used as differential relays. Predominantly, differential relays are used to protect the standard and non-standard transformers, and their operation highly depends on appropriate analysis of different electromagnetic transient events \cite{chtransformerguide}.

\section{Background}
Electrical power transfer between two points changes when the phase difference ($\delta$) between the sending end voltage ($V_S$) and the receiving end voltage ($V_L$) is changed. PARs are used to get this required phase difference between $V_S$ and $V_L$. They can control the active power flow in branches in meshed networks and connect two otherwise independent grids in a power system by changing the phase angle ($\delta$) \cite{harlow}. 

PAR inserts a variable quadrature voltage to line to neutral voltage of source which is derived from the phase-to-phase voltages of the remaining two phases and thereby realizing the required phase shift ($\alpha$). Phase angle shift for each phase is obtained by inserting a quadrature voltage derived from the other two phase voltages. The phase angle shift is varied by changing the magnitude of the quadrature voltage which is introduced between the sending and receiving end voltages with the help of a series transformer. The modified real power flow in a line with a PAR is given by
\begin{equation}
    P=\frac {V_S \times V_L}{X_{line}+ X_{PAR}} \times sin(\delta + \alpha)
\end{equation} where, $\delta$ is the phase angle difference between $V_S$ and $V_L$;  $X_{line},X_{PAR}$ are the transmission line and PAR reactance respectively; and $\alpha$ is the new constraint added which is responsible for controlling the power flow.

PARs using on-load tap changers were first introduced in the 1930s to solve power flow problems. Since then it has been an integral component of the power systems. They are classified on the basis of the number of magnetic cores and on the basis of the magnitude of source-side voltage with respect to load-side voltage.  Direct PARs consist of one 3-phase transformer. The transformer windings are connected in a specific manner to get the required phase shift. Indirect PARs consist of two separate 3-phase transformers; one exciter transformer with variable tap to regulate the amplitude of the quadrature voltage and one series transformer to inject the quadrature voltage in the required phase. Asymmetrical PARs give an output voltage with a different phase angle and amplitude compared to the input voltage. Symmetrical PARs give an output voltage with a different phase angle compared to the input voltage, but with the same amplitude. The combination of these results in four different groups of PAR:
\begin{itemize}
 \renewcommand{\labelitemi}{\scriptsize$\blacksquare$}
 \item Direct Symmetrical 
  \renewcommand{\labelitemi}{\scriptsize$\blacksquare$}
 \item Indirect Symmetrical 
  \renewcommand{\labelitemi}{\scriptsize$\blacksquare$}
 \item Direct Asymmetrical 
   \renewcommand{\labelitemi}{\scriptsize$\blacksquare$}
 \item Indirect Asymmetrical 
\end{itemize}
In two-core Symmetric PARs or Indirect Symmetrical PAR (\acrshort{ISPAR}) the secondary winding of the series unit is connected across the exciting unit secondary phase to phase voltages. The exciting unit secondary phase to phase voltage is manifested in the primary of the series transformer and is added or subtracted from the source side primary voltage to obtain the desired angular shift ($\alpha$) of the load side primary voltage. The magnitude of the required quadrature voltage ($\Delta V$) to obtain the required phase shift is given by
\begin{equation} \Delta V= V_{ph} \cdot 2 \cdot sin(\frac{\alpha}{2})
\end{equation}
In comparison to reactive compensators, PARs bring a new dimension to the control of dynamic events, the capability to exchange real power \cite{hingorani}. PARs enable cost-effective load flow management and grid asset optimization
in complex grids. On one hand, real power is controlled by quadrature voltage injection via phase shift and on the other reactive power can be controlled with in-phase voltage injection by voltage regulators (on-load tap changers). Nowadays, voltage phase angle regulators with fast electronic control can also handle dynamic system events such as improving the transient state stability, damping out power oscillations (when $\frac{d\delta}{dt}>0$ phase shift is made negative and during $\frac{d\delta}{dt}<0$ it’s made positive), minimize post-disturbance overloads and corresponding dips in voltage. 

Power flow can be enhanced in the line by increasing the voltage at one end or both ends of the line. But it has a larger impact on the reactive power flow than on the real power flow and is also constrained by the insulation requirements. Hence, unsymmetrical PAR is seldom used as compared to symmetrical PAR. ISPARs have the same source- and load-side voltages with two cores: series and exciting (Fig.\ref{ch2par23}A). They are the conventionally used PARs with higher security against high voltage levels. To regulate power flow, the exciting unit creates the required phase difference through the load tap changer (\acrshort{LTC}), and the forward/backward transition can be achieved in the series secondary with an advance-retard-switch or change-over selectors in the exciting secondary \cite{chibrahim}. Taking into account the high repair and replacement cost and to limit further damages, the PARs require a sensitive, secure, and dependable protection system. Maintaining dependability for in-zone faults and security against no-fault conditions is a challenge.

\section{Motivation}
Differential protection, being the foremost for standard and non-standard transformers, however, suffers from traditional challenges of unwanted tripping in situations of magnetizing inrush, external faults with CT saturation, and overexcitation. These problems are addressed by {current-based} methods in two ways: using harmonic restraint and waveshape identification methods \cite{chwaveshape}. The changing complexity and operating modes in the power system have threatened the reliability of these methods. Percentage differential relay with restraint actuated by restraining current and/or harmonic components of operating current is generally used in differential schemes. The second harmonic component identifies magnetizing inrush, and the fifth identifies overexcitation. The second harmonic restraint method \cite{chharmonic} used to detect magnetizing inrush may fail because of lower second harmonics in transformers with modern core \cite{chmodern_core}. Moreover, the protection system's sensitivity is compromised due to higher second harmonics during internal faults with CT saturation and the presence of distributed and series compensation capacitance \cite{chtx3pro}. The fifth harmonic restraint may also fail in case of internal faults during overexcitation. Use of fourth harmonic with second in case of inrush and adaptive fifth harmonic pick up in case of overexcitation improves the security, yet the challenges persist. External faults with CT saturation may also cause false trips if the settings of the {dual-slope} current differential relays are not set effectively \cite{chctsat1}. Differential relays also fail to detect ground faults near-neutral of grounded wye-connected transformer winding \cite{chtransformerguide}.

Besides the traditional challenges associated with transformer differential protection, high sensitivity to detect turn-to-turn (\acrshort{t-t}) and winding phase-to-ground faults, and security against series winding saturation are specific to PARs \cite{chukhan}. Also, the phase compensation techniques used in standard differential protection with fixed phase shifts cannot be applied for the compensation of the phase shift across the PARs with a non-standard phase shift \cite{chpstguide}\cite{chukhan2}. Consequently, special considerations are required while designing their protection system.

Two-element-based differential protection is proposed in \cite{chibrahim2} which performs well for internal faults and series saturation, although it suffers from other traditional and PAR specific challenges.
Phase/magnitude compensation is proposed to address the non-standard phase shift in \cite{chzgajic}. However, it requires tracking the tap positions and has a lower sensitivity for low current faults. Reference \cite{chkasztenny} proposes differential protection, which does not need the knowledge of tap positions. But it applies to hexagonal PARs only. Reference \cite{chukhan} proposes directional comparison-based protection, which provides overall protection addressing various challenges; however, it needs both current and voltage information to function. The present work attempts to provide an alternative and complete solution to the conventional and non-conventional protection challenges associated with a PAR using Machine Learning (ML).

Data Mining and ML-based methods which do not require predefined threshold values and mathematical models  have been proposed
to distinguish faults and disturbances in transformers in the last two decades \cite{chaiml}. 
Neural Networks (\acrshort{NN}) \cite{chann1}\cite{channdtwt}, Support Vector Machines (SVM) \cite{chsvm1mag} \cite{chSVM2}, Decision Tree (DT) \cite{channdtwt} \cite{chdtwt}, k-nearest neighbor (\acrshort{kNN}) \cite{chknnwt}, and Random Forest (\acrshort{RF}) \cite{chShahrfc} are some of the popular algorithms that have been used for differential protection of transformers. 
Although several such studies exist in transformer protection, this problem is insufficiently explored for PARs. Few literatures have considered using ML to detect faults and other transients in PARs. In \cite{chtencon} internal faults were differentiated from inrush currents using Wavelet Transform and classified with NN. 

\section{Contribution}
This chapter studies the suitability of time, and time-frequency domain features to discriminate faults from transient disturbances like magnetizing inrush, external faults with CT saturation, overexcitation for a PAR. The ISPAR is modeled in Power System Computer Aided Design (\acrshort{PSCAD})/ Electromagnetic Transients including DC (\acrshort{EMTDC}) using 2- and 3-winding transformers to simulate the transients. A series of time and wavelet features are extracted and then selected using feature selection algorithms. Six classifiers trained and tested on 60552 transient cases simulated by changing the system parameters demonstrate the proposed scheme's validity. The stability of the scheme is also tested during conditions such as fault during magnetizing inrush, saturation of series-winding, CT saturation, and addition of an inverter interfaced wind turbine; and with different transformer ratings, tap positions, and noise levels.

\section{Chapter Organization}
The remainder of the chapter is arranged in the following order. The 2- and 3-winding single phase transformer fault models are developed, and the transient events in the ISPAR are modeled and simulated in Section \ref{ch2sec2}. Section \ref{ch2proposed_scheme} presents the proposed differential protection scheme that includes the event detection, extraction and selection of features, and introduces the six classifiers. The performance of the classifiers for detection of faults, localization of faulty units, and classification of transients are presented in Section \ref{ch2results}. Section \ref{ch2impacts} includes the assessments for various non-conventional challenges that the PAR may encounter. The last section concludes the chapter. 

\section{Modeling and Simulation}\label{ch2sec2}
\begin{figure}[ht]
\centerline{\includegraphics[width=4.7in, height= 3.0in]{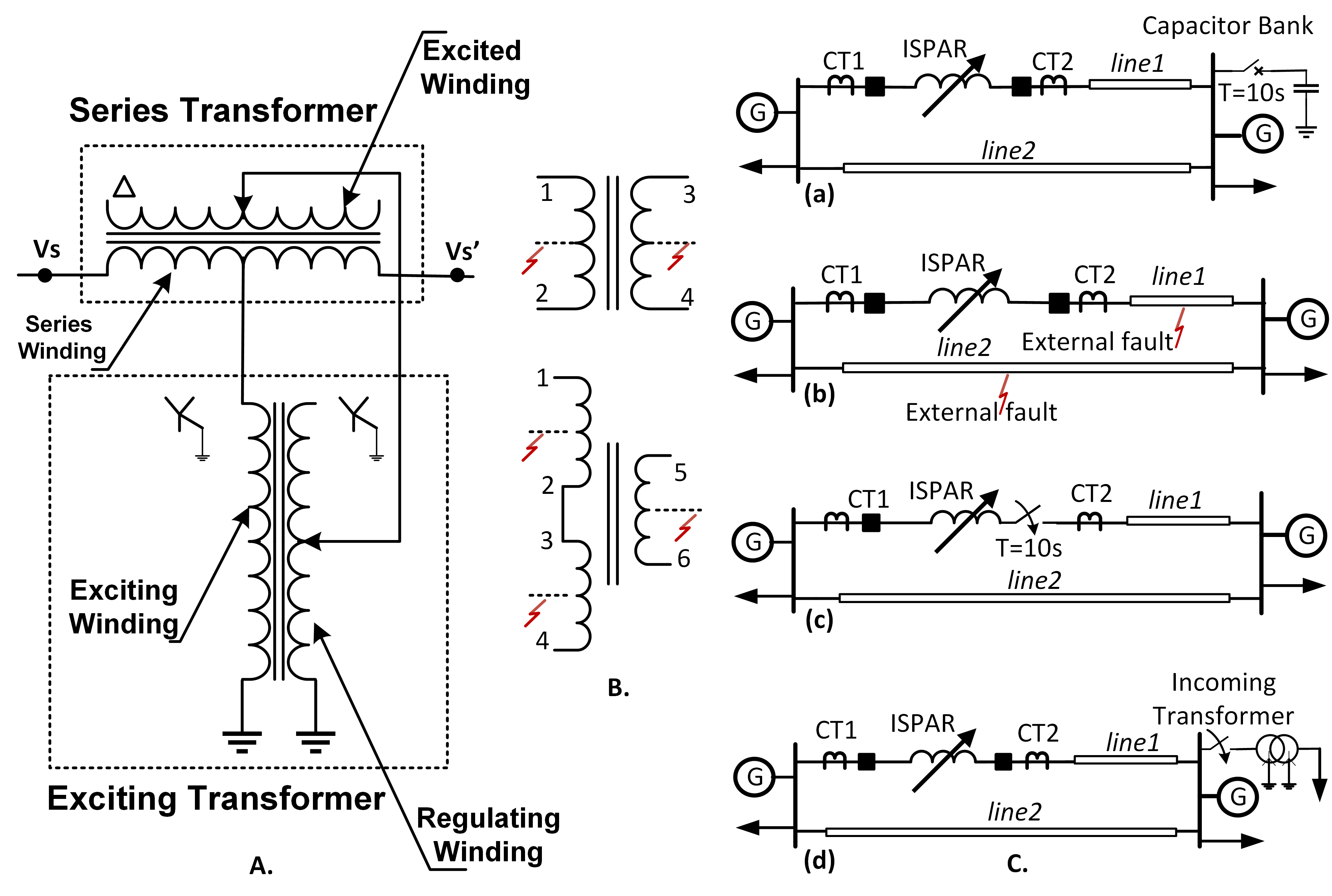}}
\caption{ (A) ISPAR model, (B) 2-\& 3-winding transformer fault model, (C) Simulation models for overexcitation, external fault with CT saturation, magnetizing inrush, and sympathetic inrush (top to bottom) }
\label{ch2par23}
\end{figure}

PSCAD/EMTDC is used to model the ISPAR and simulate the electromagnetic transients.
The rating of the ISPAR are: $S_n$=500MVA, $V_n$=230kV, maximum phase shift = $\pm25^{\circ}$. CT1 and CT2 are located on the two sides of the PAR. The fault model of ISPAR is not available in most simulation software. The single-phase 2-winding transformer fault model needed for faults in the exciting unit and the single-phase 3-winding transformer fault model needed for faults in the series unit (Fig.\ref{ch2par23}B) are designed in PSCAD/EMTDC with Fortran. The voltage-current relationship for the four-coupled coils of the 2-winding transformer and the six-coupled coils of the 3-winding transformer are described in equation\ref{matrix} and equation\ref{ch2matrix}. The self inductance (\textit{Li}) and mutual inductance (\textit{Mij}) of the 4$\times$4 matrix of the 2-winding transformer in equation\ref{matrix} and \textit{Li} and \textit{Mij} of 6$\times$6 matrix of the 3-winding transformer in equation\ref{ch2matrix} are computed from the voltage ratios, reactive part of the no-load current ($I_m$), and short-circuit tests.
The saturation characteristics, percentage of turns faulted, and other parameters can be changed in the developed 2-and 3-winding transformers. The Appendix Section includes the Fortran script for the 1-phase 2-winding transformer and the 1-phase 3-winding transformer.


\begin{equation}\label{matrix}
\footnotesize
\begingroup 
\setlength\arraycolsep{1.7pt}
\begin{bmatrix}V1 \\ V2\\ V3\\V4
\end{bmatrix}=
\begin{bmatrix}
L1 & M12 & M13& M14\\
M21 & L2 & M23& M24\\
M31 & M32 & L3& M34\\
M41 & M42 & M43& L4\\

\end{bmatrix}
\cdot\dfrac{d}{dt}
\begin{bmatrix}I1 \\ I2\\I3\\I4
\end{bmatrix}
\endgroup
\end{equation}

\begin{equation}\label{ch2matrix}
\footnotesize
\begingroup 
\setlength\arraycolsep{1.7pt}
\begin{bmatrix}V1 \\ V2\\ V3\\V4\\V5\\V6
\end{bmatrix}=
\begin{bmatrix}
L1 & M12 & M13& M14&M15&M16\\
M21 & L2 & M23& M24&M25&M26\\
M31 & M32 & L3& M34&M35&M36\\
M41 & M42 & M43& L4&M45&M46\\
M51 & M52 & M53& M54&L5&M56\\
M61 & M62 & M63& M64&M65&L6\\
\end{bmatrix}
\cdot\dfrac{d}{dt}
\begin{bmatrix}I1 \\ I2\\I3\\I4\\I5\\I6
\end{bmatrix}
\endgroup
\end{equation}

In the present analysis, the internal faults, overexcitation, external faults with CT saturation, and magnetizing and sympathetic inrush conditions for ISPAR are considered. These scenarios are studied successively in the sections that follow. In the simulations, the total run-time is 10.2s, switching time (\acrshort{ST}) is 10.0s, and the duration of faults is 0.05s (3 cycles). The multi-run component in the master library is employed as needed during the simulations.

\subsection{Internal Faults}
The internal faults are simulated in primary (P) and secondary (S) sides of exciting and series units in the ISPAR. They include the faults occurring inside the enclosure and inside the CT locations. They are usually caused by insulation breakdown and require faster action by protective relays to limit the extent of the damage.
The basic internal faults include short circuits and phase (\acrshort{ph}) faults, \textit{t-t}, and winding-winding faults. 46872 faults are simulated by varying the percentage of turns shorted (\acrshort{PTS}), fault resistance (\acrshort{FR}), faulty unit (\acrshort{FU}), fault type (\acrshort{FT}), fault inception time (\acrshort{FIT}), phase shift (\acrshort{PS}): forward \& backward, and the PAR tap positions.

\textit{Phase \& ground faults:}
These include winding ph-g faults (a-g, b-g, c-g), winding ph-ph-g faults (ab-g, ac-g, bc-g), winding ph-ph faults (ab, ac, bc), 3-ph and 3-ph-g faults. The values of different parameters of the ISPAR used to simulate 33264 instances are shown in Table \ref{ch2tab_ww_tt}a.

\begin{table}[ht]
\renewcommand{\arraystretch}{1}
\footnotesize
\centering
\caption{Parameters: (a) internal ph and g faults, (b) \textit{t-t}
and \textit{w-w} faults}\label{ch2tab_ww_tt}
\subfloat[{\vspace{-1mm}}]{\begin{tabular}{ll}\toprule
\textit{Variable}  & \textit{Values} \\ \midrule
\textit{FR} &  0.01, 0.1 \&   1 $\Omega$     (3)   \\
\textit{PTS} & 20\%, 50\%, 70\%  (3)\\
\textit{FT}     & lg, llg, ll, lll \& lllg (11) \\
\textit{FIT}      & 10s to 10.0153s  (12)  \vspace{0.5mm}\\
\textit{FU}          &     \begin{tabular}{@{\extracolsep{\fill}}l}Exciting (P \& S) (2)\\ \& Series (P \& S) (2) \end{tabular}  \vspace{0.5mm} \\
\textit{PS} & forward \& backward  (2)\\ 
\textit{tap}  & .2,.4,.6,.8,1[1\&0.5 in exciting] \\ \hline
\end{tabular}}
\quad
\renewcommand{\arraystretch}{1}
\subfloat[{\vspace{-1mm}}]{\begin{tabular}{ll}\toprule
\textit{Variable}  & \textit{Values} \\ \midrule
\textit{FR} &  0.01, 0.5 \&   1 $\Omega$     (3)   \\
\textit{PTS} & 20\%, 50\%, 70\% (3)\\
\textit{FIT}       & 10s to 10.0153s (12)  \vspace{0.5mm}\\
\textit{FU}          &     \begin{tabular}{@{\extracolsep{\fill}}l} Exciting ph A,B,C (P \& S) (6)\\ \& Series ph A,B,C (P \& S) (6) \end{tabular}  \vspace{0.5mm} \\
\textit{PS} & forward \& backward  (2)\\ 
\textit{tap}  & .2,.4,.6,.8,1 [1\&0.5 in exciting] \\ \hline
\end{tabular}}
\end{table}

\textit{Turn-to-turn (\textit{t-t}) faults:} 
Insulation failures are responsible for a major percentage of faults in a transformer. The insulation degrades over time with thermal, electrical, and mechanical stresses causing \textit{t-t} faults which can develop into serious faults if go undetected \cite{chturn}. They are challenging to detect, particularly when the PTS is low. The values of different parameters resulting in 9072 cases are displayed in Table \ref{ch2tab_ww_tt}b.

\textit{Winding-to-winding (\acrshort{w-w}) faults:}
The electrical, thermal, and mechanical stress due to short circuits and transformer aging reduces the mechanical and dielectric strength of the winding and results in degradation of the insulation between LV and HV winding and may damage the winding eventually \cite{chturn}. The values of different parameters used to obtain 4536 cases are listed in Table \ref{ch2tab_ww_tt}b.

\subsection{Overexcitation}
Faults due to over fluxing develop slowly and cause deterioration of insulation and may lead to major faults. They cause heating and vibration and can damage the transformer \cite{choverexc}. Since it is difficult for differential protection to control the amount of overexcitation a transformer can tolerate, tripping of the differential element during overexcitation is undesirable. Generally, the 5th harmonic restraint is used to restrain the operation of differential relays \cite{ch5th}. Several conditions may lead to overexcitation in electrical systems. Here, two such situations have been modeled: overvoltage during load rejection and capacitor switching (Fig.\ref{ch2par23}C). The typical differential current waveforms for these are shown in Fig.\ref{ch2ext}a and Fig.\ref{ch2ext}b. Parameter values are listed in Table \ref{ch2inrush}a.
\begin{table}[h]
\renewcommand{\arraystretch}{1}
\footnotesize
\centering
\caption{Parameters: (a) Overexcitation, (b) Magnetizing inrush and Sympathetic inrush}
\vspace{-1mm}
\label{ch2inrush}
\subfloat[{\vspace{-1mm}}]{\begin{tabular}{ll}\toprule
\textit{Variable}  & \textit{Values} \\ \midrule
\textit{switch} & load (3) \& capacitor (3)\\
\textit{ST  }     & 10s to 10.0153s (12)     \\
\textit{tap}  & 0.2, 0.4, 0.6, 0.8, 1 (5)                       \\
\textit{PS} & forward \& backward (2)\\ \midrule
\multicolumn{2}{l}{Total=6$\times$12$\times$5$\times$2=720}\\
\hline
\end{tabular}}
\quad
\subfloat[{\vspace{-1mm}}]{\begin{tabular}{ll}\toprule
\textit{Variable}  & \textit{Values} \\ \midrule
\textit{RFD} & $\pm80,\pm60,\pm40,0\% $ in 3-phs (21)\\
\textit{ST}      & 10s to 10.0153s (12)     \\
\textit{tap}  & 0.2, 0.4, 0.6, 0.8, 1 (5)                       \\
\textit{PS} & forward \& backward (2)\\ \midrule
\multicolumn{2}{l}{Total = 21$\times$12$\times$5$\times$2 = 2520}\\
\hline
\end{tabular}}
\end{table}

\subsection{External fault with CT saturation}

\begin{figure}[ht]
\centerline{\includegraphics[width=3.5 in, height= 3.2 in]{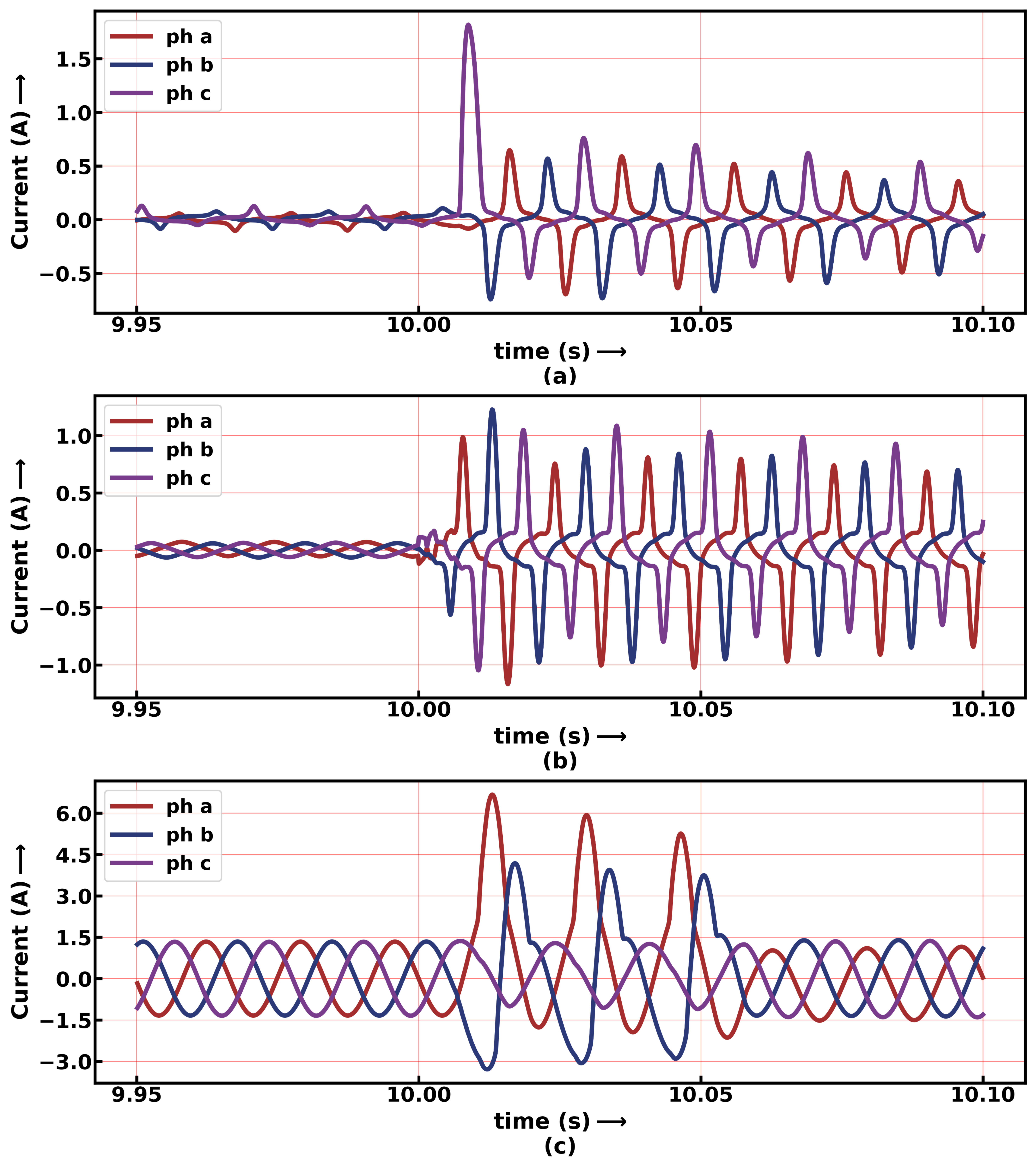}}
\caption{Differential currents for (a) load rejection, (b) capacitor switching, (c) CT saturation during external faults}
\label{ch2ext}
\vspace{2mm}
\end{figure}

External short circuit stresses the PAR and reduces the transformer life.
The differential currents become non-zero due to CT saturation in case of heavy through faults and may lead to a false trip \cite{chctext}. While raising the bias threshold ensures security (i.e. no mal-operation), the dependability for in-zone resistive faults gets reduced. 
The fault location (\acrshort{FL}) is varied while simulating these cases (Fig.\ref{ch2par23}C) besides FR, FT, FIT, tap position, and PS (Table \ref{ch2externaltab}). {Fig.\ref{ch2ext}c shows the differential current for an external \textit{lg} fault with PS=forward, FIT=10.0083s, FL=line1, and FR=1$\Omega$ at full tap.}

\begin{table}[ht]
\renewcommand{\arraystretch}{1}
\footnotesize
\centering
\caption{Parameters for External faults on \textit{line1 \& line2}}
\label{ch2externaltab}
\begin{tabular}{ll}\toprule
\textit{Variable } & \textit{Values} \\ \midrule
\textit{FR} & 0.01, 0.1 \&   1 $\Omega$     (3)   \\
\textit{FT}     & lg, llg, ll, lll \& lllg in 3 phs (11) \\
\textit{FIT}     & 10s to 10.0153s in steps of 1.38ms (12)    \\
\textit{tap}  & 0.2 to full tap in steps of 0.2 (5)\\
\textit{PS} & forward \& backward  (2)\\ 
\textit{FL}& line1 \& line2 (2)\\ \midrule
\multicolumn{2}{l}{Total = $3\footnotesize{\times}11\footnotesize{\times}12\footnotesize{\times}5\footnotesize{\times}2\footnotesize{\times}2$ = 7920}\\
\hline
\end{tabular}{}
\end{table}


\subsection{Magnetizing inrush} 
\begin{figure}[ht]
\centerline{\includegraphics[width=2.1 in, height= 1.7 in]{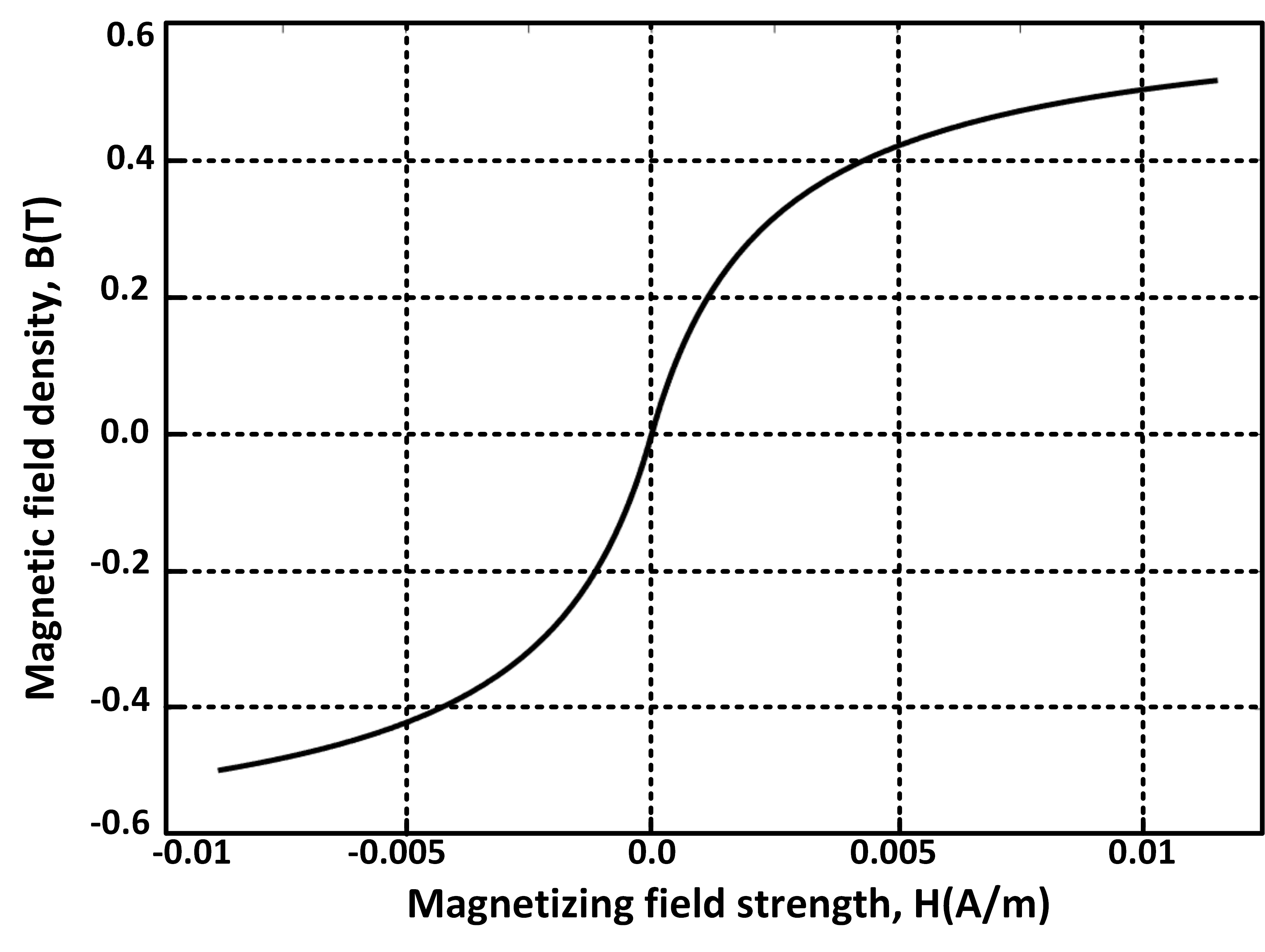}}
\caption{B-H curve of exciting transformer unit}
\label{ch2bh}
\end{figure}

\begin{figure}[ht]
\centerline{\includegraphics[width=3.5 in, height= 2.6 in]{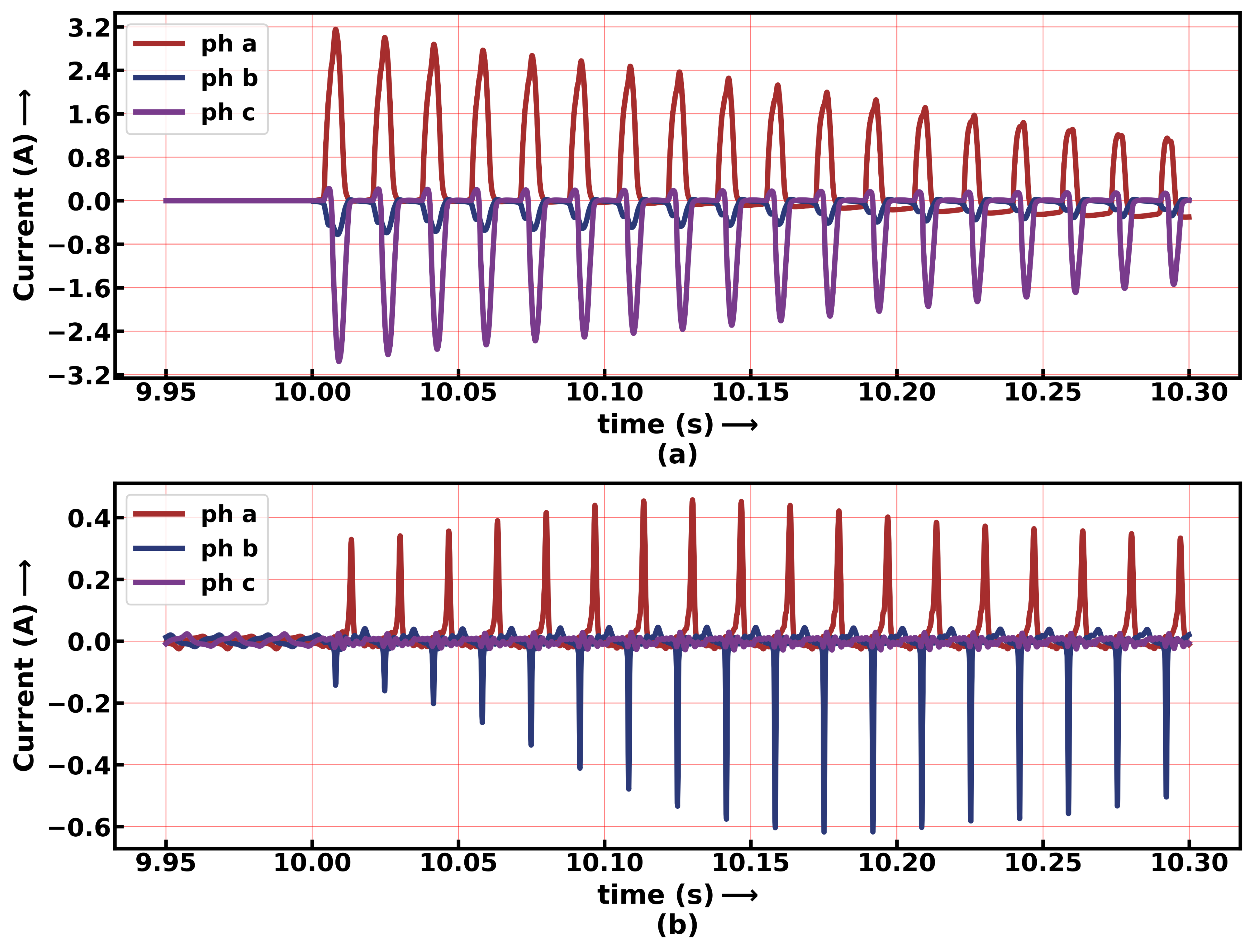}}
\caption{3-phase differential currents for (a) Magnetizing inrush, and (b) Sympathetic inrush}
\label{ch2symmag}
\vspace{1mm}
\end{figure}

Transients caused by the energization of transformers are common. Discriminating inrush from fault currents has been studied since the 19th century. 
When a transformer is energized, a high inrush current of the order of 10-15 times of normal current flows because of the saturation of the transformer core. 
Second harmonic restraint relays may fail to detect inrush currents in modern transformers having high flux density.
The flux in a transformer core just after switching is expressed as:
\begin{equation}\label{ch2eq_mag}
\Phi=\Phi_R+\Phi_m cos\omega t'-\Phi_m cos\omega(t+t')
\end{equation}
where $\Phi_R$, $\Phi_m$, and t$'$  representing the residual flux density (RFD),   the maximum flux, and the switching time (ST) respectively are the important parameters \cite{chmag}. 
The transformer draws a high peaky non-sinusoidal current to meet the high flux demand when switched on. Since this current flows only on one side of the transformer the differential scheme mal-operates.
 DC sources are used to get the desired $\phi_R$ in the single-phase transformers. The values for the DC currents in phase-a, b, and c are obtained from the x-coordinates of the B-H \footnote{B-H curve represents the curve characteristic of the magnetic properties of a material or element or alloy. It describes how a material reacts to an external magnetic field and is critical information for magnetic circuit design.} curve (Fig.\ref{ch2bh}). Table \ref{ch2inrush}b shows the values of RFD, ST, PAR taps, and PS used to obtain the 2520 cases. Fig.\ref{ch2symmag}(a) shows typical differential currents for a magnetizing inrush with tap=full, ST=10s, PS=forward, and RFD =0 in all phases. The exciting transformer unit in the ISPAR is considered to be responsible for the inrush currents\cite{chpstguide}.
 

\subsection{Sympathetic Inrush} Sympathetic inrush can occur when a transformer is switched on in a power network with already {energized} PARs  (Fig.\ref{ch2par23}C). {The flux change per cycle which drives the PAR to saturation is given by equation (\ref{ch2eq_sym}).
\begin{equation}\label{ch2eq_sym}
\Delta\Phi=\int_{t}^{2\pi + t}[(R_{sys}+ R_{par})i_1 +R_{sys}\cdot i_2]
\end{equation}} 
\noindent where $R_{sys}$ = system resistance , $R_{par}$ = resistance of PAR, and $i_1$ and $i_2$ are magnetizing currents of PAR and the incoming transformer \cite{chsym}. This interaction between the incoming transformer and the PAR may lead to failure of the harmonic restraint relays and may cause prolonged harmonic over-voltages \cite{chsym2}. Some factors responsible for such mal-operations are: cores with soft magnetic material, application of superconducting technology in windings, and CT partial transient saturation \cite{chmodern_core}\cite{chctsat}.  Sympathetic inrush is influenced by the residual flux ($\phi_R$) of the incoming transformer, switching time (t$'$), and the system resistance \cite{chkumbhar}. It can happen with the incoming transformer energized in series or parallel. Here the incoming transformer is energized at t=10s and the values of $\Phi_R$ and t$'$ are varied (See Table \ref{ch2inrush}b). 
Fig.\ref{ch2symmag}(b) shows the differential currents for tap=0.2, ST =10.0069s, PS = backward, and no RFD. 

{The differential currents obtained from the 60552 transient cases of internal faults, overexcitation, external faults, and inrush currents simulated in this section will be preprocessed to obtain the relevant time and time-frequency features and used as inputs to classifiers for detection and classification of the transients in the succeeding sections.} The entire dataset is available on IEEE Dataport \cite{chdata2}.

\begin{figure}[ht]
\centerline{\includegraphics[width=4.5 in, height= 3.0 in]{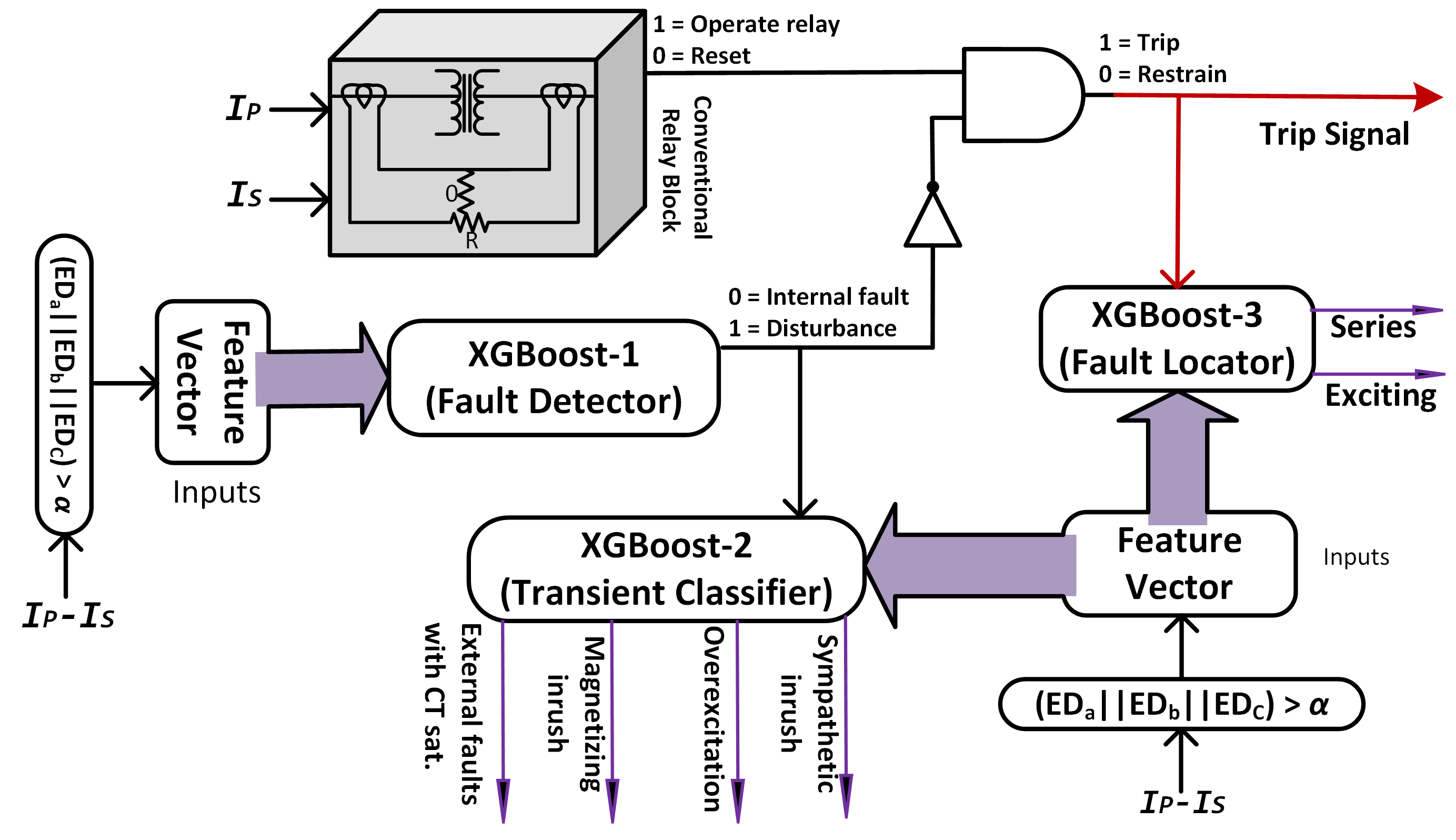}}
\vspace{0mm}\caption{Flowchart for fault detection and localization }
\label{ch2flowchart}
\end{figure}

\section{{Proposed PAR Differential Protection}}\label{ch2proposed_scheme}
Fig.\ref{ch2flowchart} depicts the time and time-frequency domain based proposed protection and classification scheme having three applications: detection of internal faults (\acrshort{DIF}), localization of faulty unit (\acrshort{LFU}), and classification of transient disturbances (\acrshort{CTD}).
The event detector {(\acrshort{ED})} detects the change in the differential currents ($I_P$-$I_S$) if the ED index in any phase is more than the threshold, $\alpha$ and registers one cycle of post transient 3-phase differential currents. These currents are preprocessed to obtain detailed wavelet-coefficients (\acrshort{WC}), wavelet-energy (\acrshort{WE}), and time-domain (\acrshort{TD}) features.
The proposed scheme can be seen as a design having three classifiers. The fault detector is the first classifier (Xgb-1). 
It recognizes the internal faults with \say{0} and transient disturbances with \say{1}. Thus, Xgb-1, together with the NOT gate regulates the operation of the trip/restrain function block by obstructing the transient disturbances and allowing internal faults. 
The transient classifier is the second classifier (Xgb-2), which examines an event further if the output of Xgb-1 is \say{1}.
It can identify the disturbance responsible for faulty operation of the conventional relay block (\acrshort{CRB})
(Xgb-1 is \say{1} \& CRB is \say{1}). 
The fault locator is the third classifier (Xgb-3). It locates the defective transformer core unit: series or exciting (Xgb-1 is \say{0} \& CRB is \say{1}).

\subsection{Event Detection}

The differential currents become non-zero when a power system transient occurs. {The ED which detects this change and computes the fractional increase between cumulative sum of modulus of samples of two successive cycles is defined by equation (\ref{ch2eq_ed}).}
\begin{equation} \label{ch2eq_ed}ED (t) = \frac{\sum_{i=t}^{n_c+t}|Id_\phi(i)|-\sum_{i=t}^{n_c+t}|Id_\phi(i-n_c)|}{\sum_{i=t}^{n_c+t}|Id_\phi(i)|}
\end{equation}
where $n_c$ is number of samples in one cycle, 
$Id$ is differential current, $\phi$ denotes the 3-phases, and i is the sample number starting at the second cycle.
{The 3-phase differential current samples are recorded by the ED filter from the time instant:  \begin{equation}\label{ch2ed}
 ED(t) \geq \alpha = 0.05   
\end{equation} in any of the three phases.} In the absence of transients, ED(t) values are negligibly small \cite{chDharmapandit2017}. These recorded samples are used for the feature extraction.

\subsection{Feature Selection Methods}\label{ch2}
The success of any classification algorithm highly depends on the input features. Feature selection is critical in reducing the classification error. Given a dataset with features X=\{$x_j$; j=1,..,N\} and target \textit{y}, feature selection obtains a subset of \textit{S} features from the N-dimensional space to distinguish \textit{y}, boosting the interpretability and reliability of predictions, and reducing the time complexity.

\subsubsection{Maximum Relevance Minimum Redundancy (\acrshort{mRMR})}
Feature selection methods based on mutual information, F-test select the top features without considering the relationship among the selected features. They calculate the mutual information as a score between the joint distribution of all features \textit{($x_i$)}, and target \textit{y} and select the features with the largest score. However, the selected features might be correlated and not cover the whole space. mRMR penalizes a feature's relevancy using the mutual information score by its redundancy when other features are also present. It searches for features, \textit{S} satisfying {equation (\ref{ch2maxrel})} to select the features with highest mutual information \textit{I($x_i$;y)} to target variable \textit{y} and satisfying  {equation (\ref{ch2minred})} to reduce the redundancy of the features selected using maximum relevance {(equation (\ref{ch2maxrel})) }\cite{chmrmr}.
\begin{equation} \label{ch2maxrel}
max D(S,y), D= \frac{1}{|S|}\sum_{x_i\in S}I(x_i;y)
\end{equation}
\begin{equation} \label{ch2minred}
min R(S), R= \frac{1}{|S|^2}\sum_{x_i,x_j \in S}I(x_i;x_j)
\end{equation}
Here \textit{I($x_i$;y)} and \textit{I($x_i$;$x_j$)} are mutual information {that} determine the amount of difference between the joint distribution and product of marginal distributions of the pair of random variables involved. 
\subsubsection {Random Forest Feature Selection}\label{ch2rfsel}
Random forest as a classifier performs implicit feature selection during training for classification, which results in higher accuracy. This implicit feature selection is utilised to rank a feature $x_i$ {which adds} the impurity decrease $\Delta i(\tau,T)$ for all nodes $\tau$ where $x_i$ is used and is averaged over all trees, T \cite{chBreiman2001}. {The feature importance is defined by equation (\ref{ch2imp})}.
\begin{equation} \label{ch2imp}
Imp(x_i)= \frac{1}{T}\sum_{T}\sum_\tau \Delta i(\tau,T) 
\end{equation}
Here $i(\tau)$ is the `gini impurity' at node $\tau$, {expressed as: 
\begin{equation}
i(\tau) = 1-\sum_{i}^{c}(p_i|\tau)^2\end{equation}} where $p_i$ is the fraction of samples that belong to the $i$th class of the $c$ classes. 

 The input features for the six classifiers are obtained using the feature selection methods, considering time-domain and time-frequency domain features.

\subsection{Features Selected}\label{ch2features selected}
The composition of a signal can be analyzed by different time-domain statistics and frequency components. Time-domain analysis provides the transitory response of a system and allows a better understanding of the flow of electrical quantities. Wavelet transform is suitable for decomposing an aperiodic signal into frequency bands, and their time-frequency analysis has been used in several applications that require time and frequency information simultaneously: gait analysis, fault detection, ultra-wideband wireless communications, etc. 

\subsubsection{Wavelet Coefficients (WC)} \label{ch2wfeatures}
Discrete Wavelet Transform (\acrshort{DWT}) quantifies the similarity between the original signal and the wavelet function by the detail ($d_l$) and approximation ($a_l$) coefficients \cite{chWAVELET}. The low and high-frequency components are obtained at each decomposition level $l$ using
{equation (\ref{ch2wt1}) and equation (\ref{ch2wt2}).
\begin {equation}\label{ch2wt1}
\footnotesize
a_l(k)=\sum_{l_k} w_\varphi(l_k\!-\!2k)a_{l-1}(l_k)   
\end {equation}}
{
\begin {equation}\label{ch2wt2}
\footnotesize
d_l(k)=\sum_{l_k}  w_\psi(l_k\!-\!2k)a_{l-1}(l_k)
\end {equation}
}

where $w_\varphi$, $w_\psi$ are the low and high pass filters. The mother wavelet and decomposition level {influence} the detail coefficients and thus the classification accuracy. However, researchers \cite{chSVM2,chann1,chdtwt,channdtwt,chknnwt} have arbitrarily chosen the wavelet function and decomposition level without justifying their use.
To address this issue, \cite{choptimalwavelet} used Particle Swarm Optimization to obtain the optimal wavelet functions combination to extract the most prominent features for classification of faults and \cite{choptimalwavelet2} used harmony search algorithm to determine the suitable wavelet functions and decomposition levels. 

Here multilevel 1D DWT is used with wavelet families `Daubechies', `Symlets', `Coiflets', `Biorthogonal', `Reverse biorthogonal', and `Discrete Meyer' to extract the WCs. The wavelet functions in each wavelet family (`Daubechies'- db1 to db38, `Symlets'- sym2 to sym20, `Coiflets'- coif1 to coif14, `Biorthogonal'- bior1.1 to bior6.8, `Reverse biorthogonal'- rbio1.1 to rbio6.8, `Discrete Meyer'- dmey) are decomposed at different levels. The maximum useful level of decomposition chosen to avoid edge effects caused by signal extension is given by {the equation (\ref{ch2maxlvl}):
\begin {equation}\label{ch2maxlvl}\footnotesize
Maximum\ level = \lfloor  log_{2}(\frac{signal\ length}{filter\ length-1})\rfloor \end {equation}}
Features (wavelet functions + decomposition level) for DIF {are} chosen using a classifier-involved method. 
The detail coefficients of the 3-phase differential currents obtained from each of these wavelet functions at the permissible decomposition levels are used to train and test DT (the baseline classifier here), finding the one which minimizes the error rate. Five WCs with the best-balanced accuracies averaged over 10 runs are selected. 
Thus, \textbf{bior2.2 at level 3, db4 at level 4, rbio3.3 at level 3, rbio4.4 at level 4, and sym4 at level 4} are obtained for DIF. The same features are used for LFU and CTD as well.

\subsubsection{Differential Wavelet Energy (WE)}\label{ch2we features}
\begin{figure}[ht]%
\footnotesize
\centering
\subfloat{\includegraphics[width=0.6\textwidth]{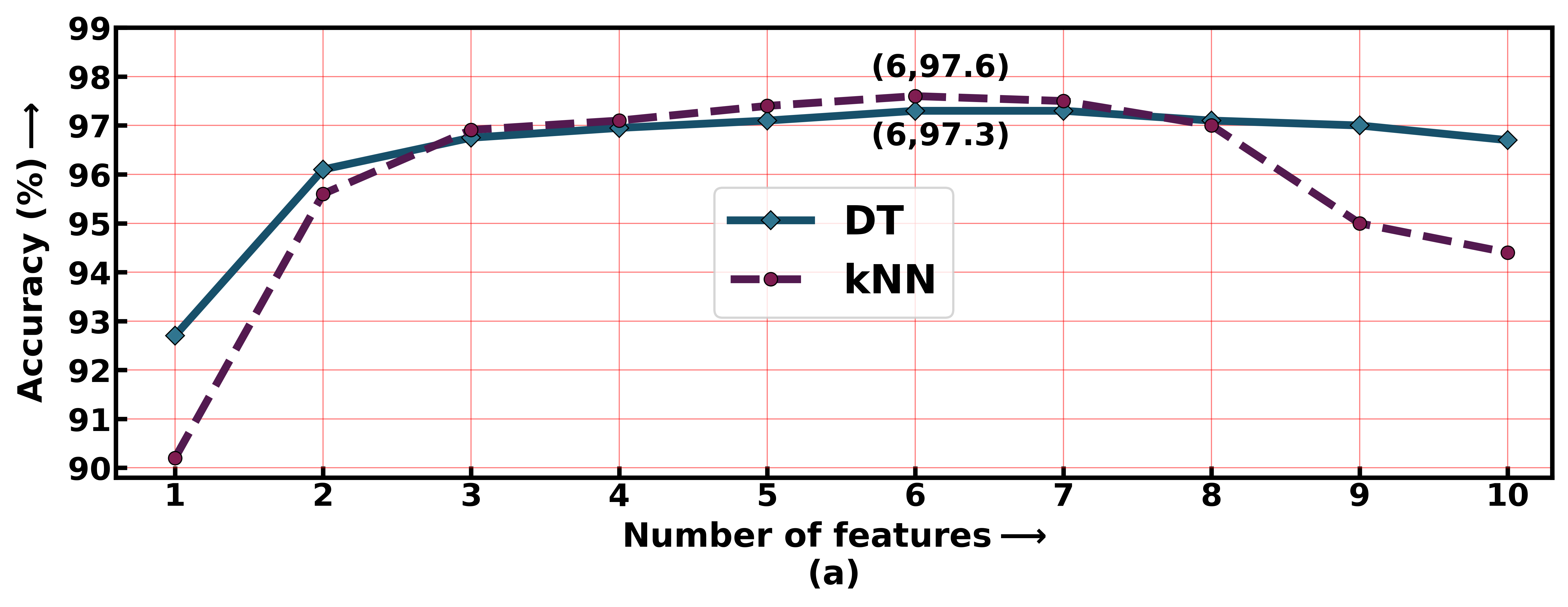}}
\qquad 
 \subfloat{\includegraphics[width=0.6\textwidth]{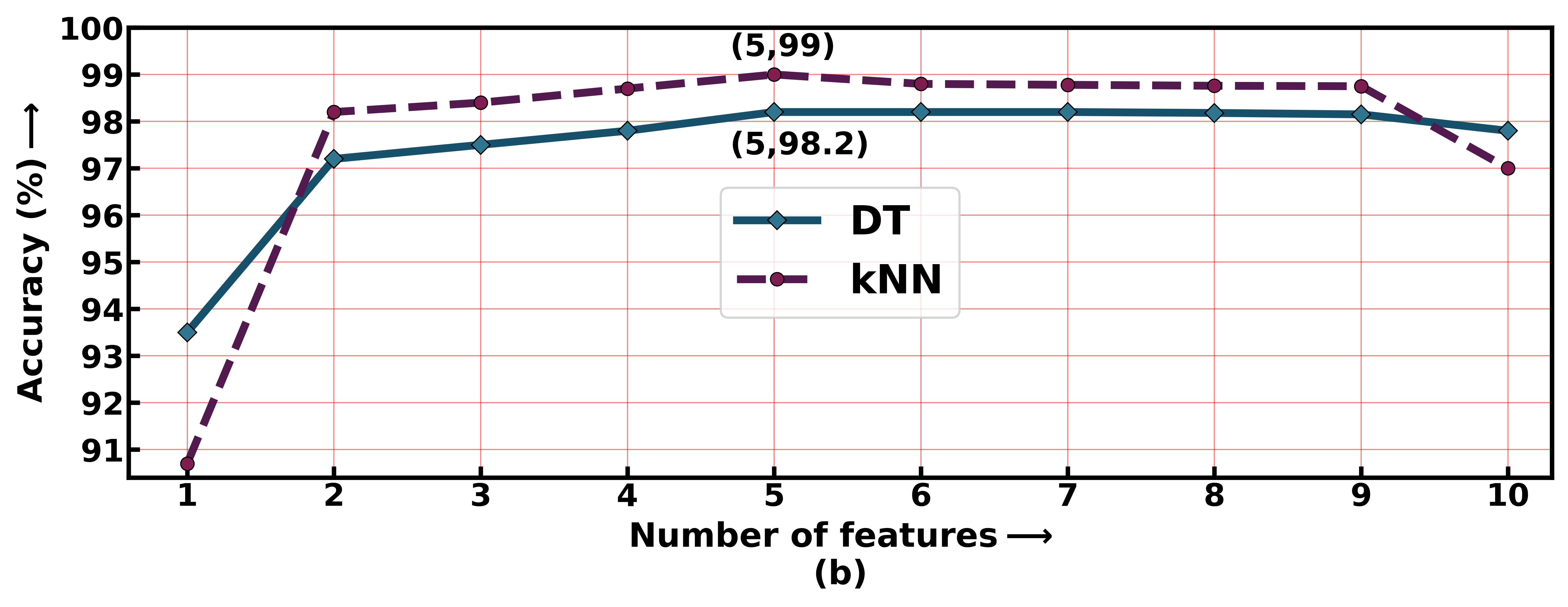}}%
\caption{Selecting optimal number of features: (a) Wavelet energy, and (b) Time-domain features}%
\label{ch2opfeature}%
\end{figure}

Wavelet energy is also a powerful tool to extract features. The differential WE is employed for differential protection of transformers in \cite{chwe1} \cite{chwe2}. The detail WC energy of the different wavelet functions which belong to the above mentioned wavelet families are combined to form a new set of inputs.
The energy associated with the WCs for each wavelet function at all permissible levels considering one cycle post-transient 3-phase differential is calculated using {equation (\ref{ch2wavener}).}
\begin {equation}\label{ch2wavener}
{E}_{dl}^w=\sum_{k}|d_l(k)|^2 
\end {equation}
The top 10 WE features are then obtained using mRMR feature selection method, which finds the optimal feature subset considering the {importance of the features} and their correlations. 
An exhaustive search over $2^{10}$-1 combinations of the 10 features {obtained with mRMR} is performed using kNN and DT as the baseline classifiers to obtain the optimum {number of} features. {It is noticed that the accuracy vs number of features curve of both kNN and DT improved up to 6 features and then started decreasing as the number of features increases (Fig.\ref{ch2opfeature}a). These 6 WE features, namely \textbf{rbio3.1, sym17, bior3.9, rbio3.9, coif13, and dmey} are thus selected and combined to form the inputs to the classifiers.}

\subsubsection{Time-Domain (TD) features}\label{ch2td features}
The 3-phase differential currents are also used to extract a comprehensive number of TD features. The entire feature list consisting of 794 features can be obtained from \cite{chtsfresh}. Random forest feature selection method is used to rank these features in order of information gain. {Subsequently, the number and combination of most relevant features are obtained by an exhaustive search over $2^{10}$-1 combinations of the top 10 features ranked by Random Forest feature selection using kNN and DT classifiers as the baseline again.}  {It is observed that the accuracy vs number of features curve of both kNN and DT improved up until 5 features and then started decreasing with any further increase in the number of features (Fig.\ref{ch2opfeature}b). These 5 TD features, namely \textbf{average change quantile, sample entropy, excess kurtosis, variance, and complexity invariant distance} are detailed in the following part.}
\begin{itemize}
 \renewcommand{\labelitemi}{\scriptsize$\blacksquare$}
 \item \textit{Average Change Quantile} = $\footnotesize{  \frac{1}{n'}{\sum_{t=1}^{n'-1} |Id_{\phi_{t+1}} - Id_{\phi_{t}}| }}$, computes mean of absolute consecutive changes in the signal inside two values: $qh$ and $ql$ having $n'$ samples.
\vspace{1.5mm}
 \item \textit{Sample Entropy} measures time complexity by computing the negative logarithm of the probability that subseries of length $m$ have distance < r, then subseries of length m+1 also have distance < r.
 \vspace{1.5mm}
\item \textit{Excess Kurtosis} = $\frac{\mu4}{\sigma^4}-3$, is the fourth standardized moment with mean $\mu$ and standard deviation $\sigma$.
\vspace{1.5mm}
 \item \textit{Variance} = $\frac{1}{n}\sum_{t=1}^{n}(Id_{\phi_{t}}-\mu)^2$ where $n$ is total samples.
 \vspace{1.5mm}
\item \textit{Complexity Invariant Distance} =\scriptsize{$\sqrt{\sum_{t=1}^{n-1} (Id_{\phi_{t+1}}-Id_{\phi_{t}})^2}$}, \normalsize estimates the time series complexity. A time series having more peaks, valleys etc. has a higher value.
\end{itemize}

Once the wavelet functions and the corresponding decomposition levels are obtained using the DT as baseline, the WCs are used to train and test RF, Xgb, NB, SVM, NN, and kNN classifiers. Similarly, the WE and TD features selected using the DT and kNN as baseline classifiers are used to train and test the six classifiers.

\subsection{Choice of Classifiers}
 Different classifiers are used to evaluate the validity of the proposed feature-based protection scheme. Tree-based ML estimators: random forest (RF), and XGBoost (\acrshort{Xgb}) having superior performance are very popular in data mining. The other classifiers used are Naive Bayes (\acrshort{NB})- a probabilistic classifier competitive in many applications; Support-vector machines (SVM)- basically a non-probabilistic classifier; Neural Networks (NN)- inspired by the human brain and adapted in a variety of applications ranging from social networking to cancer diagnosis; and k-nearest neighbors (kNN) where the system generalizes the training data after receiving a query.

\subsubsection{Decision Tree}
Decision trees are distribution-free white box Machine Learning models that learn simple decision rules inferred from the feature values. In 1984 Breiman et al. introduced Classification and Regression Trees (CART) \cite{chBreiman}. Here, the CART algorithm implemented in scikit-learn is used which constructs binary trees by splitting the training set recursively till it reaches the maximum depth or a splitting doesn't reduce the impurity measure. The candidate parent data $D_p$ is split into $D_l$ and $D_r$ at each node using a feature ($f$) and threshold that yields the largest Information Gain. The objective function IG which is optimized at each split is defined as:
\begin{equation}
IG(D_p,f)=I(D_p)-\frac{N_l}{N_p}\cdot I(D_l)-\frac{N_r}{N_p}\cdot I(D_r)
\end{equation}
where, I is impurity measure, ${N_p},{N_l}$ and ${N_r}$ are the number of samples at the parent and child nodes \cite{chQuinlan}. 

\subsubsection{Random Forest}
Random Forest (RF) classifier belongs to the family of ensemble trees which builds numerous base estimators and averages their predictions which produces a better estimator with reduced variance. Each tree constitutes a random sample (drawn with replacement) of the training set and the best split is found at each node by considering a subset of input features. The individual trees tend to overfit but averaging the predictions of all trees reduces the variance \cite{chBreiman2001}.  RF has also been used to select the important time-series features in Section \ref{ch2td features}. 
\subsubsection{Extreme Gradient Boosting (XGBoost)}
XGBoost (Xgb) is a supervised learning algorithm that sequentially combines weak learners into a stronger one, with each new model attempting to correct the previous model minimizing the objective function given by 
\begin{equation}
    J^{(t)} \approx \sum_{i=1}^{n}[a_iw_{c(x_i)}+\frac{1}{2}b_iw^2_{c(x_i)}]+\gamma L+\frac{1}{2}\lambda\sum_{i=1}^{n}[w_j^2]
\end{equation}
where a(.) and b(.) are the first and second-order derivatives of mean square error loss, c(.) assigns data to the corresponding leaf, $w$ is score vector on leaves, $\gamma$ is complexity, $\lambda$ scales the penalty, and L is the number of leaves. The regularization term {expressed as: 
\begin{equation}
\Omega=\gamma L+\frac{1}{2}\lambda\sum_{i=1}^{n}[w_j^2]
\end{equation}}present in the objective function is added as an improvement to reduce overfitting\cite{chxgboost}. Xgb is one of the best gradient boosting machine frameworks and has become popular as the algorithm of choice for many winning teams of ML competitions. 

\subsubsection{Naive Bayes} Naive Bayes (NB) is the simplest Bayesian Network model that applies Bayes' Theorem to classify the target on the basis of conditional independence of every pair of features given the value of the class variable $y$. {It is based on estimating $P(A|B)$, the probability density of features A given class B. It has lesser training time and requires smaller training data.} NB has shown good performance for applications such as text categorization, spam filtering, and medical diagnosis \cite{chnb}. 

\subsubsection{Support Vector Machines}
Support Vector Machines (SVMs) are memory-efficient classifiers that use the kernel method to create hyperplanes that separate the input data in high dimensional feature spaces\cite{chsvmc1}. {The training samples and the boundaries are called the support vectors and hyperplanes respectively.  Generally, a larger distance between the hyperplane and the nearest training sample leads to a lower generalization error of the classifier. Radial Basis Function and polynomial kernels were used in the study.} 

\subsubsection{Neural Network}
The Neural Network (NN) used is a fully connected feedforward network consisting of two hidden layers of perceptrons  {between the input and the output layer. It learns a non-linear function approximator} $f(.): R^S \xrightarrow{}R^c$ with $S$ features and $c$ outputs through back-propagation\cite{chmlpbook}. {It is an effective and efficient pattern recognition technique for ML applications.}

\subsubsection{k-Nearest Neighbor}
{k-Nearest Neighbor (kNN) is an instance-based non-parametric supervised learning algorithm used in applications of data mining, pattern recognition, and image processing,} which computes the class of an instance by majority voting of the k (an integer) nearest neighbors of each query point. The training phase involves storing the features and target labels \cite{chknn}. kNN has also been used as the baseline to select the optimum number of features in Section \ref{ch2we features} and Section \ref{ch2td features}.
 
\subsection{Bayesian Hyperparameter Optimization}
The performance of an ML algorithm depends on the choice of hyperparameters. Bayes' Theorem and Gaussian Process (\acrshort{GP}) are used to optimize the hyperparameters of the classifiers used. Specifically, to get the optimal parameters for computationally intensive training of Xgb, which has numerous hyperparameters, the Bayesian Optimization has been used. It constructs a probabilistic surrogate of the objective function from the previous observations and then generates the next candidate of parameter list $z_{i+1}$ by optimizing the surrogate function. GP is used to model prior on $f$. The acquisition function $u$ proposes the next sampling points in the search space. The Bayesian Optimization with GP is described in Algorithm \ref{ch2alg:bogp}\cite{chBOGP}. The hyperparameters and their values used in case of Xgb classifier for the search are: ``learning\_rate": [0.01, .05, 0.1], ``max\_depth": [5,10,15,20], ``min\_child\_weight":[1,10], ``subsample": [1,0.8], ``colsample\_bytree": [0.8,1], ``colsample\_bylevel": [0.5,1], ``n\_estimators": [1000,2000,5000,10000,12000].

\begin{algorithm}

\caption{{Bayesian Optimization}}\label{ch2alg:bogp}
 Collect initial observations $\mathcal{D}_n$=$\{z_i, f(z_i);i=1,...,n\}$.
 
\textbf{for} {$n=1,2,...$} \textbf{do}

\hspace{3mm}  Obtain the next sampling point $z_{n+1}$ by optimizing the acquisition function over the $GP$ : $z_{n+1}$= $arg$ $max_z$ $u(z|\mathcal{D}_{n})$.

\hspace{3mm}  Calculate $y_{n+1}=f(z_{n+1})$.

\hspace{3mm}  Augment observations $\mathcal{D}_{n+1}$=\{$\mathcal{D}_{n},(z_{n+1}, y_{n+1})$\} and update the $GP$.

\textbf{end of for}
\end{algorithm}

\section{Results}\label{ch2results}
The 3-phase differential currents of the simulated transient cases acquired from CT1 and CT2 are sampled at a frequency of 10kHz. The features extracted and selected from the 167 post transient samples per phase and registered by the ED {are} used for training the six classifiers. The input dimension of the training and testing cases varies depending on the level of decomposition and wavelet function chosen when WCs are used as features. In the case of TD features, the input dimension is 15 (5$\times$3), and with WE as feature, it is 18 (6$\times$3). To reduce the classification error and improve the generalization, 10-fold stratified cross-validation and Bayesian search are applied, which use the available data effectively and train the classifiers on optimized hyperparameters. Normally, the performance of a classifier is evaluated with the accuracy metric. However, in the case of data imbalance between classes, the results are biased. 
Since, the classes are imbalanced, balanced accuracy which is defined as mean of the accuracies obtained on all classes and expressed as (\ref{ch2acc})
\begin{equation}\footnotesize
\label{ch2acc}\bar{\eta}=\frac{1}{2}(\frac{TP}{TP+FN}+ \frac{TN}{TN+ FP})\end{equation} for binary classes is used to compute the performance measure where, TP {represents} true positive, TN {represents} true negative, FP {represents} false positive, and FN {represents} false negative \cite{chimbalance}.


\begin{table}[ht]
\centering
\footnotesize
\renewcommand{\arraystretch}{1}
\centering
\caption{Performance with WCs for DIF}\label{ch2wfaults}
\begin{tabular}{@{}ccccccc@{}}
\toprule
\multicolumn{1}{c}{\multirow{2}{*}{\begin{tabular}[c]{@{}c@{}}  \textit{Wavelet}\end{tabular}}} &  \multicolumn{6}{c}{\textit{Classifier($\bar{\eta}$)}} \\ \cmidrule(l){2-7} 
\multicolumn{1}{c}{}    & RF & \textbf{Xgb} & NN  & kNN &  NB &  SVM \\ \midrule
bior2.2        & 99.5   & 99.7     & 99.7  & 99.4 & 71.0  & 90.2\\
db4             & 99.5   & 99.7     & 99.5  & 99.4 & 77.2  & 93.0\\
rbio3.3          & 99.6   & 99.7     & 99.5  & 99.1 & 76.5  & 93.0\\
\textbf{rbio4.4}          & 99.7   & \textbf{99.8}     & 99.7  & 99.7 & 76.2  & 97.0  \\
sym4             & 99.7   & 99.7     & 99.6  & 99.5 & 77.7  & 93.7\\  \hline
\end{tabular}
\end{table}

\subsection{Detection of internal faults (DIF)}\label{ch2fault_detect}
Since the occurrence of any power system transient event is unpredictable in time, the use of an ED becomes imperative. Out of the {three mentioned} classification tasks, the correct distinction of internal faults from the other transients is the foremost. The security and dependability of the proposed method depend on the type 1 error (FP) and type 2 error (FN) of this binary classification problem. The less the classification error, the better is the performance of the entire scheme. To achieve this, the six classifiers are trained on 48442 cases and tested on 12110 cases of one cycle of the post fault differential currents simulated in section \ref{ch2sec2}.
The classifiers are trained with three sets of features, and the testing accuracies are reported. First, the selected WCs obtained using exhaustive search by training DTs are used as the inputs, and the classification performance is shown in Table \ref{ch2wfaults}. Xgb gives the best $\bar{\eta}$ of 99.8\% on `rbio4.4' at level 4.
Second, the classifiers are trained on the 6 WE features obtained by an exhaustive search of $2^{10}$-1 different combinations of the top 10 WEs ranked using the mRMR algorithm. Table \ref{ch2waveall} shows the classification performance on the 6 features of the different classifiers. Xgb overshadows the rest of the classifiers with $\bar{\eta}$ of 99.5\%.
Third, the 5 features obtained again from an exhaustive search over $2^{10}$-1 different combinations of the 10 TD features ranked using RF are put-to-use. Table \ref{ch2timeall} shows the performance of the six classifiers. Again, Xgb gives the best performance with $\bar{\eta}$ = 99.8\%.

\begin{table}[ht]
\centering
\footnotesize
\renewcommand{\arraystretch}{1}
{
\centering
\caption{Performance with WE}\label{ch2waveall}
\begin{tabular}{@{}lllllll@{}}
\toprule
\multicolumn{1}{l}{\multirow{2}{*}{\begin{tabular}[c]{@{}c@{}} \textit{Model}\end{tabular}}} & \multicolumn{6}{c}{\textit{Classifier($\bar{\eta}$)} }\\ \cmidrule(l){2-7} 
\multicolumn{1}{c}{}                                                                            & \multicolumn{1}{c}{}                                                                                
RF & \textbf{Xgb} & NN  & KNN &  NB &  SVM \\ \midrule
\textit{DIF} & 93.2 &\textbf{99.5}&  86.0& 99.2 & 78.4 &60.0\\
 \textit{LFU}   & 93.2 &  \textbf{98.3} & 82.4  & 94.0  & 57.3 & 72.5\\
  \textit{CTD}  & 95.6 & \textbf{98.7} & 96.1  & 98.7 &62.4 &88.8\\
 \hline
\end{tabular}}
\end{table}

\begin{table}[ht]
\centering
\footnotesize
\caption{Performance with TD}\label{ch2timeall}
\begin{tabular}{@{}lcccccc@{}}
\toprule
\multirow{2}{*}{\begin{tabular}[c]{@{}l@{}}
\scriptsize{\textit{Model} }
\end{tabular}} &      \multicolumn{6}{c}{\textit{Classifiers($\bar{\eta}$)} }\\ \cmidrule(l){2-7} 
                                                                                                       & RF  & \textbf{Xgb}   & NN      & kNN       & NB       & SVM      \\  \midrule     
\textit{DIF}                                                                                            &    96.2  &   \textbf{99.8}   & 94.6 & 98.6      & 77.5     & 87.0.    \\ 
\textit{LFU }     &  94.0   &  \textbf{98.8}    & 89.2 & 95.2      & 61.3  & 85.9    \\
\textit{CTD }     & 99.2  & \textbf{99.9}      & 98.8 & 99.7      & 75.2 & 95.3  \\ \hline
\end{tabular}
\end{table}

\subsection{Localization of faulty unit (LFU)} After the fault detector recognizes an internal fault, the faulty unit (exciting or series) is identified using the one-cycle of the post fault differential currents. The six classifiers are trained on 37498 fault cases and tested on 9374 cases for LFU. Table \ref{ch2wfltloc} shows the classification performance on selected WCs as features, and Table \ref{ch2waveall} shows the same for WE features. Table \ref{ch2timeall} shows the classification performance on TD.
Xgb performs better than the other classifiers with an $\bar{\eta}$ of 98.8\% obtained using TD features, $\bar{\eta}$ of 97.8\%  with `rbio4.4' at level 4, and $\bar{\eta}$ of 98.3\% with WE as feature. 

\begin{table}[ht]
\centering
\footnotesize
\renewcommand{\arraystretch}{1}
\caption{Performance with WCs for LFU}\label{ch2wfltloc}
\begin{tabular}{@{}ccccccc@{}}
\toprule
\multicolumn{1}{c}{\multirow{2}{*}{\begin{tabular}[c]{@{}c@{}}  \textit{Wavelet}\end{tabular}}} &  \multicolumn{6}{c}{\textit{Classifier($\bar{\eta}$)}} \\ \cmidrule(l){2-7} 
\multicolumn{1}{c}{}   & RF & \textbf{Xgb} & NN  & kNN &  NB &  SVM \\ \midrule
bior2.2  & 93.6 &  97.6 & 93.1 & 94.1 & 63.6 & 85.5\\
db4       & 95.2 & 97.2 & 93.2 & 94.3 & 57.2 & 88.8 \\
rbio3.3    & 95.0 & 97.7 & 92.1 & 92.9 & 64.1 & 86.9  \\
\textbf{rbio4.4}    & 95.5 & \textbf{97.8} & 92.9 & 94.4 & 57.7 & 89.7\\
sym4      & 95.9 & 97.4 & 93.7 & 94.5 & 57.0 & 89.6 \\  \hline
\end{tabular}
\end{table}

\subsection{Classification of transient disturbances (CTD)} The different transient disturbances: overexcitation, external faults with CT saturation, magnetizing and sympathetic inrush are also classified after the fault detector {identifies} them as no-fault transients. Table \ref{ch2wtransients} shows the performance on selected WCs, table \ref{ch2waveall} on WE features, and table \ref{ch2timeall} shows the same for TD features of the six classifiers. 10944 cases are used for training and 2736 cases are used for testing the classifiers. Xgb outperforms the other classifiers with an $\bar{\eta}$ of 99.9\% obtained with the TD features, $\bar{\eta}$ of 98.7\% with WE as feature, and NN gives the best $\bar{\eta}$ of 99.4\%  with `rbio4.4' at level 4.

\begin{table}[ht]
\centering
\footnotesize
\renewcommand{\arraystretch}{1}
\caption{Performance with WCs for CTD}\label{ch2wtransients}
\begin{tabular}{@{}ccccccc@{}}
\toprule
\multicolumn{1}{c}{\multirow{2}{*}{\begin{tabular}[c]{@{}c@{}}  \textit{Wavelet}\end{tabular}}} &  \multicolumn{6}{c}{\textit{Classifier($\bar{\eta}$)}} \\ \cmidrule(l){2-7} 
\multicolumn{1}{c}{}     & RF& Xgb &  \textbf{NN}  & kNN &  NB &  SVM \\ \midrule
bior2.2   & 98.6 & 98.8 & 99.2 & 99.2 & 74.1 & 96.5\\
db4       & 98.0 & 98.7 & 98.8 & 97.8 & 66.7 & 96.2 \\
rbio3.3   & 98.6 & 98.8 & 99.3 & 99.3 & 73.1 & 98.1  \\
\textbf{rbio4.4 }  & 98.6 & 98.7 & \textbf{99.4} & 99.3 & 68.4 & 97.7\\
sym4      & 98.2 & 98.9 & 98.7 & 98.9 & 67.4 & 97.3 \\  \hline
\end{tabular}
\end{table}

It is not possible to make a fair comparison of performances with \cite{chtencon} where internal faults were differentiated from inrush currents using Wavelet Transform and classified with NN
and with \cite{chisspit} where the internal faults in series and exciting transformers of the ISPAR are classified using RFC. Nevertheless, the 97.7\% accuracies in \cite{chtencon} and 98.76\% in \cite{chisspit} are cited just as a point of reference.

\subsection{Execution Time} 
\begin{figure}[ht]
\centerline{\includegraphics[width=3.6 in, height= 1.9 in]{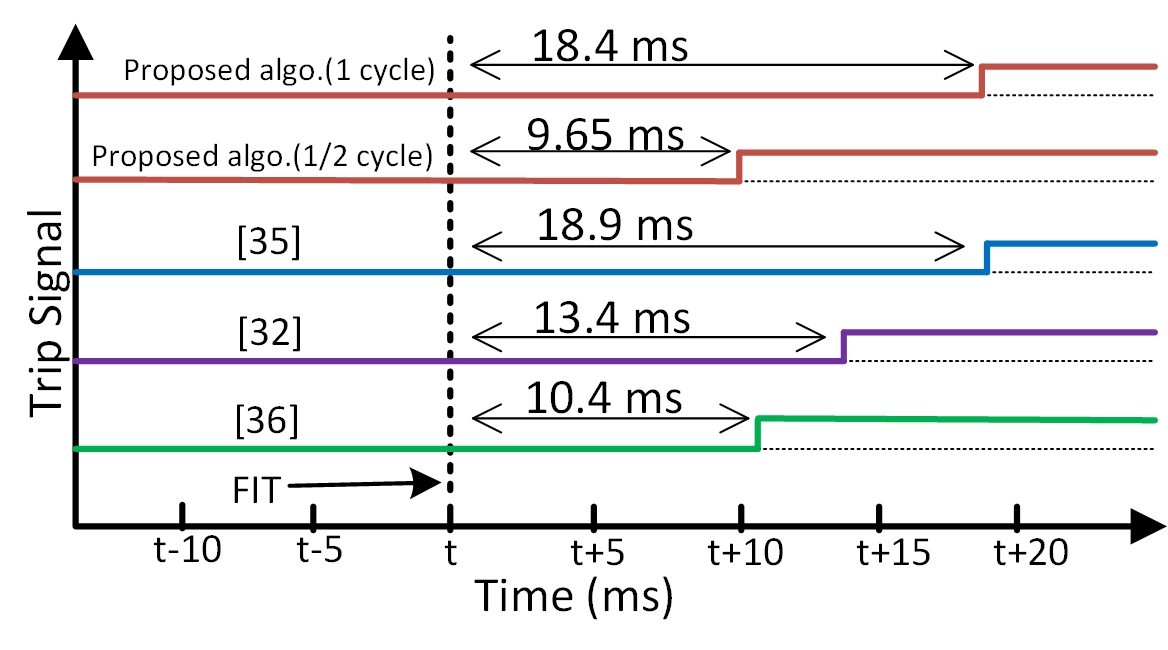}}
\caption{Operating time of different protection techniques}
\label{ch2exetime} 
\end{figure}

The proposed method can beat the operation time of 1-2 cycles of a conventional relay with harmonic blocking.
The execution time (average time of 100 runs) for the feature extraction, training, and testing of the Xgb models for the three tasks with WC, WE, and TD as features are computed on Intel Core i7-8665U CPU @1.90 GHz, 16 GB RAM (See Table \ref{ch2ptime}). The in-service operating time of the fault/no-fault decision would include time to extract the feature for a single instance and then testing it on the already trained Xgb model. Xgb trained on `rbio4.4' is the fastest taking (16.67+1.6+0.13) = 18.4ms with a $\bar{\eta}$ of 99.8\%. It takes 32.6ms with TD and 19.7ms with WE. To test the scheme for further reduction in computation time, the Xgb is trained and tested on 84 samples (1/2 cycle) on the 60552 cases. The results show that the proposed technique performs well with (8.34+1.2+0.12) = 9.65ms operating time and $\bar{\eta}$ of 99.2\%.
The time taken for LFU and CTD can be obtained from the columns `Testing time' and `Feature extract time' of the table.
Noting that computations can be further optimized, these processing times are suitable for future real-time implementation. Fig.\ref{ch2exetime} shows the operating time of the proposed technique on one cycle and 1/2 cycle, current differential-based techniques \cite{chzgajic}, and\cite{chkasztenny}; and directional-based technique \cite{chukhan}.
The computational time of 9.65ms of the proposed scheme on 1/2 cycle suggests that ML-based differential protection schemes can compete with the previously proposed techniques \cite{chzgajic,chukhan,chkasztenny}.

\begin{table}[ht!]
\centering
\footnotesize
\renewcommand{\arraystretch}{1.2}
\caption{Computational time of the three models}\label{ch2ptime}
\begin{tabular}{|c|c|c|c|c|c|c|c|c|c|}

\hline
\multirow{2}{*}{\textit{Model}} & \multicolumn{3}{c|}{\textit{Training time(s)}} & \multicolumn{3}{c|}{\textit{Testing time(ms)}} & \multicolumn{3}{c|}{\textit{Feature extract time(ms)}} \\ \cline{2-10} 
            & TD        & WC    & WE          & TD           & WC        & WE     &\hspace{0.5mm} TD  \hspace{0.5mm}     & \hspace{0.5mm}WC          \hspace{0.5mm}    & WE             \\ \hline
\textit{DIF} & 123     &   456   &    183     &   2.4        &    1.6      &  2.5     &  13.5        &   0.13          &    0.52          \\ \hline
\textit{LFU} &  90     &   407   &  128     &   2.5          &    2.1      &     2.4  &   13.5      &      0.13       &   0.52            \\ \hline
\textit{CTD} &  30    & 383     &  66      &    2.6         &   2.2      &     2.5   &  13.5       &        0.13      & 0.52       \\ \hline
\end{tabular}
\end{table}

\section{Performance Evaluation for Non-Traditional and Additional Challenges}\label{ch2impacts}
The security and dependability of the proposed method are also tested for various system conditions in addition to the aforementioned traditional challenges in Section II. These conditions, namely the integration of type-4 wind turbine, fault during magnetizing inrush, series winding saturation, change in tap positions, change in rating, saturation of CT, presence of noise, and low current faults which can jeopardize the reliability of the relay are discussed in this section. 
\subsection{Effect of integration of WTG}
\begin{figure}[ht]
\centerline{\includegraphics[width=3.0 in, height= 1.1 in]{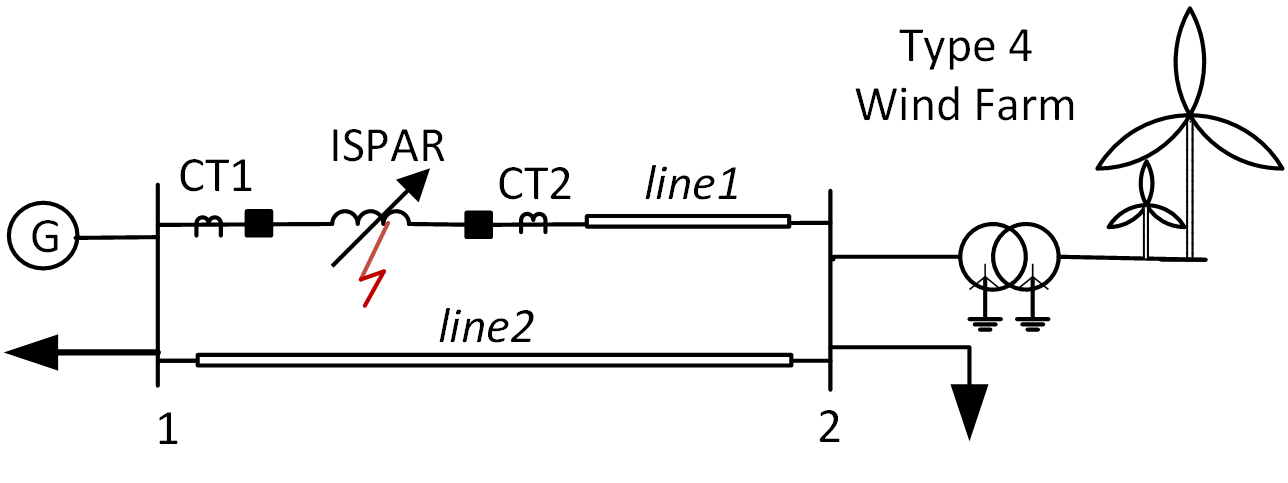}}
\caption{WTG connected at bus 2}
\label{ch2wtg} 
\end{figure}

The type 4 Wind Turbine Generators (\acrshort{WTG}) have complex fault characteristics and are very different from conventional generators. {It is also expected that systems with high wind penetration may experience larger frequency deviations after system disturbances and
in the absence of accurate modeling of its dynamics and fault behavior, the transformer differential protection may mal-operate \cite{chisgt}.}
A permanent magnet synchronous machine connected to the grid by a full-scale converter is considered in this study where the converter limits the fault current from 1.1 to 1.5 times the rated load current. The stability of the proposed scheme with the WTG is validated by the accuracy of 100\% obtained on 5049 cases of internal faults and 6360 cases of transient disturbances. The fault cases are simulated by varying the tap positions, PS, FR, FIT, and FT (Fig.\ref{ch2wtg}). Due to grid side interface similarities, this analysis is also applicable to systems with photovoltaic generations \cite{chpv}.

\subsection{Effect of Series winding saturation}
Since the voltage rating of the series winding connecting the source and load is lower than the rating of the overall system, it may saturate when subjected to considerable voltage increase. The security of traditional differential protection is tested in such conditions \cite{chibrahim2}. The stability of the proposed scheme during series winding saturation is tested by increasing the source voltage from 120\% to 150\% of the nominal voltage in steps of 10\%. 3000 cases of internal faults and 720 cases of series winding saturation are simulated by varying the tap positions, PS, and magnitude of overvoltage. Since the number of cases is imbalanced, Synthetic Minority Over-sampling Technique (\acrshort{SMOTE}) \cite{chsmote} is used to over-sample the series winding saturation cases. Xgb gives an accuracy of 100\% on 3000 cases of each class.

\subsection{Effect of Change in Tap Position}
Generally, the transformer tap changer effect is taken into account with a corrected input of the primary voltage. The proposed technique considers different tap positions and tackles possible mal-operations in case of transients due to non-standard phase shifts without tracking the tap-changer positions. 3000 cases of internal faults and 648 cases of tap-change cases are simulated by varying the tap positions, PS, and ST. It gives an accuracy of 99.9\% on 3000 cases of each obtained by oversampling the tap-change cases using SMOTE.

\subsection{Effect of Fault during Inrush}
The harmonic restraining or blocking differential relays are used to ensure security during magnetizing inrush; however, conventional relays' operation is delayed if faults occur during magnetic inrush. To ensure dependability, 12292 cases of inrush and faults during inrush are simulated by varying the parameters discussed in section II. $\bar{\eta}$ of 100\% suggests that the proposed scheme performs well in the event of faults during magnetizing inrush.

\subsection{Effect of Change in Rating}
The proposed scheme works even if an ISPAR of a different rating is considered. 6912 internal faults and other transients are simulated for an ISPAR with $S_n$=450MVA, $V_n$=345kV by varying FR, PAR tap position, FT, FIT, ST, etc. The same Xgb-1 model, which was trained on $S_n$=500MVA, $V_n$=230kV, is used to test these 6912 cases. The stability of the scheme is substantiated by $\bar{\eta}$ of 99.3\%.

\subsection{Effect of CT saturation}
The impedance of CT secondary may influence the level of harmonics in the differential currents. To study the effect of saturation of the CTs, the burden and CT secondary impedance are changed. $\bar{\eta}$ of 100\% on 6912 cases of internal faults and other transient disturbances obtained by varying the different parameters discussed in section II validate the reliability of the proposed scheme for CT saturation conditions.

\subsection{Effect of Noise}
 In the real-world presence of noise during the capture and processing of differential currents may affect the protection system's stability. The white Gaussian noise of different Signal-to-Noise-ratio (\acrshort{SNR}) is added to the data to study its effect on the proposed method. Table \ref{ch2noise} shows the accuracy of Xgb for noise levels from 40dB to 20dB. The classifier performs poorly with lower SNR, but even so always above 80.2\% \scriptsize{($\frac{93.4+67.8}{2}$)}. \normalsize The $\bar{\eta}$ varies from 99.8\% with no noise to 80.2\% for SNR of 20dB. {The accuracy dips down further to 67\% for a SNR of 10dB which is understandable as the ratio of the desired information to the undesired signal is only about 3.16.}

\begin{table}[ht]
\footnotesize
\centering
\caption{Effect of Noise\label{ch2noise}}
\footnotesize
\renewcommand{\arraystretch}{1}
\begin{tabular}{cccccc}
\toprule

\multirow{2}{*}{\begin{tabular}[|c|]{c}Fault/\\Disturbances\end{tabular}} & \multirow{2}{*}{\begin{tabular}[c]{@{}c@{}}SNR (dB)\end{tabular}} & \multirow{2}{*}{\begin{tabular}[c]{@{}c@{}}Number\\ of cases\end{tabular}} & \multicolumn{2}{c}{Predicted class} & \multirow{2}{*}{\begin{tabular}[c]{@{}c@{}}Accuracy\\ (\%)\end{tabular}} \\ \cmidrule(lr){4-5}
\multicolumn{2}{c}{}                                                                                          &                                                                            & Faults        & Disturbances        &                                                                        \\ \midrule
\multirow{4}{*}{\begin{tabular}[c]{@{}c@{}}Internal \\ faults\end{tabular}}                 & $\infty $             & 9406      &    9399       & 7               &        99.9                                                              \\
                                &40  &   9406     &     9324      &       82              &      99.1                                                               \\
                                & 30              &       9431         &      9246         &      185     &     98.0                                                                 \\
                                 & 20              &     9318         &     8700       &      618         &      93.4    \\ \midrule                                                              
\multirow{4}{*}{\begin{tabular}[c]{@{}c@{}}Other \\ disturbances\end{tabular}}              &  $\infty $             &  2705            &     13   &         2692        &      99.5                                                                 \\
                        &   40              &     2705     &    96         &       2609             &        96.5                                                          \\
                        &    30             &  2680                    &    337          &   2343             &     87.4                                                                \\
                     &  20               &  2793       &  898       &   1895            &    67.8       \\ \midrule                                                          
\end{tabular}
\end{table}

\subsection{Effect of Low current \textit{t-t} \& winding ph-g  faults}
The proposed algorithm performs well for both high resistive winding ph-g faults and \textit{t-t} faults also. To test its sensitivity, \textit{t-t}  faults with 2\% of the series winding shorted, and winding ph-g faults with high resistance of 50$\Omega$ in the series winding are simulated. The ED was able to detect all the 48 winding ph-g and 144 \textit{t-t} faults obtained by varying the tap positions, FR, and FIT.

\section{Summary}\label{ch2sec5}
The chapter addresses the problem of detection and localization of faults and classification of transients for an ISPAR. The internal faults are distinguished from overexcitation, external faults with CT saturation, and inrush conditions. After that, depending on the detection of a fault, the faulty unit (ISPAR series or exciting) is located, or the transient disturbances are classified. Wavelet and time-domain features obtained from one cycle of post transient 3-phase differential currents registered by the event detector are used to train six prominent classifiers.
Firstly, the classifiers are trained with the most important WCs obtained by exhaustive search using DT. Secondly, the top WE features obtained using mRMR are put to use. Lastly, TD features selected by maximizing the information gain are used. It is observed that overall XGBoost trained with the TD features outperforms the other models for DIF, LFU, and CTD; and when both accuracy and computation time are considered the XGBoost model trained on `rbio4.4' WC is superior for DIF. On top of fault detection with $\bar{\eta}$=99.8\%, localization with $\bar{\eta}$=98.8\%, and classification of transients with $\bar{\eta}$=99.9\%, the proposed scheme has several benefits over the conventional differential relays: 
\begin{itemize}[topsep=0pt, partopsep=0pt,itemsep=0pt,parsep=0pt]
 \renewcommand{\labelitemi}{\scriptsize$\blacksquare$}
    \item the proposed scheme is more dependable for fault during magnetic inrush and sensitive to low current turn-to-turn and winding ph-g faults.
    \item it is secure to magnetic and sympathetic inrush, overexcitation, external fault with CT saturation, series winding saturation, CT saturation, tap position changes, and integration of WTG.
    \item it takes care of the non-standard phase shift in the PAR without tracking the exciting unit tap positions.
    \item the proposed technique is robust to change in PAR ratings and noise in measurements.
    \item it does not need additional voltage or phase information. 
\end{itemize}

The protection scheme advanced in this chapter can cooperate with standard microprocessor-based differential relays and offer supervisory control over its operation, thus improving the stability of the power system and providing a complete solution to the problem of PAR protection.

\newpage

\chapter{Differential Protection of Power Transformers and Phase Angle Regulators}\label{ch3}

\section{Introduction}\label{ch3_intro}
Power Transformers are an integral part of an electrical grid and their protection is vital for the reliable and stable operation of the power system. An important requirement of the protection system is the faithful discrimination of faults from other transients. Differential protection has been the primary protection scheme in transformers because of its inherent selectivity and sensitivity. Mal-operations due to magnetizing and sympathetic inrush, and CT saturation during external faults are the major problems associated with differential protection. Second-harmonic restraint method is extensively used to distinguish internal faults from magnetizing inrush since more second-harmonic component exists in inrush currents than in internal faults \cite{chharmonic}. However, higher second-harmonics are generated during internal faults with CT saturation, presence of shunt capacitance, or because of the distributed capacitance of EHV lines \cite{chtx3pro}. In addition, the second-harmonic content in inrush currents has reduced in modern transformers with soft core material \cite{chmodern_core}. Hence, several cases of mal-operation of conventional relays in distinguishing faults and inrush have been reported \cite{chchina}.
CT saturation during external faults may also cause false trips due to the inefficient setting of commonly used dual-slope biased differential relays \cite{chctsat1}.

Phase Angle Regulators or Phase Shifters or Phase Shift Transformers as introduced in Chapter 2 are a special class of transformers used to control real power flow in parallel transmission lines. They ensure system reliability and allow easier integration of new generations with the grid. 
The PARs similar to Power Transformers require a fast, sensitive, secure, and dependable protection system. Discriminating external faults with CT saturation, magnetizing inrush, and other transient disturbances from internal faults is a challenge for the protection systems of PARs as well. Moreover, methods used to compensate the phase for differential relays in Power Transformers with a fixed phase shift are not applicable in PARs with variable phase shift \cite{chpstguide}.


\section{Motivation}
Various ML methods to distinguish internal faults and magnetizing inrush in Power Transformers have been used in the past two decades. A combination of Artificial Neural Network (ANN) and spectral energies of wavelet components was used to discriminate internal faults and inrush in \cite{chann1}. Support Vector Machines (SVM) and Decision Tree (DT) based transformer protection were proposed in \cite{chSVM1,chSVM2} and \cite{chdt1,chdt2,chdtwt} respectively.
Probabilistic Neural Network (PNN) has been used to detect different conditions in Power Transformer operation in \cite{chtripathy}. Random Forest Classifier (RFC) was proposed to discriminate internal faults and inrush in \cite{chShahrfc}. Works of literature also suggest extensive use of S-Transform, Hyperbolic S-Transform, Wavelet Transform (WT) to detect Power Quality (\acrshort{PQ}) transient disturbances and then classify them using DT, SVM, ANN, PNN \cite{chPQ1,chPQ2,chPQ3,chPQ4,chPQ5,chPQ6}. These transient disturbances are caused by variations in load, capacitor switching, charging of transformers, starting of induction machines, use of non-linear loads, etc.

There is limited literature investigating internal faults and inrush in an ISPAR. Attempts were made in \cite{chtencon} where internal faults are distinguished from magnetizing inrush using WT and then the internal faults are classified using ANN and in \cite{chisspit} where the internal faults in series and exciting transformers of the ISPAR are classified using RFC. But, the authors have predominately used an isolated and simple network having a Power Transformer \cite{chann1,chSVM1,chSVM2,chdt1,chdt2,chdtwt,chtripathy} or a PAR \cite{chtencon,chisspit} to support their proposed protection scheme. Furthermore, the transient disturbances have not been studied rigorously in these works.

\section{Contributions}
This chapter studies the use of Decision Tree-based algorithms to discriminate the internal faults and other transient disturbances including magnetizing inrush and CT saturation during external faults in a 5-bus interconnected system with Phase Angle Regulators and Power Transformers which has not been attempted before. Customized two-winding and three-winding transformers developed in Chapter 2 are used to simulate the internal faults. The change detector filter detects the faults and registers the differential currents. Five most relevant time and frequency domain features have been used to train SVM, RFC, DT, and GBC classifiers to detect, locate and identify the internal faults and classify six transient disturbances. The proposed scheme is tested on 101,088 transients cases simulated on PSCAD/ EMTDC by varying various system parameters. The entire dataset of internal faults and other disturbances is available on IEEE Dataport \cite{chdata}.

\section{Chapter Organization}
The rest of the chapter is organized as follows. Section \ref{ch3_model} illustrates the modeling and simulation of the internal faults and other transient disturbances in the power network containing Power Transformers and ISPARs. Section \ref{ch3_algo} comprises the detection of internal faults, feature extraction and selection, and the classifiers used for the detection and identification of transients. Section \ref{ch3_results} includes the results for detection of internal faults, identification of faults and transient disturbances, and evaluates the effect of noise, CT saturation, and change in transformer rating and connection on the proposed scheme. Section \ref{ch3_summary} concludes the chapter. 

\begin{figure}[htpb]
\centerline{\includegraphics[width=4.65 in, height= 3.18 in]{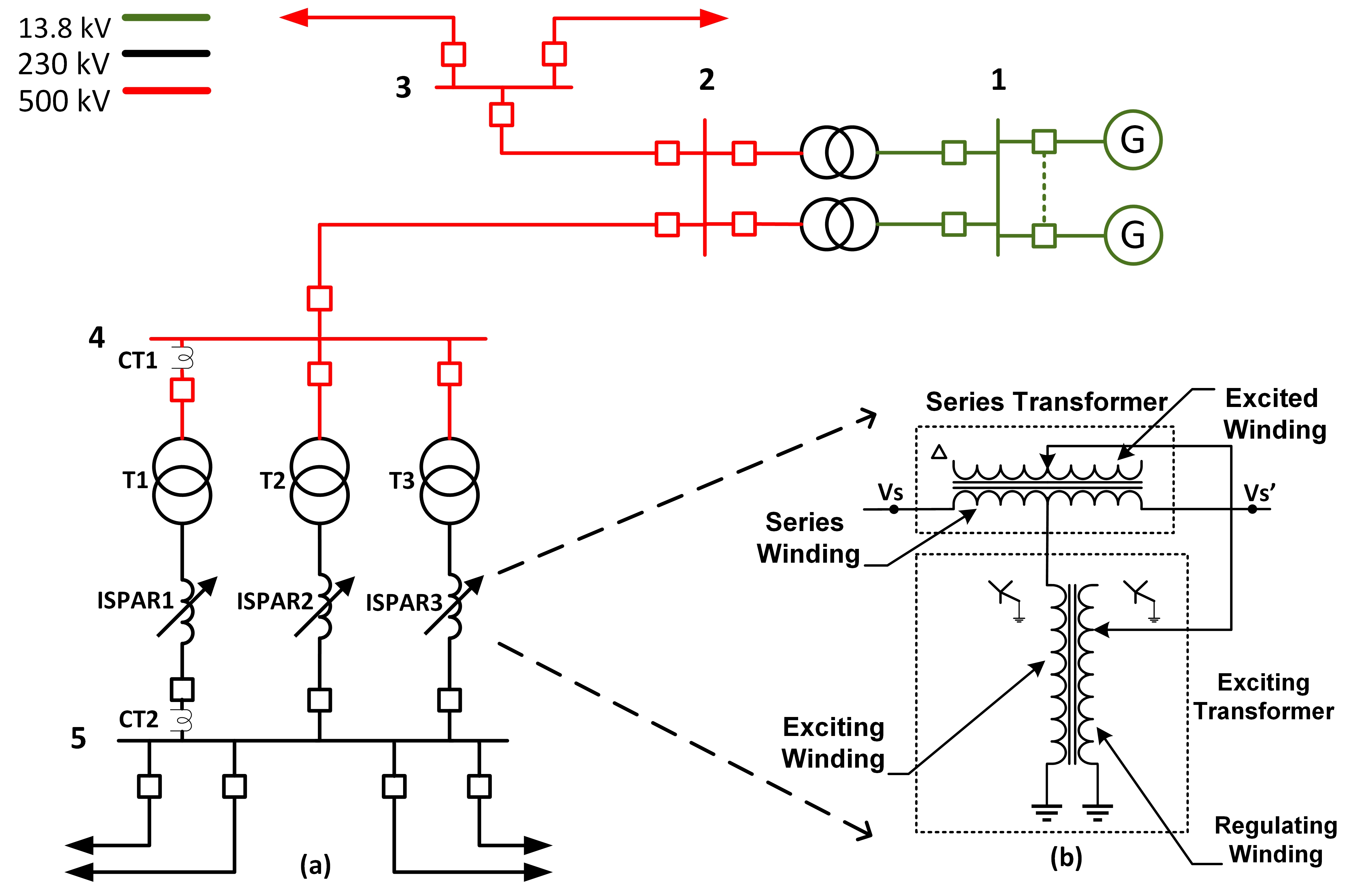}}
\vspace{0mm}\caption{(a) 5-bus interconnected system with ISPARs, Power Transformers, T-lines, and AC sources, (b) Series and exciting transformers in ISPAR}
\label{ch3leaps}
\end{figure}

\section{Modeling and Simulation}\label{ch3_model}
The power network chosen for the simulation of the internal faults and the transient disturbances is based on a proposed Pumped-storage (an efficient form of renewable storage designed to meet energy needs and reduce emissions by utilizing the energy stored in an upper water body pumped from a lower water body) project in California, USA \cite{chsiemens}.

PSCAD/EMTDC is used for the modeling and simulation of the transients in the ISPAR and Power Transformer in the chosen interconnected power system. Fig.\ref{ch3leaps}(a) shows the single-line diagram of the 5-bus interconnected model consisting of the AC source, transmission lines, ISPARs, Power Transformers, and 3-phase loads working at 60Hz. The ISPARs have a rating of 500 MVA, 230kV/230kV, with phase angle variations of $\pm25^{\circ}$ and the Power Transformers are rated at 500 MVA, 500kV/230kV. The AC source consists of 9 units of 120 MVA, 13.8kV hydro-generators. Two transformers are used in cascade to step up the voltage from 13.8kV to 500kV. 3 ISPARs (ISPAR$_1$, ISPAR$_2$, and  ISPAR$_3$) are connected between bus4 and bus5 through transformers T$_1$, T$_2$, and T$_3$. Only the internal faults in ISPAR$_1$ and T$_1$ are studied. 
The three-winding transformer required for the series units of ISPAR and the two-winding transformer required for the exciting units of ISPAR and Power Transformers for the simulation of various internal faults including turn-to-turn and primary-to-secondary winding faults were developed in Chapter 2. 


This chapter covers various internal faults in the ISPAR and Power Transformer, capacitor switching, switching of non-linear loads, magnetizing inrush, sympathetic inrush, external faults with CT saturation, and ferroresonance. In the following paragraphs, these conditions are considered one after the other. The simulation run-time, fault/disturbance inception time, and fault duration time are 15.2s, 15.0s, and 0.05s (3 cycles) respectively in all cases. Again, the multi-run component is used to change the parameter values wherever possible to get the different simulation cases and snapshots of the first simulation runs are used to start the simulation from initialized conditions to reduce the simulation time. 

\subsection{Internal Faults}
The internal faults are created in the Power Transformer, ISPAR series, and ISPAR exciting unit. 88,128 internal fault cases which include basic internal faults, turn-to-turn, and winding-to-winding faults are simulated by varying the fault resistance (\acrshort{FR}), \% of turns shorted (\acrshort{PTS}), fault inception time (\acrshort{FIT}), phase shift (\acrshort{PS}), and the LTC in the exciting unit.

\subsubsection{ Internal phase \& ground faults (ph \& g)}
Phase winding to ground ($w_{a}$-g, $w_{b}$-g, $w_{c}$-g), phase winding to phase winding to ground ($w_{a}$-$w_{b}$-g, $w_{a}$-$w_{c}$-g, $w_{b}$-$w_{c}$-g), phase winding to phase winding ($w_{a}$-$w_{b}$, $w_{a}$-$w_{c}$, $w_{b}$-$w_{c}$), 3-phase winding ($w_{a}$-$w_{b}$-$w_{c}$), and 3-phase winding to ground ($w_{a}$-$w_{b}$-$w_{c}$-g) faults are simulated in the primary (P) and secondary (S) sides of the Power Transformer and on the primary and secondary sides of exciting and series transformer units in the ISPAR. Table \ref{ch3tabintrnalfault} shows the values of different system and fault parameters in T$_1$ and ISPAR$_1$ (Fig.\ref{ch3leaps}(a)) which are varied to get the training and testing cases for the internal phase \& ground faults. 
\begin{table}[htbp]
\renewcommand{\arraystretch}{0.9}
    \centering

    \caption{Parameters for ph \& g  faults in the ISPAR and Power Transformer}
    \label{ch3tabintrnalfault}
    \begin{tabular}{ll}
\hline
Variables  & Values \\ \hline
FR &  0.01, 0.5 \&   10 $\Omega$     (3)   \\
PTS & 20\%, 50\%, 80\%  (3)\\
FT      & $w$-g, $w$-$w$-g, $w$-$w$, $w$-$w$-$w$ \& $w$-$w$-$w$-g (11) \\
FIT       & 15s to 15.0153s in steps of 1.38ms (12)\\
FL            &     \begin{tabular}{@{\extracolsep{\fill}}l}Transformer (P \& S)  (2) \\ ISPAR Exciting unit (P \& S) (2)\\ \& ISPAR Series unit (P \& S) (2) \end{tabular}  \\
PS & Forward and backward  (2)\\ 
LTC  & 0.2,0.4,0.6,0.8,1[1 \& 0.5 in ISPAR exciting] \\ \midrule
 \multicolumn{2}{l}{{\color[HTML]{141414} }}                                                 \\
\multicolumn{2}{l}{\multirow{-2}{*}{{\color[HTML]{141414} \begin{tabular}[c]{@{}l@{}} \footnotesize Transformer or ISPAR series faults = 3$\times$3$\times$11$\times$12$\times$2$\times$2$\times$5 = 23,760                                                                                                                                                                                   \\ \footnotesize ISPAR exciting faults = 3$\times$3$\times$11$\times$12$\times$2$\times$2$\times$2 = 9504\end{tabular}}}} \\ \hline
\end{tabular}
\end{table}

\subsubsection{Turn-to-turn (t-t) faults} 
\begin{figure}[htbp]
\centerline{\includegraphics[width=3.4 in, height= 3.1 in]{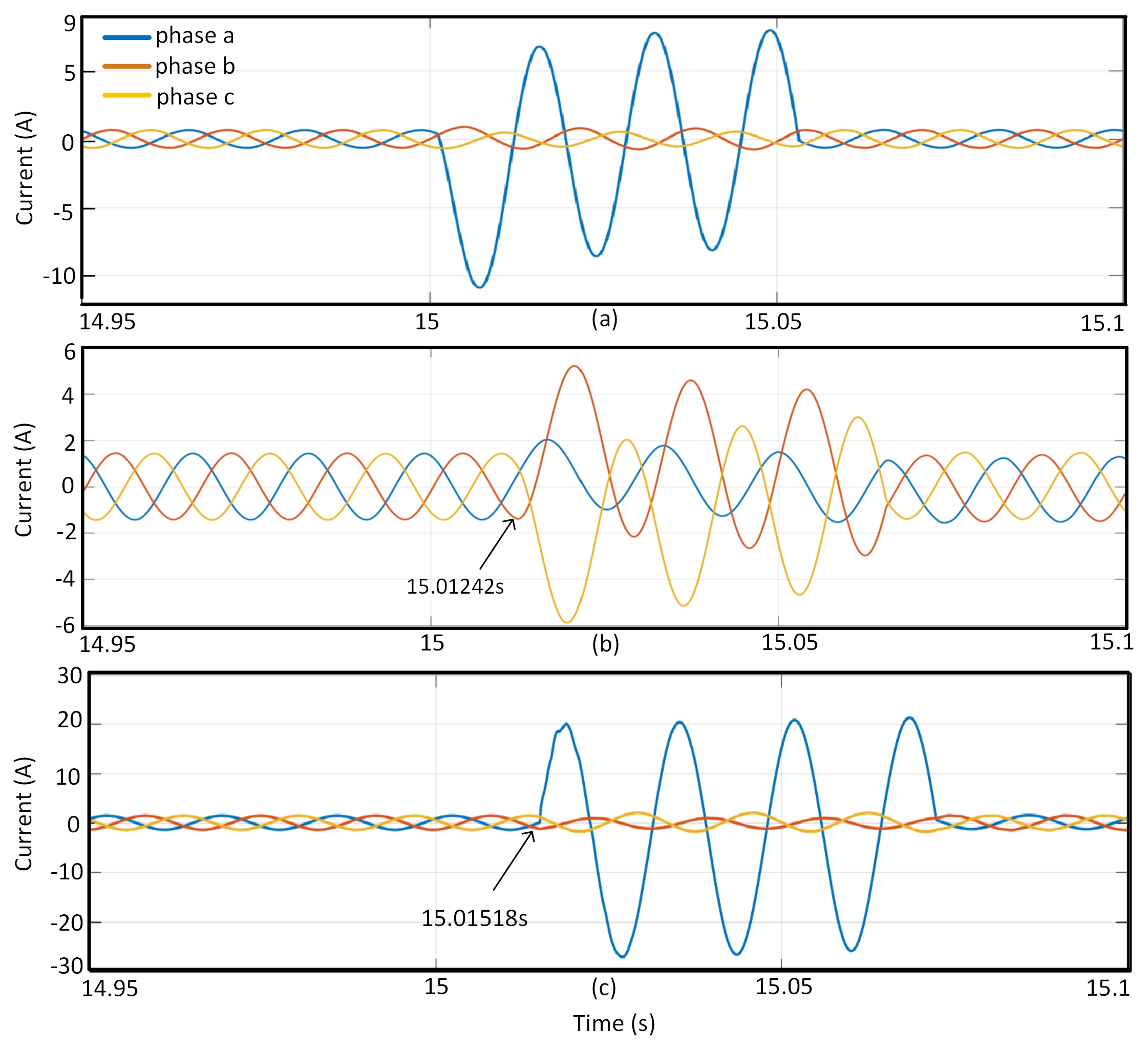}}
\vspace{0mm}\caption{3-phase differential currents for turn-to-turn faults in (a) primary of exciting unit, (b) primary of series unit, (c) primary of Power Transformer}
\label{ch3tt}
\end{figure}

Turn-to-turn faults may lead to more serious faults and inter-winding faults if not detected quickly \cite{chturn}.
Table \ref{ch3tab_ww_tt} shows the values of different parameters of the Power Transformer and the series and exciting unit of ISPAR used to simulate 20,736 turn-to-turn faults. Fig.\ref{ch3tt}(a) shows the differential currents for LTC = half, FIT = 15s, backward phase shift, FR = 0.01$\Omega$ and \% turns shorted = 20 in primary of exciting unit. Fig.\ref{ch3tt}(b) shows the differential currents for LTC = full, FIT = 15.0124s, backward phase shift, FR = 0.01$\Omega$ and \% turns shorted = 40 in primary of series unit. Fig.\ref{ch3tt}(c) shows the differential currents for LTC = full, FIT = 15.01518s, forward phase shift, FR = 0.01$\Omega$ and \% turns shorted = 60 in primary of the Power Transformer.
\begin{table}[ht]
\renewcommand{\arraystretch}{0.9}
\setlength{\tabcolsep}{10pt}
    \centering
    \caption{Parameters for winding-to-winding \& turn-to-turn faults in the ISPAR and Power Transformer}
    \label{ch3tab_ww_tt}
    \begin{tabular}{ll}
\hline
Variables  & Values \\ \hline
FR &  0.01, 0.5 \&   10 $\Omega$     (3)   \\
PTS & 20\%, 40\%, 60\%, 80\%  (4)\\
FIT       & 15s to 15.0153s in steps of 1.38ms (12)\\
FL            &     \begin{tabular}{@{\extracolsep{\fill}}l}Transformer phase a,b,c (P \& S)  (6) \\ ISPAR Exciting phase a,b,c (P \& S) (6)\\ \& ISPAR Series phase a,b,c (P \& S) (6) \end{tabular}  \\
PS & Forward and backward  (2)\\ 
LTC  & 0.2,0.4,0.6,0.8,1 [1 \& 0.5 in ISPAR exciting] \\ \midrule
\multicolumn{2}{l}  {{\color[HTML]{141414} }}                                                                                                                                                                                                                                                           \\
\multicolumn{2}{l}  {{\color[HTML]{141414} }}                                                                                                                                                                                                                                                                                \\
\multicolumn{2}{l}  {{\color[HTML]{141414} }}                                                                                                                                                                                               \\
\multicolumn{2}{l}{\multirow{-4}{*}{{\color[HTML]{141414} \begin{tabular}[c]{@{}l@{}}\footnotesize Transformer or ISPAR series(t-t) faults = 3$\times$4$\times$12$\times$6$\times$2$\times$5 = 8640\\  \footnotesize ISPAR exciting(t-t) faults = 3$\times$4$\times$12$\times$6$\times$2$\times$2 = 3456 \\  \footnotesize Transformer or ISPAR series(w-w) faults = 3$\times$4$\times$12$\times$3$\times$2$\times$5= 4320\\  \footnotesize  ISPAR exciting(w-w) faults = 3$\times$4$\times$12$\times$3$\times$2$\times$2 = 1728\end{tabular}}}} \\ \hline
\end{tabular}
\end{table}
\subsubsection{Winding-to-winding (w-w) faults}
Transformer aging and short circuits degrade the insulation between LV and HV winding and cause winding failure \cite{chturn}.
Table \ref{ch3tab_ww_tt} shows the values of different parameters of the Power Transformer and the series and exciting unit of ISPAR used to simulate 10,368 winding-to-winding faults.

Magnetizing inrush, sympathetic inrush, and external faults during CT saturation were introduced in Chapter 2. The rest of the transients are covered in detail in this chapter.

\subsection{Magnetizing inrush} 

\begin{figure}
\centerline{\includegraphics[width=2.9in, height= 2.7 in]{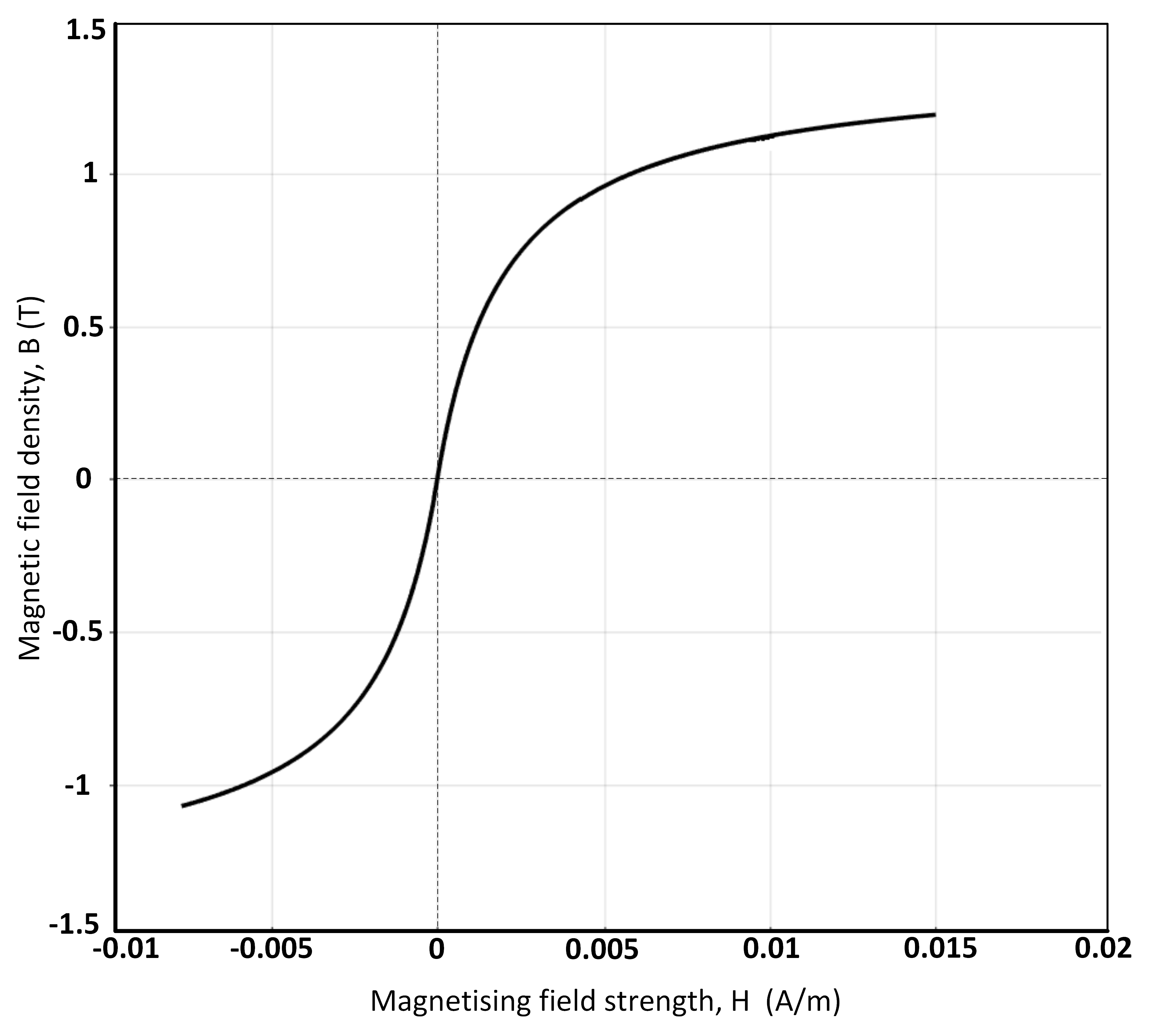}}\vspace{0mm}
\caption{B-H curve of transformer core}
\label{ch3bh}
\end{figure}

\begin{figure}[htbp]
\centerline{\includegraphics[width=3.9 in, height= 2.5 in]{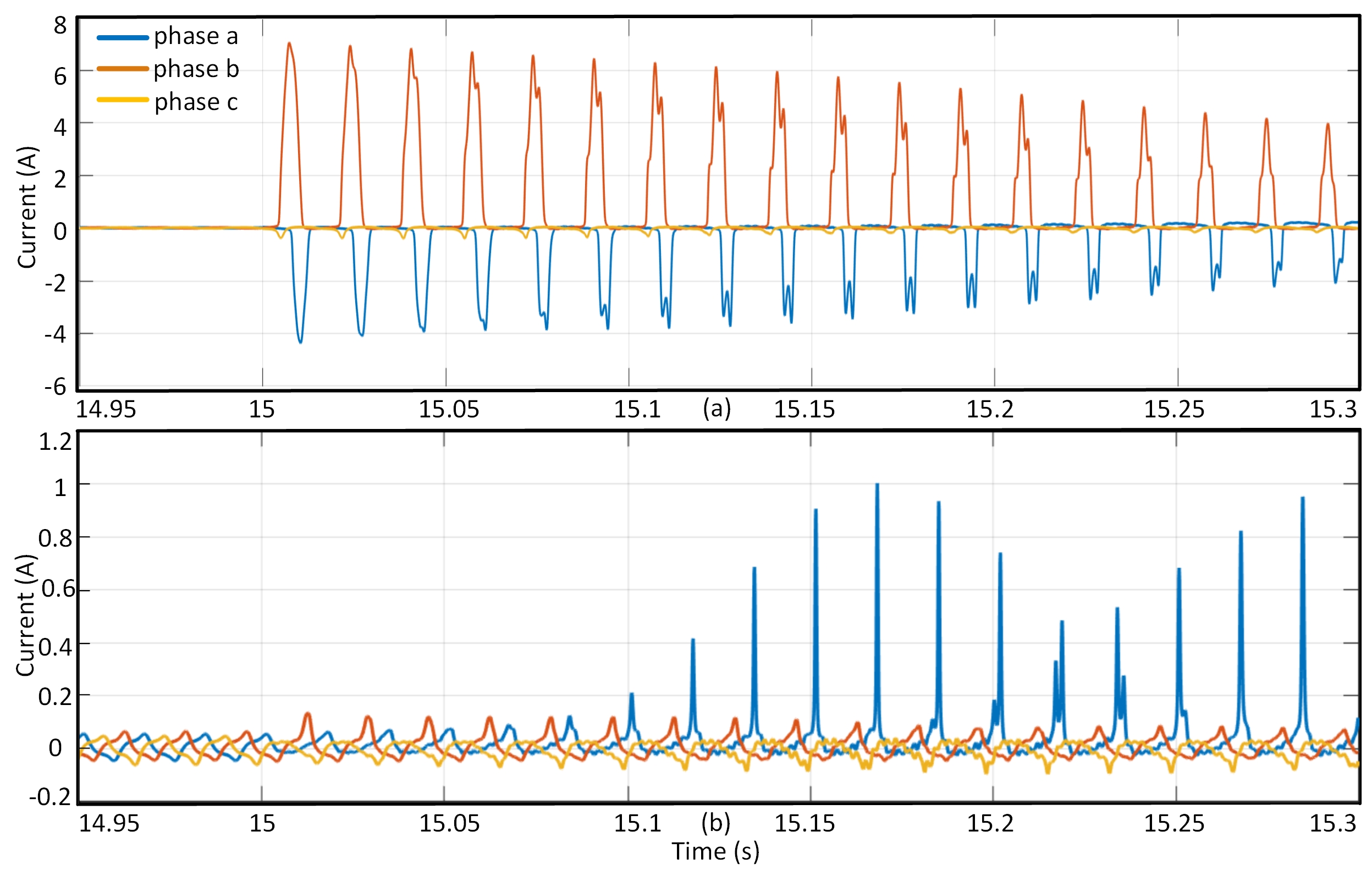}}
\vspace{0mm}\caption{ 3-phase differential currents for (a) Magnetizing inrush and (b) Sympathetic inrush}
\label{ch3inr}
\end{figure}

Power Transformer T$_1$ (Fig.\ref{ch3leaps}(a)) is chosen as the incoming transformer and 
the B-H curve of the transformer core material is shown in Fig.\ref{ch3bh}. Table \ref{ch3tab23} shows the values of different parameters including $\phi_R$ and  t$'$ used to get the data for training and testing for magnetizing inrush and Fig.\ref{ch3inr}(a) shows the 3-phase differential currents for LTC = full, ST = 15s, forward phase shift, and -80\% residual flux density (\acrshort{RFD}).

\begin{table}[ht]
\renewcommand{\arraystretch}{0.9}
\setlength{\tabcolsep}{10pt}
\centering
\caption{Parameters for Magnetizing and Sympathetic inrush\label{ch3tab23}}
{\begin{tabular}{ll}\toprule
Variables  & Values \\ \midrule
RFD &   $5\times3$ = (15)\\
ST       & 15s to 15.0153s in steps of 1.38ms   (12)     \\
LTC  & 0.2 to full tap in steps of 0.2 (5)                       \\
PS & Forward and backward (2)\\
& \color[HTML]{141414} {Total=$15\times12\times5\times2$=1800}\\
\hline
\end{tabular}}{}
\end{table}

\subsection{Sympathetic Inrush} The in-service transformer (T$_1$) experiences sympathetic inrush when the incoming transformer (T$_2$) is energized. The asymmetrical flux change per cycle during switching of T$_2$ which drives T$_1$ to saturation is expressed as:
\begin{equation}\Delta\phi=\int_{t}^{2\pi + t}[(R_{sys} + R_{T_1})i_1 + R_{sys} i_2]\end{equation}  
where $R_{sys}$ {is the} system resistance , and $R_{T_1}$ {is the} resistance of transformer T$_1$, $i_1$ and $i_2$ are magnetizing currents of T$_1$ and T$_2$.  
The magnitude and direction of $\phi_R$, and t$'$ are varied and the incoming transformer T1 is connected in parallel to simulate the scenarios. Table \ref{ch3tab23} shows the values of the different parameters used to get the training and testing cases for sympathetic inrush.
Fig.\ref{ch3inr}(b) shows the 3-phase differential currents for LTC = 0.2, ST = 15s, forward phase shift, and -80\% RFD.

\subsection{External faults with CT saturation}
\begin{figure}[htbp]
\centerline{\includegraphics[width=3.6 in, height= 3.1 in]{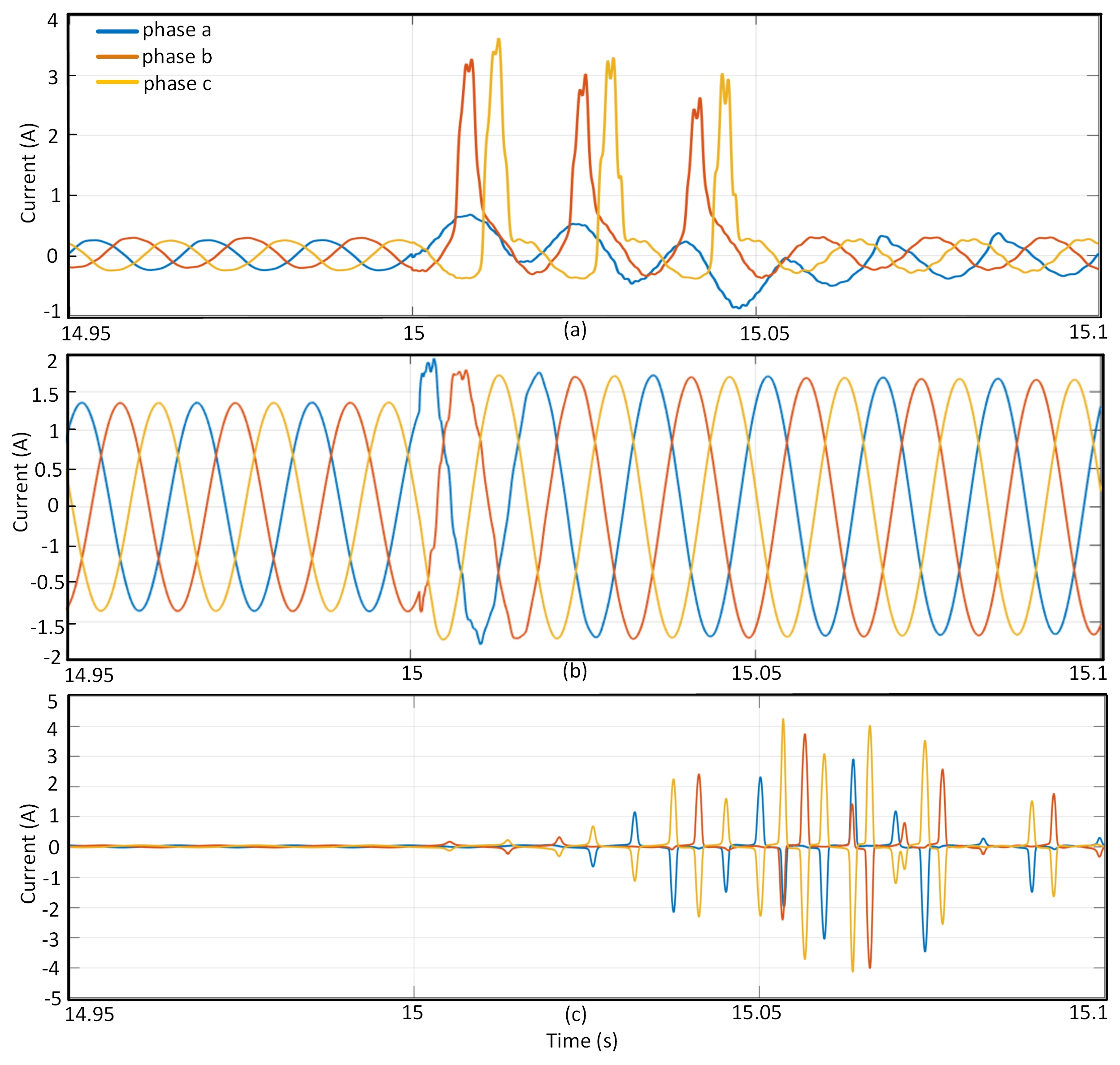}}
\vspace{0mm}\caption{3-phase differential currents for (a) External fault with CT saturation, (b) Capacitor Switching, and (c) Ferroresonance}
\label{ch3ext}
\end{figure}

The external faults with CT saturation are simulated on the 500kV and 230kV buses (bus4 \& bus5). The values for the different parameters are given in Table \ref{ch3externaltab}. Fig.\ref{ch3ext}(a) shows the 3-phase differential currents for an external line-to-ground (lg) fault when LTC = 0.2,  PS = forward, FIT = 15s, and FR = 0.01$\Omega$ on the 230kV bus.

\begin{table}[ht]
\renewcommand{\arraystretch}{0.9}
\setlength{\tabcolsep}{10pt}
\centering
\caption{Parameters for External faults on 230kV \& 500kV bus}
\label{ch3externaltab}
\begin{tabular}{ll}\toprule
Variables  & Values \\ \midrule
FR & 0.01, 0.5 \&   10 $\Omega$     (3)   \\
FT      & $l$g, $ll$g, $l$$l$, $l$$l$$l$ \& $lll$g (11) \\
FIT       & 15s to 15.0153s in steps of 1.38ms    (12) \\
LTC  & 0.2 to full tap in steps of 0.2 (5)\\
PS & Forward and backward  (2)\\ 
FL & 230kV \& 500kV bus (2)\\
& {Total=$3\times11\times12\times5\times2\times2$=7920}\\
\hline
\end{tabular}{}
\end{table}

\subsection{Non-linear Load Switching} 
\begin{figure}[htbp]
\centerline{\includegraphics[width=3.6 in, height=2.1 in]{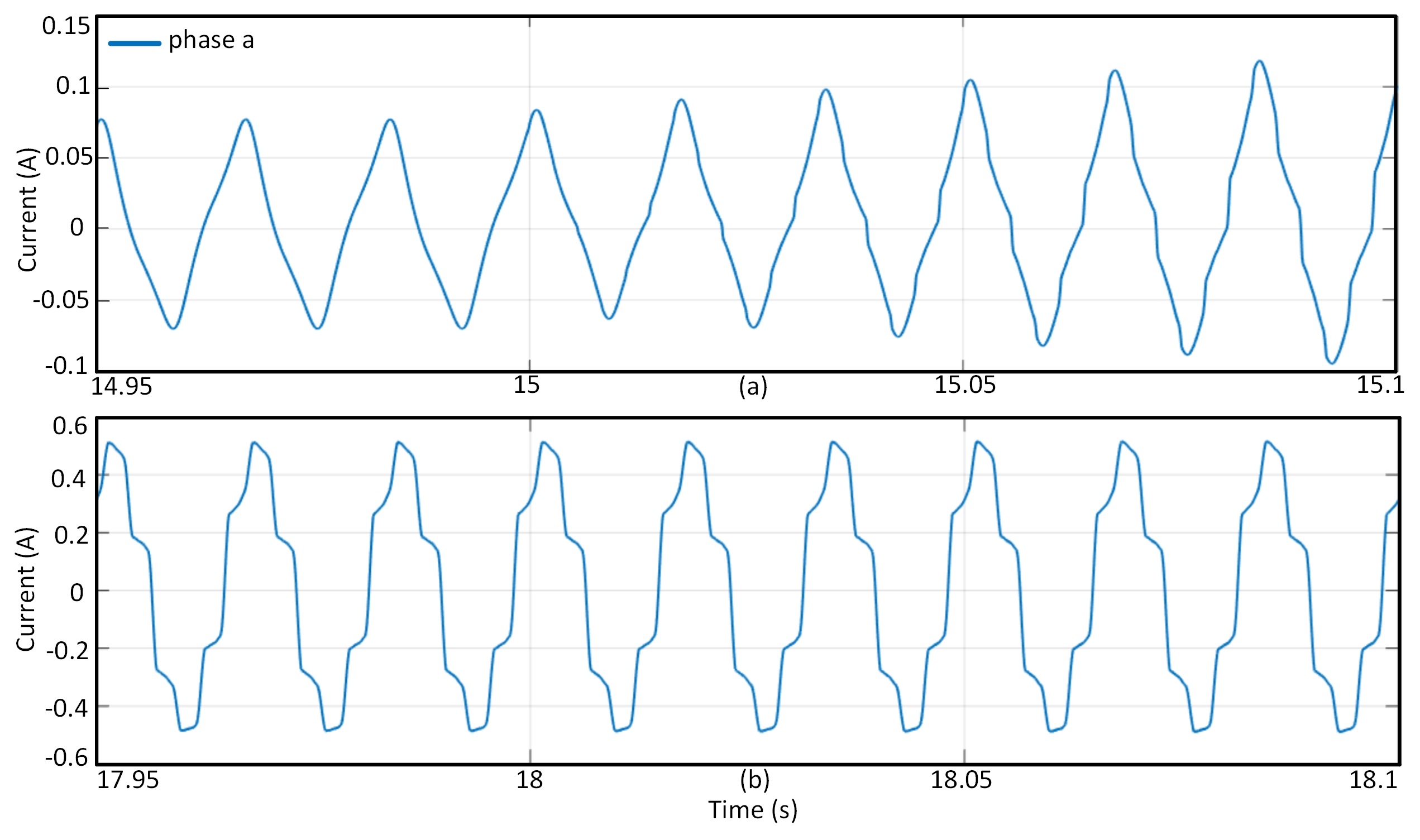}}
\vspace{0mm}\caption{Non-linear Load Switching (a) Transient and (b) Steady-state differential currents}
\label{ch3nl}
\end{figure}

With the advancement in semiconductor technology and the use of non-linear loads with power converters, harmonic contents in the line currents have increased. The differential relays may mal-operate when non-linear loads e.g steel furnaces are switched in a network containing transformers because of mutual enhancement effects between the transformer core and the load causing extreme saturation of the transformer core for several cycles \cite{chnon_linear_load_switch}. The harmonic information has been used to discriminate faults from other disturbances and locate the faults in the transmission line using SVM and ANN \cite{chkoley}. A thyristor-based 6-pulse bridge rectifier with a wye-delta transformer as the non-linear load is connected to the 230 kV bus to obtain the training and testing cases for load switching. The values for the different parameters are given in table \ref{ch3tab33} and Fig.\ref{ch3nl} shows the phase-a differential current for LTC = full, ST = 15s and firing angle (\acrshort{FA}) of 0$^{\circ}$. Fig.\ref{ch3nl}(a) shows the transient and Fig.\ref{ch3nl}(b) shows the steady-state differential current after the switching.

\begin{table}[ht]
\renewcommand{\arraystretch}{0.9}
\setlength{\tabcolsep}{10pt}
\centering
\caption{Parameters for Non-linear Load Switching \label{ch3tab33}}
{\begin{tabular}{ll}\toprule
Variables  & Values \\ \midrule
FA & 0$^{\circ}$, 10$^{\circ}$, 20$^{\circ}$, 30$^{\circ}$, 40$^{\circ}$, 50$^{\circ}$ (6)                         \\
ST      & 15s to 15.0153s in steps of 1.38ms    (12)   \\
LTC  & 0.2 to full tap in steps of 0.2 (5)\\
& \color[HTML]{141414}{ Total=$6\times12\times5$=360}\\

\hline
\end{tabular}}{}
\end{table}

\subsection{Capacitor Switching} Capacitor banks are used to improve voltage profile, reduce losses, and enhance power factor. Mal-functioning of customer equipment due to voltage magnification coinciding with capacitor switching is common. \cite{ch5051} used WT to detect high transient inrush currents from capacitor-bank switching to avoid malfunctioning of instantaneous and time overcurrent relays (50/51). Capacitor banks having 3 Legs with capacitor bank rating (\acrshort{CBR}) of 500 MVAr each is connected to the 230kV bus. Capacitor bank reactors and resistors are used in each Leg to reduce the effect of transients in voltages. Table \ref{ch3tab143} shows the different parameters and their values used to get the data for training and testing for capacitor switching. Fig.\ref{ch3ext}(b) shows the 3-phase differential currents for LTC = full, ST = 15.00138s, and switching of 3 Legs of the capacitor bank.
\begin{table}[ht]
\renewcommand{\arraystretch}{0.9}
\setlength{\tabcolsep}{10pt}
\centering
\caption{Parameters for Capacitor Switching\label{ch3tab143}}
{\begin{tabular}{ll}\toprule
Variables  & Values \\ \midrule
CBR & 500,1000,1500 MVAr (3)                  \\
ST  & 15s to 15.0153s in steps of 1.38ms (12)    \\
PS & Forward and backward (2)\\
LTC  & 0.2 to full tap in steps of 0.2 (5)\\
& \color[HTML]{141414}{Total=$3\times12\times2\times5$=360}\\
\hline
\end{tabular}}{}
\end{table}
\subsection{Ferroresonance}  Initiated by faults and switching operations, ferroresonance causes harmonics and overvoltages and may lead to mal-operation of protective relays and damage of power equipment \cite{chferro2}. Mal-operation of the differential relay occurs because of the higher magnitude of current in the HV side than the LV side \cite{chferromanitova}. Besides, the low loss, amorphous core transformer increases the intensity and occurrence of ferroresonance \cite{chFERRO3}. Several configurations may lead to ferroresonance in electrical systems. In this chapter, one such arrangement has been modeled when one of the phases of a 3-phase transformer is switched off. The parameters and their values for ferroresonance conditions are presented in Table \ref{ch3tab123}.
Fig.\ref{ch3ext}(c) shows the 3-phase differential currents for ST = 15s and grading capacitance (\acrshort{GC}) = 0.2$\mu$F simulated between bus2 and bus4.

\begin{table}[ht]
\renewcommand{\arraystretch}{0.9}
\setlength{\tabcolsep}{10pt}
\centering
\caption{Parameters for Ferroresonance\label{ch3tab123}}
{\begin{tabular}{ll}\toprule
Variables  & Values \\ \midrule
GC & 0.02$\mu$F to 0.2$\mu$F in steps of 0.02$\mu$F (10)\\
Location & a,b,c phases (3)\\
ST         & 15s to 15.016s in steps of 0.69ms   (24) \\
& \color[HTML]{141414}{Total=$10\times3\times24$=720}\\
\hline
\end{tabular}}{}
\end{table}

\section{Proposed Differential Protection Scheme}\label{ch3_algo}
Internal fault detection, feature extraction and selection, classifiers for transient detection and identification, and a proposed method are all covered in this section. 

\subsection{Change detection filter (\acrshort{CDF}) for transient detection}
The change in the differential currents in case of transients is detected by a change detection filter (CDF) which calculates the difference between the cumulative sum of modulus of two consecutive cycles.
\begin{equation} CDF (t) = \sum_{x=n_c+t}^{2n_c+t}|Id(x)|-\sum_{x=n_c+t}^{2n_c+t}|Id(x-n_c)|
\end{equation}
where x is sample number beginning at the second cycle, $n_c$ is  number of samples in a cycle, n is total number of samples, and $Id$ {represents} a, b, and c phase differential currents.

The change detection filter starts logging the data from the instant CDF(t) is greater than a threshold, $th$ in any one of the 3-phases. In normal conditions when there is no transient, the values of CDF(t) are nearer to zero \cite{chDharmapandit2017}. 

\subsection{Feature Extraction $\&$ Selection }\label{ch3feature selection}
Time series analysis of the differential currents helps in the classification and characterization of power system events. Features extracted from these time series are used as input to the ML algorithms. Informative and distinctive features that help to classify the events may range from simple statistical functions to complex ones. Researchers have used time-frequency representations like Wavelet Transform \cite{chSVM1,chSVM2,chann1,chdtwt,chPQ2,chPQ3,chPQ6} and Stockwell Transform \cite{chPQ1,chPQ4,chPQ5,chPQ6}  to extract features from the non-stationary transients to discriminate inrush and internal faults and for classification of PQ disturbances. In this chapter, to differentiate the faults from the other transient disturbances, three time-domain features
and  two frequency-domain features 
have been used.

A comprehensive number of features (794) from different domains are extracted from the 3-phase differential currents obtained from the current transformers, CT1 and CT2 located near bus4 and bus5. The complete list of the features extracted can be found in \cite{chtsfresh}. Out of these 794 features, Random Forest feature selection (Refer Chapter 2) is used to rank and select the features with maximum Information Gain to distinguish between the different classes. The most relevant and common features for each of the classification tasks obtained after performing feature ranking belong to the set \textbf{F} = \{F1, F2, F3, F4, F5\}  where, F1 {is} average change quantile, F2 {is} fast Fourier transform (\acrshort{FFT}) coefficients, F3 {is} aggregate linear trend, F4 {is} spectral welch density, and  F5 {is} autoregressive coefficients.
Only those features of set \textbf{F} which are present in each of the 3-phase differential currents
are used for training the classifiers to detect the faults, localize the faulty units, identify the fault type, and identify the disturbance type (Table \ref{ch3featuretab}). The feature set \textbf{F} is detailed in what follows.
\begin{itemize}
 \renewcommand{\labelitemi}{\scriptsize$\blacksquare$}
 \item F1, \textit{average change quantile} calculates the average of absolute values of consecutive changes of the time series inside two constant values $qh$ and $ql$.
\begin{equation} Avg.\  change\ quantile =\frac{1}{n'}\cdot{\sum_{t=1}^{n'-1} |Id_{t+1} - Id_{t}| }
\end{equation} where, $n'$  {equals} number of sample points in the differential current between $qh$ and $ql$, $Id$ {is} a, b, and c phase differential currents with n sample points. Average change quantile  was also used in the previous chapter as one of the features. This points to the possibility that few features are more useful than other while classifying power system events. 

  \item F2, \textit{FFT coefficients}, (X$|$k) returns the Fourier coefficients of 1-D discrete Fourier Transform for real input using fast FT.
\begin{equation} (X|k)= {\sum_{t=0}^{n-1}Id_t\cdot e(-  \frac{j2 \pi kt}{n} ), k\in Z}
\end{equation}

  \item F3, \textit{aggregate linear trend} calculates the linear least-squares regression for values of the time series over windows and returns aggregated values of either intercept or standard error. 

  \item F4, \textit{spectral welch density} uses Welch’s method to compute an estimate of the power spectral density by partitioning the time series into segments and then averaging the periodgrams of the discrete Fourier transform of each segment \cite{chwelch}.

 \item F5, \textit{autoregressive coefficients} are the least-square estimates of $\varphi_{i's}$ which are obtained by minimizing Eq.\ref{ch3ar} with respect to $\varphi_0, \varphi_1..., \varphi_P$ and lag P. \begin{equation}\label{ch3ar} \sum_{t=p+1}^{n} [Id_t- \varphi_0 - \varphi_1\cdot Id_{t-1} - ...-\varphi_P \cdot Id_{t-P}]^2 \end{equation}
\end{itemize}

More than one feature can be extracted from the above time and frequency domain functions by varying their parameters. e.g ($qh,ql$) = (0.8,0.4) \& (0.8,0.2) yields 2 features from change quantile and window length = 5, 10, and 15 would return 3 features of  linear trend.
\subsection{Choice of Classifiers}
Tree-based learning algorithms like decision trees, random forest, and gradient boosting are considered among the best and predominantly used supervised learning methods in problems related to data science. This is already established in Chapter 2 where decision tree, random forest, and gradient Boosting have been used for detection and classification of transients in PARs. These estimators have higher accuracy, stability and are easy to interpret. They can also handle non-linear relationships quite well. In this chapter Gradient Boosting Classifier (GBC) has been used along with DT, RFC, and SVM  to detect and classify the transients.

\subsubsection{Decision Tree}
The DT hyperparameter impurity measure: Gini, classification error, and entropy is selected using Grid Search to optimize the performance. The default parameters are used for the rest of the hyperparameters. (For more information see Chapter 2)

\subsubsection{Random Forest}

The RF hyperparameters no\_of\_estimators (number of trees in the forest), max\_depth (tree depth), and max\_features (feature size to consider when splitting a node) are chosen using Grid Search. The no\_of\_jobs parameter was also used to parallelize the construction of trees and computation of predictions by using more processing units.
RFC has also been used during feature selection and ranking  (\ref{ch3feature selection}) to get the relative importance of the features which is measured by the fraction of samples a feature contributes to and the mean decrease in impurity from splitting the samples \cite{chphdthesis}. (For more information see Chapter 2)

\subsubsection{Gradient Boosting Classifier}
GBC like RFC belongs to the class of ensemble trees which builds the base estimators from weak learners ($w_{p}(x)$) sequentially in a greedy manner which results in a strong estimator \cite{chfriedman2001} \cite{chMason}. The newly added $w_{p}$ tries to minimize the loss function given $f_{p-1}$, step length ($\lambda_p$), and input ${(x_i,y_i)}_ {i = 1}^  n$.

\begin{equation} \label{ch3eq3}
\begin{split}
f_p(x)=f_{p-1}(x)+ \lambda_p w_p(x)\\
w_p = arg\ \stackunder{min}{w} \sum_{i=1}^{n}L(y_i,f_{p-1}(x_i) +w(x_i))
\end{split}
\end{equation}

The minimization problem is solved by taking the negative gradient of the negative multinomial log-likelihood loss function, L for mutually exclusive classes.
\begin{equation}
f_p(x)=f_{p-1}(x)-\lambda_p\sum_{i=1}^{n}\nabla_{f} L(y_i,f_{p-1}(x_i))
\end{equation} GBC uses shrinkage which scales the contribution of the weak learners by the learning rate and sub-sampling of the training data (stochastic gradient boosting) for regularization. The important hyperparameters of the different GBC classifiers are the results of Grid Search on no\_of\_estimators = [5000, 7000, 10000, 12000, 15000], max\_depth = [3,5,7,10,15], and learning\_rate = [0.01, 0.05, 0.07, 0.1]. 

\subsection{Proposed scheme}
\begin{figure}[htb]
\centerline{\includegraphics[width=5in, height= 4.3 in]{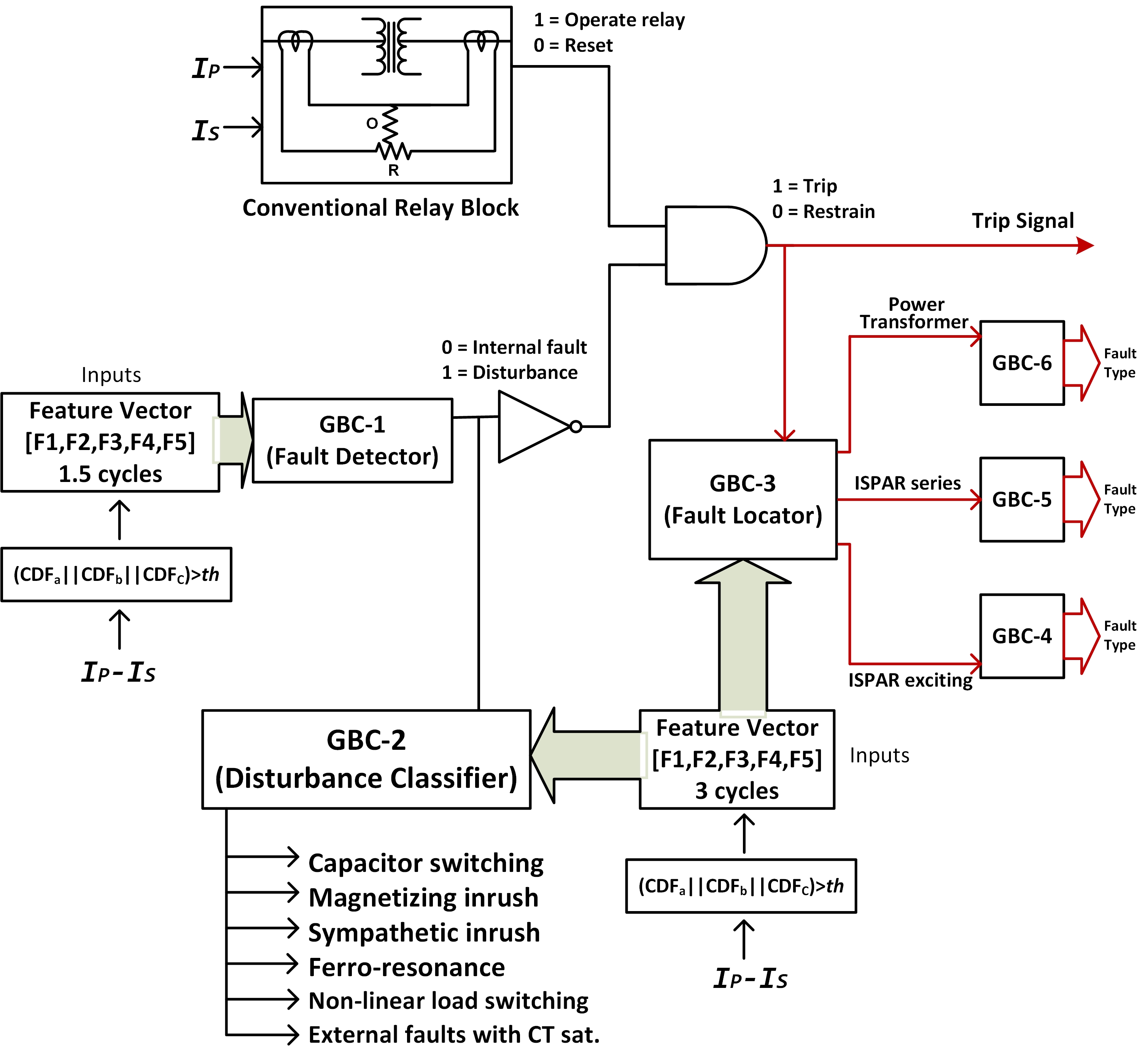}}
\vspace{0mm}\caption{Proposed transient detection and classification algorithm }
\label{ch32flowchart}
\end{figure}
The block diagram description of the CDF and GBC-based proposed internal fault detection, fault localization, and transient disturbance classification algorithm is shown in Fig.\ref{ch32flowchart}.
The change detector discovers the change in the 3-phase differential currents ($I_{P}$-$I_{S
}$) if the CDF index in any phase is greater than the threshold, $th$ = 0.05. 1/2 cycle pre-transient and 1 cycle post-transient differential current samples are used to detect an internal fault and 3 post-transient cycles are used for localization of faults and classification of transient disturbances. The scheme consists of a four-level classifier design. 
The level-1 classifier (GBC-1) consists of the fault detector, which can apply supervisory control over the operation of the conventional differential relay. GBC-1 identifies an internal fault with \say{0} and other transient disturbances with \say{1}. Hence, it governs the working of the trip/restrain function by blocking all other power system transients but an internal fault. The level-2 classifier (GBC-2) does further analysis of the power system events in case the output of GBC-1 is \say{1}.
The GBC-2 can identify the transient disturbance responsible for the mal-operation of the conventional differential relay
(GBC-1 is \say{1} \& Operate relay is \say{1}). The level-3 classifier (GBC-3) locates the faulty transformer unit (Power Transformer, ISPAR series, and ISPAR exciting) if the output of GBC-1 is \say{0}. The level-4 classifiers (GBC-4, GBC-5, and GBC-6) further identifies the internal faults in the ISPAR exciting, the ISPAR series and the Power Transformer.

\section{Results and Discussion}\label{ch3_results}
1.5 cycles of 3-phase differential currents are used for detection, and 3 cycles are used for localization and identification of transients from the time of their inception. Thus, at a sampling rate of 10 kHz, 167 sample data per cycle are analyzed. Several factors influence the classification accuracy of an algorithm. Cross-validation and grid search helps in using the data effectively and training the classifier with the best combination of hyperparameters. The data is split randomly into training and test set in a 4:1 ratio. To avoid the problem of overfitting and underfitting of the estimator on the test set, cross-validation is applied on the training data and the hyperparameters are optimized using grid search over a parameter grid. Grid search comprehensively searches for the parameters over the subset of the hyperparameter space of the estimator. The performance of the selected hyperparameters is then tested on the unseen test data that is not used during the training process. Ten-fold stratified cross-validation (rearrangement of the training data in ten folds such that each fold represents every class well) is used to select the model as it is better both in terms of bias and variance \cite{chKohavi}. 
\begin{table}[ht]
\renewcommand{\arraystretch}{0.9}
\setlength{\tabcolsep}{10pt}
\centering
\caption{Internal fault detection with CDF \& GBC-1} \label{ch3tabcdf}
{\begin{tabular}{lllll}\toprule
Fault/Disturbances  & Total & TP  & FN & FP \\
\midrule
Internal Faults   &2107 & 2105 & 2 & 0\\
Disturbances  &1852 & 1852 & 0 & 2\\
\hline
\vspace{3mm}
\end{tabular}}{}
\end{table}
\begin{table}[ht]
\renewcommand{\arraystretch}{0.9}
\setlength{\tabcolsep}{10pt}
\centering
\caption{Comparison of performances with and without CDF}\label{ch3tabcompare}
\begin{subtable}[b]{.53\linewidth}
\centering
\caption{Internal fault detection with CDF\label{ch3tabcdf2}}
{\begin{tabular}{ll}\toprule
Classifier & $\bar{\eta}$ \\
\midrule
GBC-1 & \textbf{99.95 \%}\\
DT  & 99.5\%\\
SVM & 99.7\%\\
RFC & 99.9\%\\
\hline
\end{tabular}}
\end{subtable}{}\hfill
\begin{subtable}[b]{.45\linewidth}
\centering
\caption{Internal fault detection without CDF\label{ch3tabwithoutcdf}}
{\begin{tabular}{ll}\toprule
Classifier & $\bar{\eta}$ \\
\midrule
GBC &  \textbf{98.5\%}\\
DT  & 95.3\%\\
SVM & 89.2\%\\
RFC & 94.6\%\\
\hline
\end{tabular}}{}
\end{subtable}{}
\end{table}
\subsection{Internal fault detection}
The detection of internal faults is performed using GBC in two ways, one with the CDF and the other without it. Most authors haven't considered using some technique to detect the change in differential currents in case a transient occurs. Rather they fixed the time of occurrence of the transient events and used this specified inception time to store the disturbance and fault data. However, faults and disturbances are highly unpredictable in time. In this chapter, both methods, one considering a specified time (without the use of CDF) and the other with CDF are used to register the data after the inception of transients. The CDF detects the change and registers 1/2 cycle of pre-transient and 1 cycle of post-transient samples. This 1.5 cycle (250 samples) is used to extract the relevant features which are then fed to GBC, SVM, DT, and RFC classifiers. 
Since the classes are not balanced, balanced accuracy ({$\bar{\eta}$}) computed as $
\bar{\eta}=\frac{1}{2}(\frac{TP}{TP+FN}+ \frac{TN}{TN+ FP})$ for a two-class problem is used to compute the performance.

The performance of the fault detection scheme composed of the GBC-1 and CDF is shown in Table \ref{ch3tabcdf}. $\bar{\eta}$ of 99.95\% is obtained on a training data of 15,835, testing data of 3959, and hyperparameters: learning\_rate = 0.1, max\_depth = 5, and no\_of\_estimators = 7000. The performance of the four classifiers with CDF is shown in Table \ref{ch3tabcdf2}. One cycle of post-fault data is used for training the classifiers for fault detection without the CDF. $\bar{\eta}$ of 98.52\% is obtained with GBC for max\_depth = 7,  no\_of\_estimators = 5000, and learning rate = 0.07. The balanced scores of the four classifiers trained on 80,870 cases and tested on 20,218 cases are shown in Table \ref{ch3tabwithoutcdf}. 18 features from the 3-phase differential currents (Table \ref{ch3featuretab}) 
are used as the input to the classifiers for training the fault detection models with and without CDF. GBC with CDF performed better than without CDF (Table \ref{ch3tabcompare}) as the CDF filtered out the cases where there is no appreciable change in differential currents although a transient event occurred. It is noticed that the CDF could detect the change in differential currents in all internal fault cases except turn-to-turn faults with Rf = 10$\Omega$, LTC = 0.2, and percentage of winding shorted = 20\%. Also, it detected the change for all transient disturbances except sympathetic inrush cases for switching angles from 120$^{\circ}$ to 330$^{\circ}$. On exploring the data it is observed that there is almost no change in the differential currents for these instances. The $w$-g faults for LTC = 0.2, and percentage of winding shorted = 20\% which needs higher sensitivity were detected. It proves the dependability of the scheme for ground faults near-neutral of wye grounded transformers (Power Transformer and ISPAR exciting) which is again a challenge for conventional differential relays \cite{chpstguide}.

\subsection{Identification of faulty unit $\&$ internal fault type}

Once it is confirmed that an internal fault has been detected, the locations of those internal faults are determined. 3 cycles of post-fault differential current samples are used to locate the faulty transformer unit (Power Transformer or ISPAR Exciting or ISPAR Series) and determine the type of fault. GBC, SVM, DT, and RFC are used to identify the faulty unit and further locate and pinpoint the type of fault in the Power Transformer and ISPAR units. 
$\bar\eta$ and accuracy computed as $\eta$ = $\frac{(TP+TN)}{(TP+FN+TN+FP)}$, are used as the metrics to measure the performance of the estimators for localization of faulty unit and identification of internal fault type, respectively.

\subsubsection{Localization of faulty unit}
To locate the faulty transformer unit 70,502 fault cases are trained and 17,626 cases are tested. 18 features are used to train the classifiers (Table \ref{ch3featuretab}).
GBC-3 with hyperparameters: no\_of\_estimators = 5000, learning\_rate = 0.07, and max\_depth = 10 gives $\bar\eta$ of 99.48\%. Table \ref{ch3tabfaulty_trans1} shows the localization results using GBC-3 and Table \ref{ch3tabfaulty_trans2} compares the $\bar\eta$ of the four different classifiers.

\subsubsection{Identification of internal fault type}
The internal faults in the ISPAR series, ISPAR exciting and the Power Transformer are further classified into $w_{a}$-g, $w_{b}$-g, $w_{c}$-g, $w_{a}$-$w_{b}$-g, $w_{a}$-$w_{c}$-g, $w_{b}$-$w_{c}$-g, $w_{a}$-$w_{b}$, $w_{a}$-$w_{c}$, $w_{b}$-$w_{c}$, turn-to-turn, winding-to-winding, and very rare $w_{a}$-$w_{b}$-$w_{c}$ and $w_{a}$-$w_{b}$-$w_{c}$-g faults.
21 features from 3 cycles of the 3-phase differential currents are used as the input to the estimators (Table \ref{ch3featuretab}).
Tables \ref{ch3tabexc}, \ref{ch3tabseries}, and \ref{ch3tabpt} compare the performances of GBC, RFC, DT, and SVM classifiers for ISPAR exciting, ISPAR series, and the Power Transformer respectively.

To identify the internal faults in ISPAR exciting 14,688 fault cases are used to train and test the four classifiers. GBC-4 trained with hyperparameters of max\_depth = 5, no\_of\_estimators = 7000, and learning\_rate = 0.1 achieved the best accuracy of 99.18\%. 
For the identification of internal faults in ISPAR series 36,720 cases are used to train and test the classifiers. GBC-5 trained with learning\_rate = 0.05, max\_depth = 7, and no\_of\_estimators = 5000 gives an accuracy of 98.0\%. 
Similarly, for Power Transformer the classifiers are trained \& tested on 36,720 fault cases. GBC-6 achieved the best accuracy of 99.2\% obtained by training the hyperparameters on learning\_rate = 0.05, no\_of\_estimators = 5000, and max\_depth = 5.
The identification accuracy obtained in the ISPAR series is lower than in Power Transformer and ISPAR exciting because the secondary side of the ISPAR series is delta connected. Hence, one type of fault on the primary side confuses with another type on the secondary side. 


\begin{table}[ht]
\renewcommand{\arraystretch}{0.9}
\centering
\caption{Localization of faulty transformer unit}
\setlength{\tabcolsep}{5pt}
\begin{subtable}[l]{.55\linewidth}
\centering
\vspace{-5pt}
\caption{ Localization with GBC-3\label{ch3tabfaulty_trans1}}
{\begin{tabular}{lllll}\toprule
  Transformer  &   Total &   TP  &   FN &   FP\\
\midrule
  ISPAR Exciting&  2937 &   2899 & 38 &  8\\
  ISPAR Series & 7402 & 7383 &  19 &  17\\
 {Power Transformer} & 7287 & 7287 & 0&  32\\
\hline
\end{tabular}}{}
\end{subtable} \hfill
\begin{subtable}[r]{.39\linewidth}
\centering
\vspace{4pt}
\caption{Comparison of performances\label{ch3tabfaulty_trans2}}
{\begin{tabular}{ll}\toprule
  Classifier & $\bar{\eta}$ \\
\midrule
  GBC-3 & \textbf{99.5\%}\\
  DT  &98.6\%\\
  SVM & 88.9\%\\
  RFC & 98.7\%\\ \hline
\end{tabular}}
\end{subtable}
\end{table}

\begin{table}[ht]
\renewcommand{\arraystretch}{0.9}
\setlength{\tabcolsep}{10pt}
\centering
\caption{Comparison of identification performances of internal fault type}
\begin{subtable}[b]{.3\linewidth}
\centering
\caption{Exciting unit \label{ch3tabexc}}
{\begin{tabular}{ll}\toprule
Classifier & $\eta$ \\
\midrule
GBC-4 & \textbf{99.2\%}\\
DT  & 98.6\%\\
SVM & 94.8\%\\
RFC & 98.9\%\\
\hline
\end{tabular}}{}
\end{subtable}{}\hfill
\begin{subtable}[b]{.3\linewidth}
\centering
\caption{Series unit\label{ch3tabseries}}
{\begin{tabular}{ll}\toprule
Classifier & $\eta$ \\
\midrule
GBC-5 &  \textbf {98.0\%}\\
DT  &94.7\%\\
SVM & 90.7\%\\
RFC &96.9\%\\
\hline
\end{tabular}}
\end{subtable}{}\hfill
\begin{subtable}[b]{.3\linewidth}
\centering
\caption{Power Transformer\label{ch3tabpt}}
{\begin{tabular}{ll}\toprule
Classifier & $\eta$ \\
\midrule
GBC-6 & \textbf{99.2\%}\\
DT  & 98.9\%\\
SVM & 94.0\%\\
RFC & 97.8\%\\
\hline
\end{tabular}}{}
\end{subtable}{}
\end{table}

\subsection{Identification of disturbance type} 
\begin{figure}[htb]
\centerline{\includegraphics[width=5.8 in, height= 3.7 in]{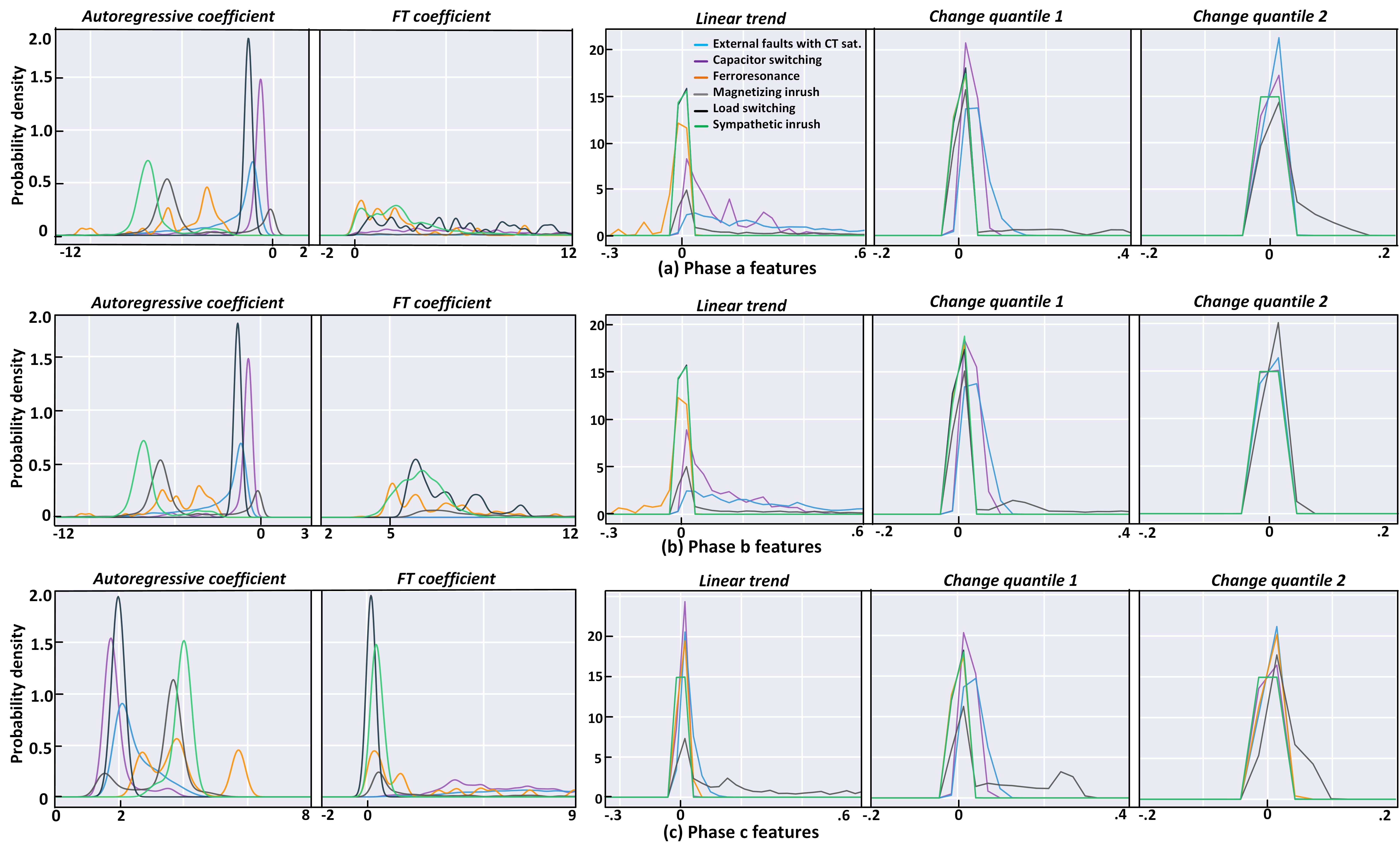}}
\vspace{0mm}\caption{Kernel Density Estimate plots showing the probability distribution of the 5 selected features for the 6 transient disturbances in (a) phase a, (b) phase b, and (c) phase c.}
\label{ch3kdeplot}
\end{figure}

\begin{figure}[htb]
\centerline{\includegraphics[width=4.0 in, height=2.5 in]{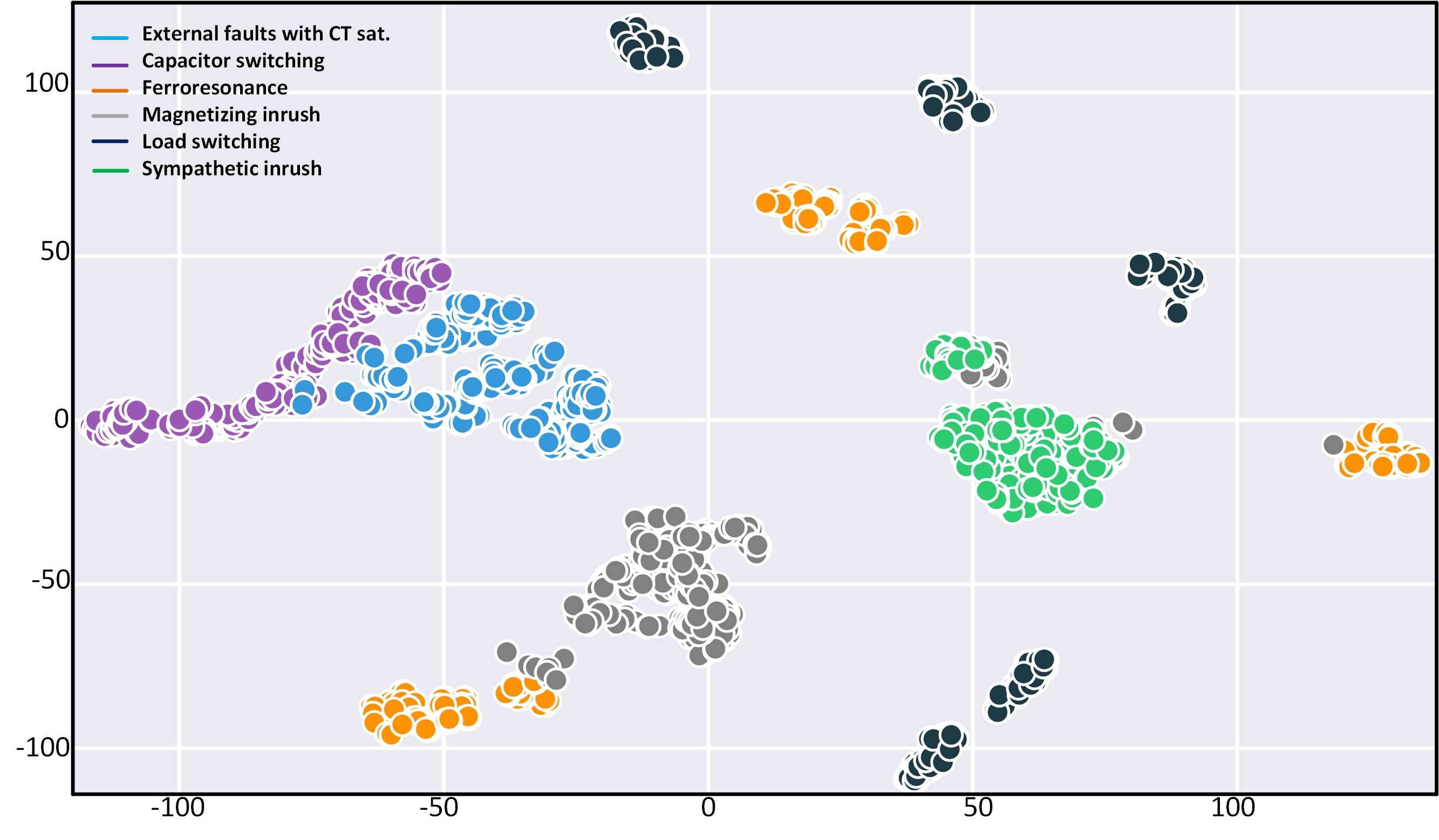}}
\vspace{0mm}\caption{2D scatterplot of the input features for transient disturbances}
\label{ch3scatter}
\end{figure}

The various disturbances: magnetizing inrush, sympathetic inrush, ferroresonance, external faults with CT saturation, capacitor switching, and non-linear load switching are also classified using 3 cycles of post-transient samples after they are differentiated as no-fault by the fault detection scheme. 
15 features are used as input to the classifiers in this case (Table \ref{ch3featuretab}).
{It's always useful to know the probabilities of the input features taking on various real values. Parzen–Rosenblatt window method is used to estimate the underlying probability density of the 5 features for the six different disturbances in phases a, b, and c. Fig.\ref{ch3kdeplot} shows the kernel density estimation plots for the chosen features. Gaussian is used as the kernel function to approximate the univariate features with a bandwidth of 0.2 for autoregressive coefficient, and FFT coefficient and a bandwidth of 0.01 for aggregate linear trend, and avg. change quantile 1 and avg. change quantile 2. It is observed that probability density functions of autoregressive and FFT coefficients are a mixture of multiple normal distributions with varying standard deviation and mean whereas linear trend and change quantiles are unimodal with means near zero and smaller standard deviations. Table \ref{ch3feature1} and \ref{ch3feature2} shows the values of mean ($\mu$), variance ($\sigma^2 $), skewness ($\Tilde{\mu_3}$), and kurtosis ($\kappa$) of the 5 features for magnetizing inrush and CT saturation during external faults respectively in phases a, b and c. The feature statistics of only these two transients are shown. 
Furthermore, to visualize the 15-dimensional input data in a 2-dimensional plane, the T-distributed Stochastic Neighbor Embedding dimensionality reduction technique has been used which preserves much of the significant structure in the high-dimensional data in the 2-dimension while mapping \cite{chtsne}. Fig.\ref{ch3scatter} shows the clusters of similar transients (300 instances each) and also the relationships between different groups of transients as a scatter plot.}

The Table \ref{ch3tabdisturbance1} shows the classification results using GBC-2. Table \ref{ch3tabdisturbance2} compares the results of GBC with RFC, DT, and SVM. The classifiers are trained on 10,368 cases and tested on 2592 cases. $\bar{\eta}$ of 99.28\% is obtained with GBC-2 having hyperparameters: no\_of\_estimators = 5000, learning\_rate = 0.7, and max\_depth = 3.

\begin{table}[ht]
\renewcommand{\arraystretch}{0.9}
\setlength{\tabcolsep}{5pt}
\centering
\caption{Low and high-order statistics of the 5 selected features}
\begin{subtable}[b]{\linewidth}
\centering
\caption{Magnetizing inrush\label{ch3feature1}}
\scriptsize
\begin{tabular}{@{}lllllllllllllllllllll@{}}
\toprule
\multirow{2}{*}{} & \multicolumn{4}{c}{\begin{tabular}[c]{@{}c@{}}autoregressive\\ coefficient\end{tabular}} & \multicolumn{4}{c}{\begin{tabular}[c]{@{}c@{}}FT\\ coefficient\end{tabular}} & \multicolumn{4}{c}{\begin{tabular}[c]{@{}c@{}}linear \\ trend\end{tabular}} & \multicolumn{4}{c}{\begin{tabular}[c]{@{}c@{}}change\\ quantile 1\end{tabular}} & \multicolumn{4}{c}{\begin{tabular}[c]{@{}c@{}}change \\ quantile 2\end{tabular}} \\ \cmidrule(l){1-21} 
                  & $\mu  $ & $\sigma^2$  &      $\Tilde{\mu_3}  $    & $\kappa$         &  $\mu  $ & $\sigma^2$  &      $\Tilde{\mu_3}  $    & $\kappa$&$\mu  $ & $\sigma^2$  &      $\Tilde{\mu_3}  $    & $\kappa$                  &  $\mu  $ & $\sigma^2$  &      $\Tilde{\mu_3}  $    & $\kappa$                 &   $\mu  $ & $\sigma^2$  &      $\Tilde{\mu_3}  $    & $\kappa$                                        \\ \midrule
\it {ph a}                  &       -4.6                 &        5.9               &     .74               &        -.35            &       1e3            &       2e6            &       1.3            &        .74          &        2.6           &    14               &      1.9           &     3.9             &      .02            &        9e-4            &        1.9           &        2.9           &       .04             &    .008                &  2.5                  &      5.1             \\
\it {ph b}                 &        -4.7                &    6.2                  &.65                    &   -.13                &      502             &     3e5              &      1.0             &       -.06           &       2.6            &    13               &      1.8            &         3.2         &         .01           &        8e-5            &        1.5           &         1.06          &     .04               &         .01           &          2.5          &           5.7        \\

\it {ph c}                &      3.3                  &  .72                     &      -1.1              &      .33              &       33            &        1e3          &     1.1              &        .25          &      .19             &    .05               &      1.0           &    -.35              &        .02            &       5e-4             &    .90               &     -.45              &        .08            &   .009                 &         .88           &     -.84              \\ \bottomrule
\end{tabular}
\end{subtable}
\begin{subtable}[b]{\linewidth}
\vspace{2 mm}
\centering
\caption{ CT saturation during external faults\label{ch3feature2}}
\setlength{\tabcolsep}{5pt}
\scriptsize
\begin{tabular}{@{}lllllllllllllllllllll@{}}
\toprule
\multirow{2}{*}{} & \multicolumn{4}{c}{\begin{tabular}[c]{@{}c@{}}autoregressive\\ coefficient\end{tabular}} & \multicolumn{4}{c}{\begin{tabular}[c]{@{}c@{}}FT\\ coefficient\end{tabular}} & \multicolumn{4}{c}{\begin{tabular}[c]{@{}c@{}}linear \\ trend\end{tabular}} & \multicolumn{4}{c}{\begin{tabular}[c]{@{}c@{}}change\\ quantile 1\end{tabular}} & \multicolumn{4}{c}{\begin{tabular}[c]{@{}c@{}}change \\ quantile 2\end{tabular}} \\ \cmidrule(l){1-21} 
                  & $\mu  $ & $\sigma^2$  &      $\Tilde{\mu_3}  $    & $\kappa$         &  $\mu  $ & $\sigma^2$  &      $\Tilde{\mu_3}  $    & $\kappa$&$\mu  $ & $\sigma^2$  &      $\Tilde{\mu_3}  $    & $\kappa$                  &  $\mu  $ & $\sigma^2$  &      $\Tilde{\mu_3}  $    & $\kappa$                 &   $\mu  $ & $\sigma^2$  &      $\Tilde{\mu_3}  $    & $\kappa$                                        \\ \midrule
\it {ph a}                  &       -2.3                 &        3.1               &     -1.6               &        1.97            &      190            &       4e4            &       1.6            &        2.5         &        .47           &    .23              &      1.64          &     2.9             &      .04            &        6e-4            &        1           &        .98           &       .003             &    9e-6                &  1.8                  &     3.6           \\
\it {ph b}                 &        -2.2               &    3                 &-1.6                   &   1.95               &      261             &     2e4              &      .81            &      .35          &     .47 & .24&1.55&2.3&7e-4           &    9e-7               &     3.5            &        19      &         .04          &        5e-4            &        .6          &       -.42            \\

\it {ph c}                &     2.5                  &  .4                     &      1.1            &      .27              &       16           &        150         &     1.7             &        2.9          &      .02            &       5e-4           &    1.5 & 2.4            &        .003           &       8e-6             &    1.7               &     4.7              &        .04           &   4e-4                 &         .67           &     -.14              \\ \bottomrule
\end{tabular}
\end{subtable}
\vspace{5mm}
\end{table}

\begin{table}[ht]
\renewcommand{\arraystretch}{0.9}
\setlength{\tabcolsep}{10pt}
\centering
\caption{Identification of transient disturbances with GBC-2\label{ch3tabdisturbance1}}
{\begin{tabular}{lllll}\toprule
Disturbances   & Total & TP  & FN &   FP\\
\midrule
Magnetizing inrush &  365& 357 & 8 &0\\
Sympathetic inrush  &336 & 336 &0 &8\\
Capacitor switching  & 73 & 72 &1 & 1\\
Ferroresonance & 133 & 132 & 1 & 0\\
Load switching & 69 & 69 & 0 & 0\\
External faults  & 1616 & 1615 & 1 & 2\\
\hline
\end{tabular}}{}
\vspace{5mm}
\end{table}

\begin{table}[ht]
\centering
\noindent\begin{minipage}{2.9 in}
\centering
\setlength{\tabcolsep}{7 pt}
\renewcommand{\arraystretch}{0.9}
\caption{Performance comparison of identification  of transient disturbances}\label{ch3tabdisturbance2}
\begin{tabular}{lc}\toprule
Classifier & $\bar{\eta}$ \\
\midrule
GBC-2 & \textbf{99.28\%}\\
DT  & 98.09\%\\
SVM & 98.23\%\\
RFC & 98.89\%\\ 
\hline
\end{tabular}
\end{minipage}
\begin{minipage}{3 in}\centering
\setlength{\tabcolsep}{5 pt}
\renewcommand{\arraystretch}{0.9}
\caption{Misclassification between ISPAR \& Power Transformer}\label{ch3tabptpst}
\begin{tabular}{lcccc}\toprule
Faults   & Total & TP  & FN & FP \\\midrule
PT   &7262 & 7262 & 0 & 13\\
ISPAR  &10364 & 10351  & 13 & 0\\\hline
\end{tabular}
\end{minipage}
\end{table}

\subsection{Discriminate faults in Power Transformer \& ISPAR}
The PAR controls the power flow through a line and when connected with a Power Transformer, it reduces the magnitude of the differential currents and their harmonic contents and alters the wave shapes due to the additional phase shift for the Power Transformer in case of external faults. With internal faults, such changes in the differential currents are lesser. 70,502 internal fault cases are trained and 17,626 cases are tested on 18 features to verify how effectively the GBC differentiates the internal faults in the Power Transformer (PT) and the ISPAR.
The table \ref{ch3tabptpst} shows the classification errors with learning\_rate of 0.05, max\_depth = 9, and no\_of\_estimators = 5000. The balanced accuracy of 99.9\% shows that the GBC is capable of distinguishing these faults even in an interconnected network.

\subsection{Performances on balanced and imbalanced data}
Machine learning algorithms are more reliable when they operate on a balanced dataset. To adjust the data distribution of classes and remove class imbalance, under-sampling of majority classes and over-sampling of minority classes is performed. Synthetic Minority Over-Sampling Technique (SMOTE) \cite{chsmote} is used to create minority synthetic data considering k-nearest neighbors and NearMiss algorithm is used for under-sampling the majority classes avoiding information loss. Table \ref{ch3featuretab} shows the balanced accuracy/ accuracy for detecting the internal faults, identifying the faulty units and type of faults in those units, and identifying the disturbances with and without using SMOTE and NearMiss. It is observed that the accuracies obtained with SMOTE and NearMiss algorithm for the different classification tasks are similar to those obtained by training the GBCs without them. Table \ref{ch3featuretab} also gives the information about the time and frequency domain features $({\{F_i\}}_{i=1}^{5})$ that has been used to train the different GBC classifiers for the different classification tasks. 

\begin{table}[ht!]
\renewcommand{\arraystretch}{0.9}
\setlength{\tabcolsep}{6pt}
\centering
\caption{Input features and performance of different GBC classifiers with and without SMOTE analysis\label{ch3featuretab}}
\begin{tabular}{lcccccccc}\toprule
 Classification task  &F1&F2&F3&F4&F5& $\sum Fi$$\times3$& \multicolumn{1}{c}{\begin{tabular}[c]{@{}c@{}}$\bar{\eta}$\textbackslash${\eta}$ \\  (\%)\end{tabular}}   &  \multicolumn{1}{c}{\begin{tabular}[c]{@{}c@{}}\textit{$\bar{\eta}$ using}\\ \textit{SMOTE} \end{tabular}}  \\
 \midrule
Detect  Faults  &2&1&1&1&1&18& 99.9&99.9\\ 
Locate Faulty Units &2  &2&2&-&-&18& 99.5&99.6\\ 
 Identify Faults (series)  &3&1&2&1&-&21& 98.0&98.2\\ 
 Identify Faults (exciting)  &3&2&2&-&-&21& 99.2&99.1\\ 
 Identify Faults (PT)  &3&2&2&-&-&21&99.2&99.1\\ 
 Identify Transients   &2&1&1&-&1&15& 99.3&99.4\\ \bottomrule
\end{tabular}{}
\end{table}

\subsection{Effect of different ratings and transformer connections}
It is not necessary to train the fault detection scheme for different ratings and connections of the Power Transformers, rating of ISPAR, and variation in other parameters. In order to validate the effectiveness of the proposed scheme with variation in different system parameters, new internal faults and other transient cases are simulated again with 400 MVA, Y$\Delta$ connected Power Transformers and 400 MVA ISPARs. The FR, LTC, FT, and FIT are altered to generate the internal fault cases and ST, FA, LTC, etc. are altered to generate the transient cases to test the same GBC-1 model trained using 500 MVA, YY connected Power Transformers and 500 MVA ISPARs. It is observed from Table \ref{ch3dif connection} that the proposed scheme gives a balanced accuracy of 99.3\% which is compatible with the accuracy obtained when trained and tested at 500 MVA and YY connection.

\begin{table}[]
\renewcommand{\arraystretch}{0.9}
\setlength{\tabcolsep}{7pt}
\centering
\caption{Performance for 400-MVA \& Y$\Delta$ connection\label{ch3dif connection}}
\begin{tabular}{@{}clcccc@{}}
\toprule
\begin{tabular}[c]{@{}c@{}}Fault/\\ Disturbances\end{tabular}                           & \multicolumn{1}{c}{\begin{tabular}[c]{@{}c@{}}Faults/\\ Abnormalities\end{tabular}} & Total & TP   & FN & \multicolumn{1}{c}{\begin{tabular}[c]{@{}c@{}}$\bar{\eta}$\\ (\%)\end{tabular}} \\ \midrule
\multirow{4}{*}{\begin{tabular}[c]{@{}c@{}}Internal \\ faults\\ (3072)\end{tabular}} & (a) ph \& g, t-t, w-w faults (PT)                                                         & 1200  & 1200 & 0  &      100               \\
                                                                                        & (b) ph \& g, t-t, w-w faults (Series)                                                     & 1200  & 1200 & 0  &  100                   \\
                                                                                        & (c) ph \& g, t-t, w-w faults (Exciting)                                                   & 672   & 672  & 0  &     100                \\ \cmidrule(l){2-6}
                                                                                        & (d) Total  =  (a)+ (b)+ (c)                                                               & 3072  & 3072 & 0  &    100                 \\ \midrule
\multirow{6}{*}{\begin{tabular}[c]{@{}c@{}}Other \\ disturbances\\ (876)\end{tabular}}  & (e) Capacitive switching                                                                & 60    &   51   & 9   &     85                \\
                                                                                        & (f) External faults with CT saturation                                                & 528   &   525   &  3  &     99.4                \\
                                                                                        & (g) Ferroresonance                                                                      & 24    &  24    &   0 &    100                \\
                                                                                        & (h) Magnetizing inrush                                                                  & 60    & 60     & 0  &      100               \\
                                                                                        & (i) Load switching                                                                      & 144   &  144    &  0  &      100               \\
                                                                                        & (j) Sympathetic inrush     &   60   &  60  &    0 & 100                 \\  \cmidrule(l){2-6}
\multicolumn{1}{l}{}                                                                    & (k) Total = (e)+ (f)+ (g)+ (h)+ (i)+ (j)                                                                               & 876   &   864   &  12  &     98.6                \\ \midrule
Total (3948)                                                                             & Total faults and disturbances =   (d)+(k)                                                               &  3948     &  3936    &  12  &   99.3                  \\ \bottomrule
\end{tabular}
\end{table}

\subsection{Effect of Signal Noise}
\begin{figure}[h]
\centerline{\includegraphics[width=6.3 in, height= 1.6 in]{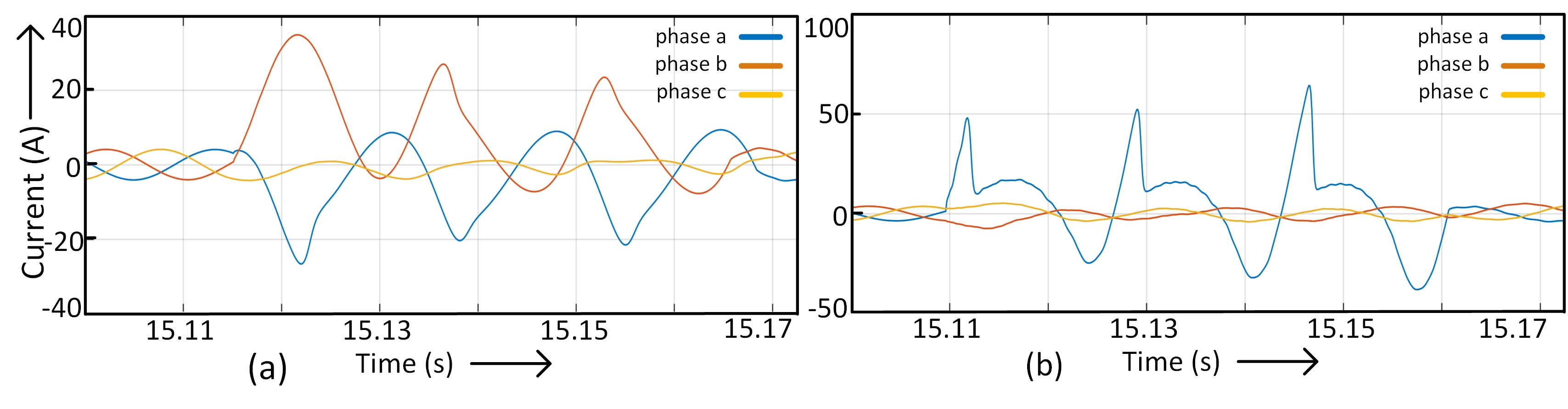}}
\vspace{0mm}\caption{3-phase differential currents with CT saturation for (a) $w_{a}$-$w_{b}$-g fault in transformer primary and (b) $w_{a}$-g fault in series primary}
\label{ch3cts}
\vspace{5mm}
\end{figure}

{In order to analyze the effect of noise in the differential currents on the proposed fault detection scheme white Gaussian noise of different levels measured in terms of Signal-to-Noise-ratio (SNR) are added to the training and testing cases for fault detection \cite{chPQ2,chPQ4,chPQ5,chPQ6}. Table \ref{ch3noise} shows the accuracy of the GBC for different levels of noise on 5000 cases of internal faults and other disturbances each. It is observed that as the level of noise increases the ${\eta}$ of the classifier dips, but still always above 93.8\% \footnotesize{($\frac{90.4+97.2}{2}$)}. \normalsize The ${\eta}$ changes from 99.4\% to 93.8\% as the SNR is varied from $\infty$ to 10dB. It is also observed from the table that the misclassification of internal faults increases as the SNR is decreased whereas the misclassifications are nearly the same for other disturbances as SNR is decreased from 30dB to 10dB. 
\subsection{Effect of CT Saturation}
To examine the effect of CT saturation the secondary side impedance (burden and CT secondary impedance) which has the major influence over the level of saturation is changed. $\eta$  of 99.5\% is obtained with GBC on 5000 cases of internal faults and other disturbances each. Fig.\ref{ch3cts} shows the 3-phase differential currents with CT saturation for faults in T1 and ISPAR1.}


\begin{table}[]
\renewcommand{\arraystretch}{0.9}
\setlength{\tabcolsep}{6pt}
\centering
\caption{Effect of Noise\label{ch3noise}}
\begin{tabular}{@{}cccccc@{}}
\toprule
\multirow{2}{*}{\begin{tabular}[c]{@{}c@{}}Fault/\\Disturbances\end{tabular}} & \multirow{2}{*}{\begin{tabular}[c]{@{}c@{}}SNR (dB)\end{tabular}} & \multirow{2}{*}{\begin{tabular}[c]{@{}c@{}}Number\\ of cases\end{tabular}} & \multicolumn{2}{c}{Predicted class} & \multirow{2}{*}{\begin{tabular}[c]{@{}c@{}}Accuracy\\ (\%)\end{tabular}} \\ \cmidrule(lr){4-5}
\multicolumn{2}{c}{}                                                                                          &                                                                            & Faults        & Disturbances        &                                                                        \\ \midrule
\multirow{4}{*}{\begin{tabular}[c]{@{}c@{}}Internal \\ faults\end{tabular}}                 & $\infty $             & 1010                                                                       &    1001         & 9                &         99.2                                                               \\
                                & 30             &   1035                                                                         &      1005       &       30               &    97.1                                                                  \\
                                                                                            & 20              &       1008                                                                     &      934         &       74              &  92.7                                                                      \\
                                                                                            & 10              &     984                                                                      &     891        &      93              &    90.4       \\ \midrule                                                              
\multirow{4}{*}{\begin{tabular}[c]{@{}c@{}}Other \\ disturbances\end{tabular}}              &  $\infty $             &   990                                                                     &       3       &         987            &    99.7                                                                    \\
                                 &   30              &     965                                                                      &     24          &       941               &    97.6                                                                   \\
                                                                                            &    20             &  992                                                                         &    26          &    966             &   97.4                                                                     \\
                                                                                            &  10               &  1016                                                                         &   28         &    988               &      97.2         \\ \midrule                                                          
\end{tabular}
\end{table}

\subsection{Execution Time}
The execution time-averaged over 10 runs for the feature extraction, training, and testing of the GBC classifiers for detection of internal faults, identifying the faulty unit and type of fault, and identifying the transient using one CPU core is reported using the in-built library in python (Table \ref{ch3ptime}).
The fault/no-fault decision includes the time to compute the feature and testing a single instance with GBC-1 which adds to 8.7ms with the CDF. Thus, the proposed scheme has a processing time of 25.37ms (16.67+1.7+7) or $\approx 1\frac{1}{2}$ cycle to detect a fault. Considering that these computations can be further optimized for example by converting Python and MATLAB code to a compiled low-level language such as C, the fault detection and localization, and transient identification schemes are suitable for future real-time implementation.{The DT, SVM, RFC, and GBC classifiers are built-in Python 3.7 using Scikit-learn framework \cite{chscikit2} while the CDF is implemented in MATLAB 2017. The pre-processing of the data is done in Python and MATLAB. All PSCAD simulations are carried out on Intel Core i7-6560U CPU @ 2.20 GHz and 8 GB RAM  and the classifiers are run on Intel Core i7-8700 CPU @ 3.20 GHz and 64 GB RAM.}

\begin{table}[ht!]
\renewcommand{\arraystretch}{0.9}
\setlength{\tabcolsep}{6pt}
\centering
\caption{Execution time of the GBC classifiers (in seconds)}\label{ch3ptime}
\begin{tabular}{lcccccc}
\toprule
\multirow{2}{*}{Classification task} & \multirow{2}{*}{\begin{tabular}[c]{@{}c@{}}Training \\ instances\end{tabular}} & \multirow{2}{*}{\begin{tabular}[c]{@{}c@{}}Testing\\ instances\end{tabular}} & \multirow{2}{*}{\begin{tabular}[c]{@{}c@{}}Training\\ time\end{tabular}} & \multicolumn{2}{l}{Testing time} & \multirow{2}{*}{\begin{tabular}[c]{@{}c@{}}Feature\\ extraction \\ time\end{tabular}} \\ \cmidrule(lr){5-6}
                                     &                                                                                &                                                                              &                                                                          & One             & All            &                                                                                       \\ \addlinespace[1mm]

\midrule

 {Detect Faults  }&  80870 & 20218 & 1506 &   0.0024 & 1.1&  0.007      \\
 Detect Faults with CDF &  15835 & 3959 & 19 &   0.0017 &0.06&  0.007       \\
 {Locate Faulty Units }&  70502 & 17626 & 1232  & 0.0049 & 1.2 & 0.0088            \\
 {Identify Faults (series) }& 29376 & 7344 &  1341  &0.016 & 2.1 &  0.0111            \\
 {Identify Faults (exciting) }&  11750 & 2938 &  357 &0.013 & 0.48 & 0.0108             \\
 {Identify Faults (PT)}& 29376 & 7344 & 2712  &0.026 &5.1 & 0.0108           \\

 {Identify Transients }  & 10368 & 2592 & 72.3  & 0.004 &  0.09   & 0.0065     \\ \midrule
\end{tabular}
\end{table}

\subsection{Comparison with Previous Works}
 This work distinguishes faults from the six transients ($\bar{\eta}$ = 99.95\%), locates the faulty unit ($\bar{\eta}$ = 99.5\%), identifies the fault type (${\eta}$ $\approx$ 99\%) and six other transients ($\bar{\eta}$ = 99.3\%) for the ISPAR and Power transformer in an interconnected system, whereas the publications \cite{chisspit} \cite{chtencon}  in the literature focused only on the ISPAR. In \cite{chisspit}, only the internal faults in ISPAR were identified and in \cite{chtencon}, the internal faults were differentiated from magnetizing inrush using WT and then the internal faults were identified. In addition to its broader functionality, the current work improves the accuracy from an average of 98.76 \cite{chisspit} and 97.7\%  \cite{chtencon} to 99.2\%.
 
\section{Summary}\label{ch3_summary}
This chapter differentiates the internal faults from the other transient disturbances in a 5-bus interconnected power system with Power Transformers and Phase Angle Regulators. The internal faults including turn-to-turn and winding-to-winding faults in the ISPAR and the Power Transformer are distinguished from magnetizing inrush, sympathetic inrush, ferroresonance, external faults with CT saturation, capacitor switching, and non-linear load switching transients. A change detector is used to detect the change in the 3-phase differential currents in case a transient event occurs and registers the current samples for detection and classification purposes. Five most relevant time and frequency domain features, selected from the differential currents on the basis of Information Gain are used to train the DT, RFC, GBC, and SVM classifiers. The fault detection scheme comprising of the CDF and GBC gives an accuracy of 99.95\% on 19,794 transient cases obtained by varying different parameters for the internal faults and other transient disturbances confirming its dependability for internal faults and security against transient disturbances. Once an internal fault is detected and a trip signal is issued using 1.5 cycles, the faulty transformer unit (Power Transformer, ISPAR series, or ISPAR exciting unit) and type of internal faults in those units are also identified in 3 cycles. Furthermore, the type of transient disturbance is determined in case the fault detection scheme detects a transient other than internal faults. {The validity of the scheme is also established for different rating and connection of the transformers, CT saturation, and SNR ratio of 30dB to 10dB in the differential currents.}
The proposed fault detection strategy can work together with a conventional differential relay offering supervisory control over its operation and thus avoid false tripping. The transient detection and identification accuracies obtained are among the best even when compared with results from works on isolated and simple networks.

\newpage
\chapter{Distance Protection of Transmission Lines Connected to Wind Farms}\label{ch4}

\section{Introduction}

Power system components are associated with protection challenges that are unique to them. Since most equipment are linked to transmission lines, a better understanding of protection-related analysis results from studying transmission line protection. The selection of the protection system depends on several factors. Security, dependability, sensitivity, selectivity, coordination, fault clearing speed, simplicity, and economic evaluation are some of the most important factors to consider \cite{chstdtlines}. 

Recent research efforts aimed at increasing the usage of wind energy have resulted in the rapid integration of WFs in distribution and transmission networks.  However, the growth of wind-based distributed generations (\acrshort{DG}) brings new challenges to the existing complex distribution and transmission line protection system.  Type-3 wind turbine generators (\acrshort{WTG}),
 also known as doubly-fed induction generators (\acrshort{DFIG}) are widely employed in wind energy generation, with the variable frequency converter using 25\%-30\% of the rated power to achieve full control of the DFIG \cite{chqiao}. Since the WFs must remain connected even during faults in the grid to maintain stability, the WF protection system has received more attention in the past decade. Different fault ride-through (\acrshort{FRT}) techniques have been developed to protect the DFIGs during faults, despite their limitations. The miscoordination of over-current relays in the distribution systems using FRT schemes has been studied in \cite{chZeineldin}\cite{chwan}. Generally, impedance type distance relay is used in the primary and/or the backup protection of HV lines \cite{chstdtlines}. However, distance relays may operate unreliably in the case of lines connected to WFs, posing a threat to the security of the protection systems.

To have a better knowledge of the protection requirements, it is critical to examine the fault characteristics of these wind turbines before they are connected to the major grid. The fault studies aid protection engineers in determining the circuit breaker and relay specifications, as well as the protection strategies to be employed. Unlike type-1 and type-2 WTGs where the fault characteristics are governed by the system, and WTG impedances, the fault profiles in type-3 and type-4 are complex and depend on the controllers used \cite{chchen}. In type-3 generators, the crowbar circuit is activated during severe faults (e.g. balanced faults) to protect the rotor side converter from high currents. The short circuit studies for such WTGs are not universal and are design specific. As a result, the protection of lines connected to such WTGs adds to the already difficult task. 

An adaptive distance relay setting is proposed for protecting the transmission system connected to a wind farm using local information of current, voltage, and number of wind farm units in \cite{chpradhan}.
Reference \cite{chsauvik} studied the impact of UPFC connected TLs on distance relay performance and proposed a fault detection and classification scheme based on the positive-sequence current.
An improved scheme based on time delay and zero-sequence impedance is proposed in \cite{chfang}. 
A 750-kV power transmission line protection is studied under different situations when a fault occurs in the presence of a DFIG \cite{chcanada_j}.
The impact of penetration levels of wind power in distribution systems on the dynamic performance of distance relays that use prefault voltages (memory voltage) as polarizing quantity was analyzed in \cite{chkhalil}.

Previously, the waveshape of faults was employed to protect T-lines connected to WFs. Support Vector Machine based differential protection was proposed to distinguish normal operation, internal and external faults correctly in \cite{chrezaei}. Reference \cite{chhosiyar1} suggested a modified permissive overreach transfer trip scheme for proper and speedier operation of distance relays in lines connected to Type-3 WTGs over the entire length of line without requiring substantial communication bandwidth.
The percentage decline if the current phasor was used to discriminate the faults in the Type-3 WTGs based substation in \cite{chhosiyar2}.

This chapter, which was published in \cite{peci} presents a protection strategy based on auto-regressive (\acrshort{AR}) coefficients of 3-phase current measurements from one end to protect lines connected to Type-3 WTGs and identify the faults fed by the WF and the grid. It studies the uncertain operation of distance relay connected to the WF in the 4-bus test system during balanced faults. The AR coefficient based method is verified on different test systems and scenarios simulated by considering various parameters that may affect the fault characteristics. In addition, the impact of active and reactive power flow control devices on the proposed scheme is analyzed.

The following is how the rest of the chapter is structured:
Section II describes the modeling and simulation of the 4-bus test system.
 The proposed Auto-regressive coefficient based protection scheme is developed in section III. Section IV summarizes the findings and validates the scheme on IEEE-9 and IEEE-39 bus systems. This section also evaluates the effect of series capacitors and phase shifters. Section V draws the conclusions.

\section{Modeling and Simulation}
We used a 4-bus test system with two external sources, transmission lines, and distance relay, connected to a WF, modeled in PSCAD/EMTDC (See Fig.\ref{ch4ckt}). It is described in the Appendix. Frequency-dependent T-lines are used for the overhead lines. Five different locations are chosen to place the balanced faults. The relay currents are pre-processed in MATLAB and Python. 
The impedance plane of the quadrilateral relay connected to the WF reflects the sequence of changes in the fault current. The current and voltage phasors are sampled at 1920 Hz and fifth-order low pass Butterworth filters are used in the relay to obtain the current and voltage.

\begin{figure}[ht]
\centerline{\includegraphics[width=3.9 in, height= 1.2 in]{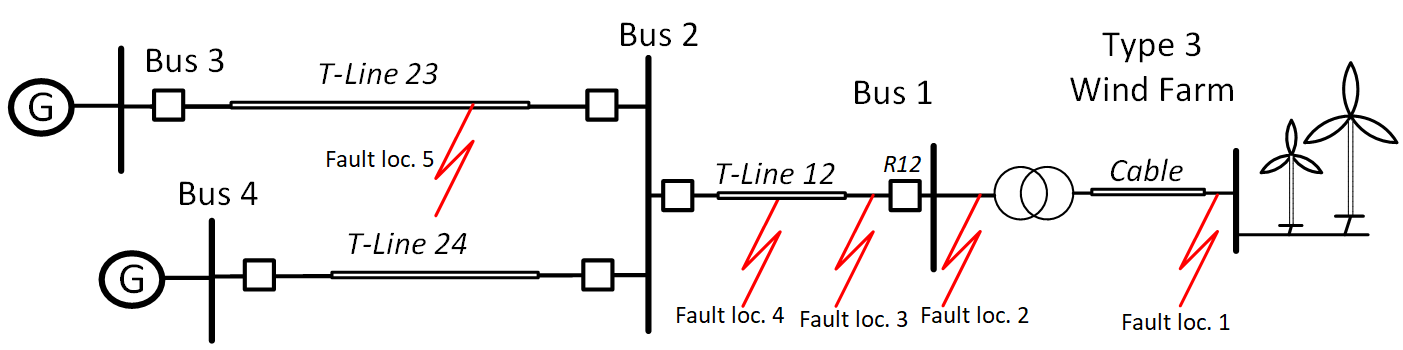}}
\caption{Single line diagram of the test system in PSCAD with the generations, transmission lines, and relays}
\label{ch4ckt}
\end{figure}

\begin{figure}[ht]
\centerline{\includegraphics[width=3.7 in, height= 4.1 in]{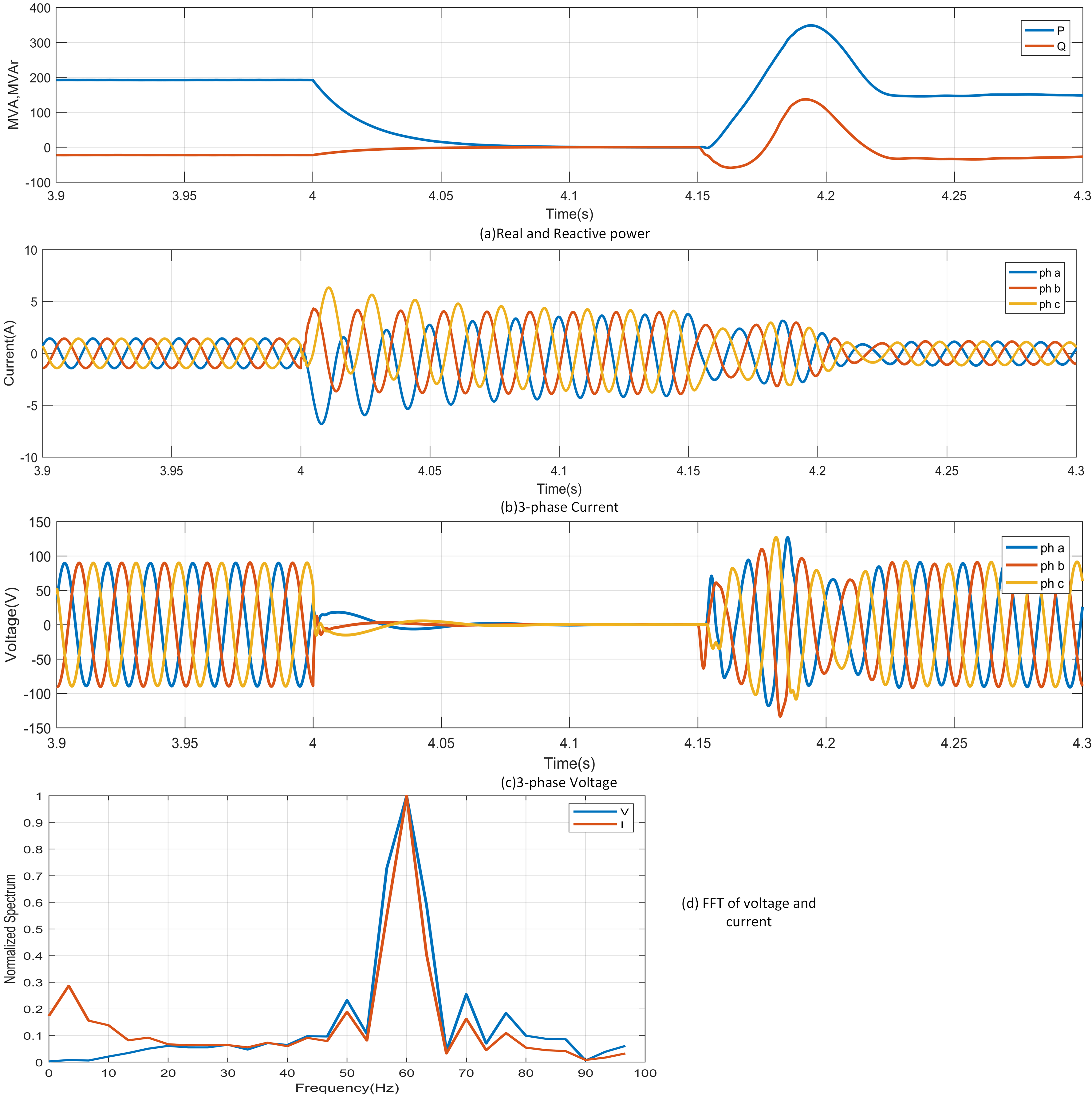}}
\caption{Characteristics for fault at Location 2}
\label{ch4loc2}
\end{figure}

\begin{figure}[ht]
\centerline{\includegraphics[width=3.7 in, height= 4.1 in]{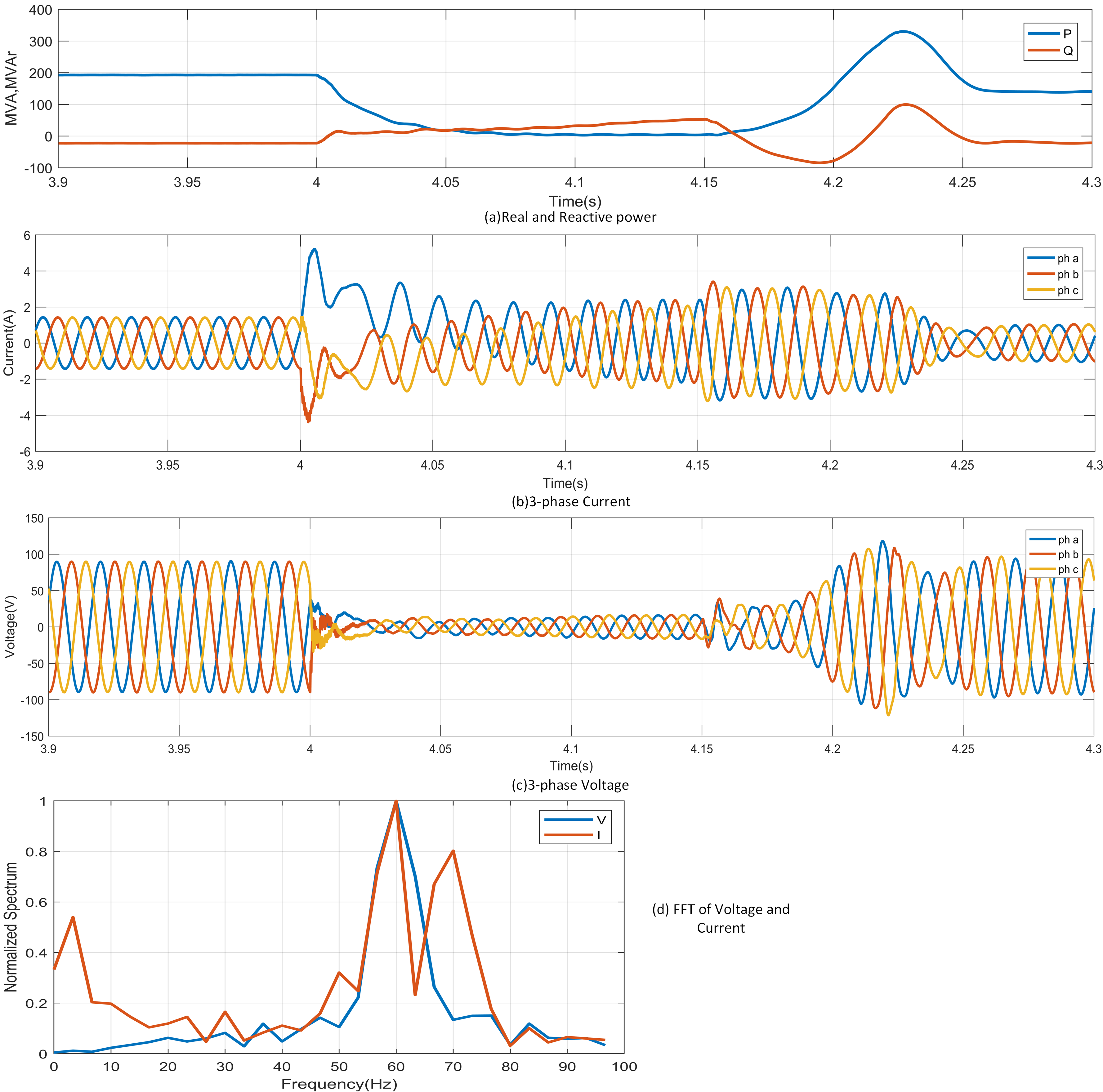}}
\caption{Characteristics for fault at Location 4}
\label{ch4loc4}
\end{figure}

\begin{figure}[ht!]
\vspace{-0.1cm}
\centering
\includegraphics[width=2.7 in, height= 1.4 in]{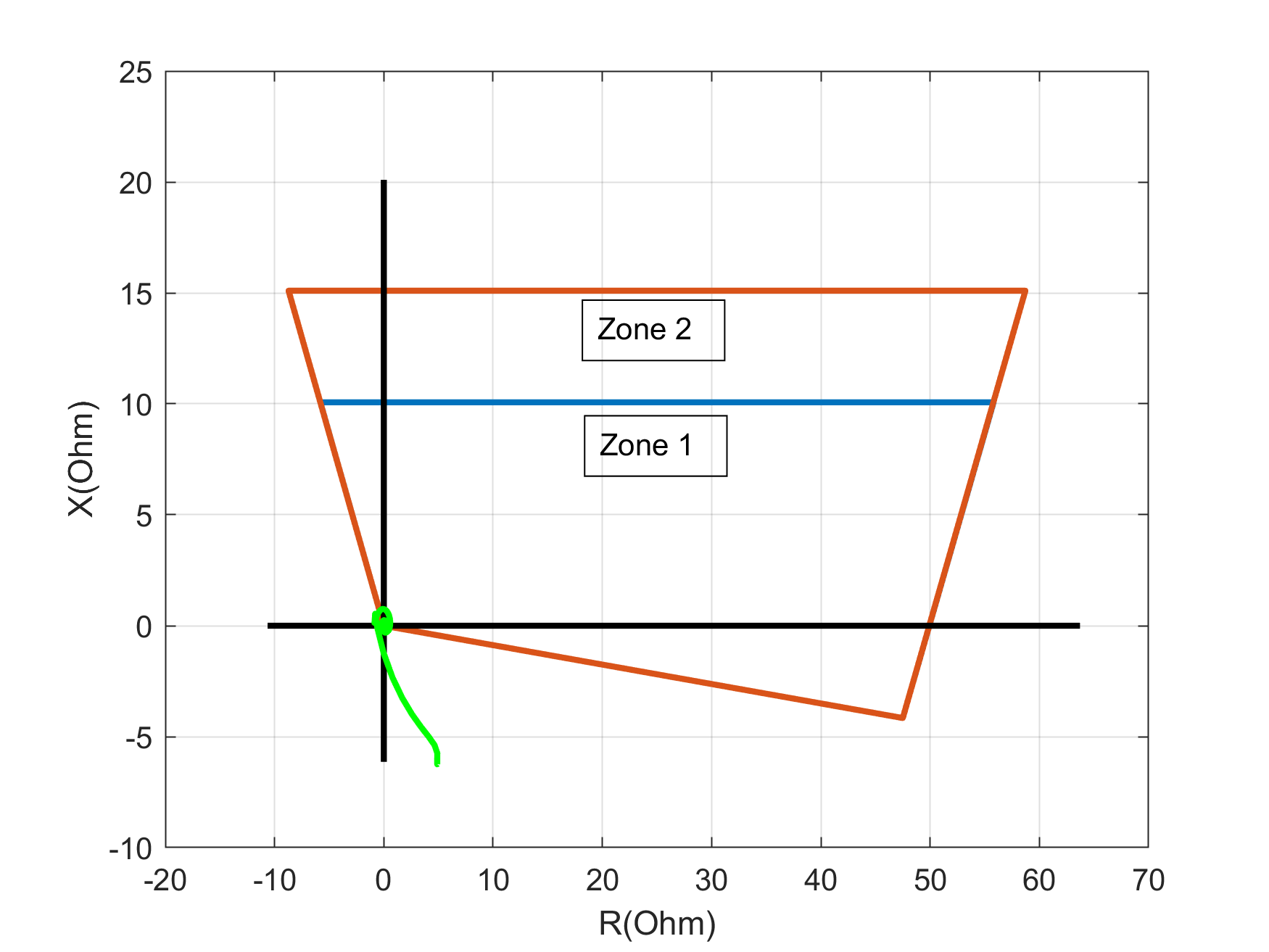}
\vspace{-0.1cm}
\caption{Operating point in impedance plane for phase C distance element for fault at location 2}
\label{ch4dr2}
\vspace{-0.1cm}
\end{figure}

\begin{figure}[ht!]
\vspace{-0.1cm}
\centering
\includegraphics[width=2.7 in, height= 1.4 in]{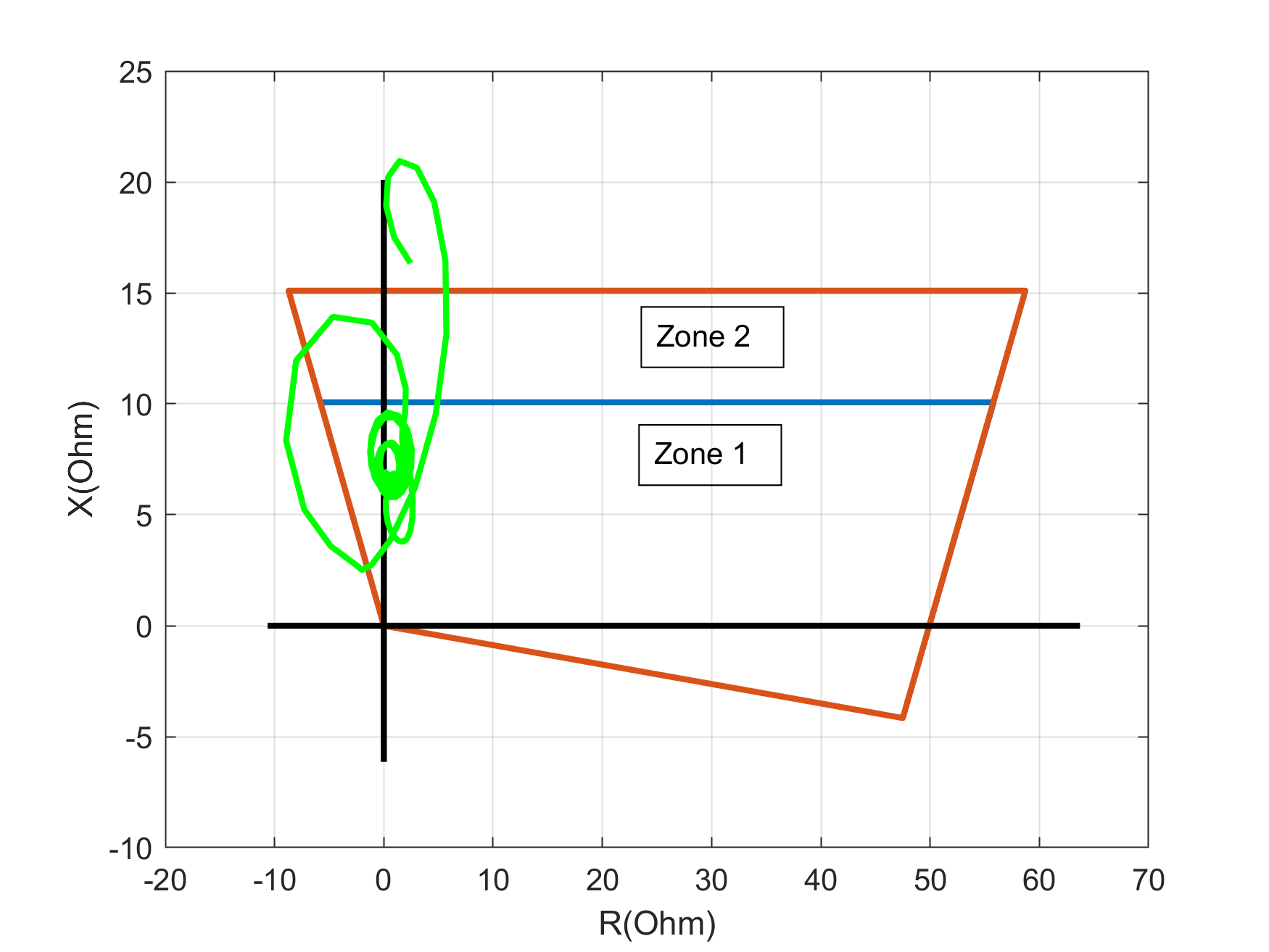}
\vspace{-0.1cm}
\caption{Operating point in impedance plane for phase C distance element for fault at location 4}
\label{ch4dr4}
\vspace{-0.1cm}
\end{figure}

This chapter examines balanced faults, which are among the most severe faults. The slip in a DFIG results in off-nominal frequency which leads to failure of distance relays. Generally, the distance relays operate on the basis of the fundamental component of voltage and current. Distance relays have frequency tracking, which allows them to reliably detect current and voltage even when there is a frequency excursion. When balanced faults occur in DFIG-based WFs, however, the frequency of the current and voltage deviates, and the impedance measured becomes unreliable. Also, the phase angle comparison method fails to detect the fault direction in such events \cite{chhosiyar1}. The operation of the distance relay, R12 is analyzed for the balanced faults at locations 2 and 4. Fig.\ref{ch4loc2} and  Fig.\ref{ch4loc4} display the current, voltage, real power, and reactive power for faults at these locations and Fig.\ref{ch4loc2}(d) and  Fig.\ref{ch4loc4}(d) show the frequency spectra of the current and voltage phasors obtained. The current spectrum has another peak at 70 Hz other than the nominal 60Hz.
 The common frequency tracking method either fails to track the frequency or fails to measure the impedance accurately. The unpredictable behavior of the relay, R12 in case of balanced faults can be adjudged from the impedance trajectories for faults at location 2 and location 4 with its zones 1 and 2 (See Fig.\ref{ch4dr2} and Fig.\ref{ch4dr4}). Fig.\ref{ch4dr4} where the impedance enters the zone 1 after time t = 4.17 ms of a balanced fault at location 4 depicts correct operation of the relay for a fault in zone 1, however, Fig.\ref{ch4dr2} where the impedance enters the zone 1 after time t = 7.29 ms of a balanced fault at location 2 depicts false tripping of the distance relay for a fault outside zone 1. The impedance measurement is discussed in the Appendix.
 The impedance trajectory is calculated by fixed point FFT of 32 samples per cycle after an interval of 4 samples.

\section{Proposed AR based Scheme}
The current and voltage phasors are time series which can be differentiated using Euclidean distance or dynamic time warping \cite{chDTW}; first and second-order statistics of mean, standard deviation, skewness, and kurtosis \cite{chNanopoulosTimeSeriesFeat}; classical and statistical features like trend, seasonality, periodicity, auto-correlation, skewness, kurtosis, nonlinearity, chaos, and self-similarity \cite{chWangTimeSeriesFeat}. 
 The IG-based and grid-based faults can be distinguished by the waveshape recognition technique \cite{chhosiyar1}. The ac components of the IG-based faults are fast decaying and the dc component of the faults fed by the grid are slowly decaying. This property was used to define the relative difference in the first consecutive peaks and was used to differentiate these faults in \cite{chhosiyar1}. Again, the relative decline in the magnitude of the measured phasor for the fault current in the first-half cycle from the second-half cycle was used in \cite{chhosiyar2}.
The power system transients can also be differentiated with other time, time-frequency, and frequency domain properties. Aggregate linear trend, FFT coefficient, average change quantile, AR coefficient, sample entropy, absolute energy, complexity invariant distance, kurtosis, wavelet coefficients, variance, spectral welch density, peak, number of peaks were used in \cite{chsystempallav}\cite{chpedes}\cite{chietpallav}. To differentiate the IG-based and grid-based faults, AR coefficients of the 3-phase currents are selected from the list of features mentioned above in \cite{chsystempallav}\cite{chpedes}\cite{chietpallav} using feature selection based on Random Forest classifier and `Gini impurity' as an indicator of the feature importance \cite{chBreiman2001}.

\begin{table}[]
\centering
\captionsetup{justification=centering}
\caption{Different parameters and their values }\label{ch4parameters}
\begin{tabular}{|c|c|c|c|c|}
\hline
\begin{tabular}[c]{@{}c@{}}Fault \\ location\end{tabular} & \begin{tabular}[c]{@{}c@{}}Fault \\ resistance\\ (ohm)\end{tabular} & \begin{tabular}[c]{@{}c@{}}Fault \\ inception time\\ (s)\end{tabular}      & \begin{tabular}[c]{@{}c@{}}Crowbar \\ resistance\\ (ohm)\end{tabular} & \begin{tabular}[c]{@{}c@{}}Wind \\ speed\\ (m/s)\end{tabular} \\ \hline
1, 2, 3, 4 ,5                                               & \begin{tabular}[c]{@{}c@{}} 2, 20,\\ 50, 200\end{tabular}      & \begin{tabular}[c]{@{}c@{}}4 to 4.0149 in\\ steps of 0.0016\end{tabular} &0, 0.01, 0.1                                                         & 11, 20                                                        \\ \hline
\end{tabular}
\end{table}
\subsection{Fault Data Simulation}
Various scenarios are evaluated to validate the efficacy of the proposed technique. The parameters that may affect the fault profiles and hence the chosen statistics are the fault resistance, fault location, fault inception time, crowbar resistance, and wind speed. The parameters and their values used to obtain the 1200 cases are given in table \ref{ch4parameters}.
Although a crowbar with non-zero resistance is preferred for DFIG based WFs connected to the HV primary grid because of positive effects on the fault current, zero crowbar resistance is also considered during the fault simulations.

\begin{figure}[ht]
\centerline{\includegraphics[width=3.9 in, height= 2.7 in]{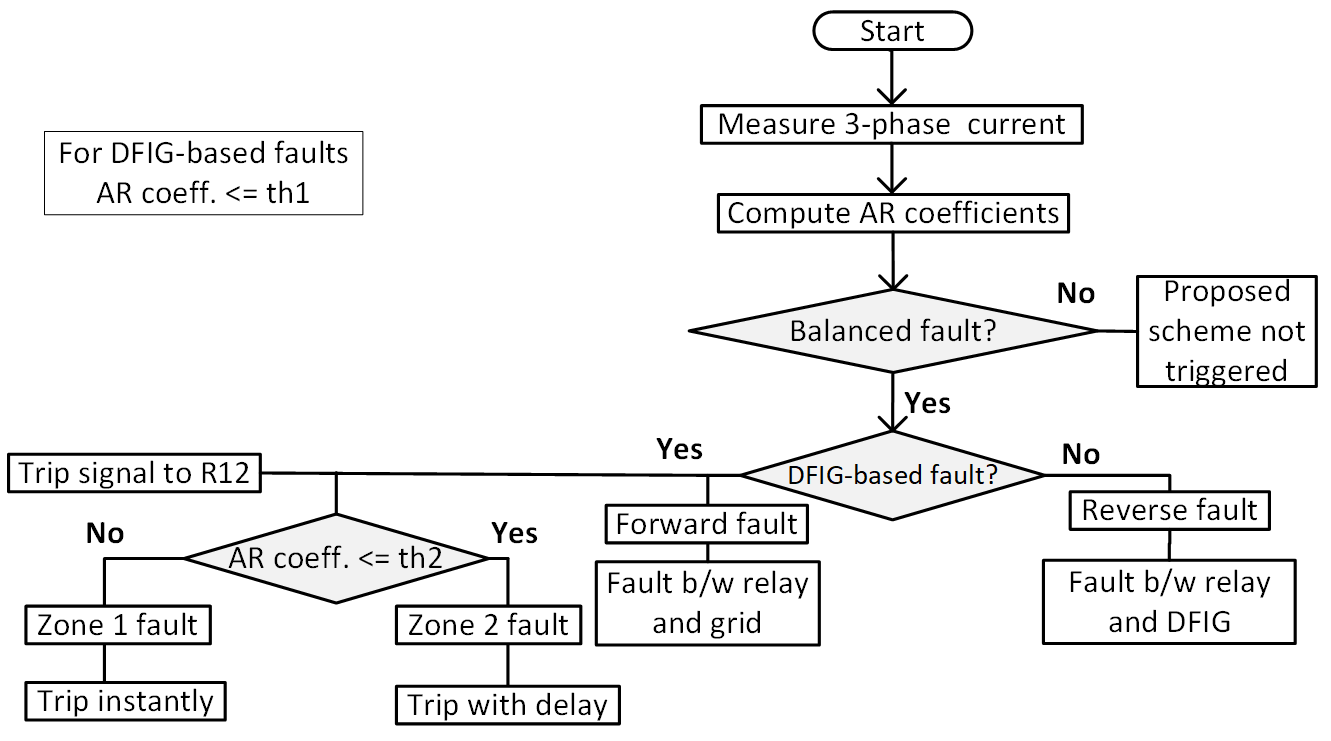}}
\caption{Flowchart of the proposed protection scheme}
\label{ch4fc}
\end{figure}
\subsection{Proposed Scheme}
Fig.\ref{ch4fc} illustrates the flowchart of the proposed mechanism. The 3-phase currents measured by the CT are used to find the AR coefficients. The balanced faults are differentiated between DFIG-based (forward) and grid-based (reverse) faults depending on  $ Ar\ coeff.\ (\varphi_2)$ $<= th1$. If the fault is in the forward direction of R12, the line 12 is tripped instantly or adjacent lines are tripped depending on $Ar\ coeff. (\varphi_7) <= th2$. 

\subsection{Auto-regressive (AR) coefficients}
An AR model represents random processes and is used to describe time-varying phenomena in signal processing, economics, and statistics.
 AR process is a linear regression model that calculates the present value from the previous ones \cite{chAR}.
 It computes the observation $X_t$ at time $t$ using the constant ($\varphi_0$), AR coefficients ($\varphi_{i's}$),
 previous observations ($X_{t-i}$), random noise ($\eta_t$), and lag $k$.
 \begin{equation}\label{ch4ar} X_t=\varphi_0 + \varphi_1\cdot X_{t-1} +\varphi_{2}\cdot X_{t-2}+...+\varphi_{k}\cdot X_{t-k}+\eta_t
 \end{equation}

 The least-square method is used to estimate the coefficients.
 The lag $k$ and Ar coefficient, $\varphi_i$ are chosen using Random Forest feature selection and `Gini impurity' as indicator for distinguishing the faults as forward and reverse, and for identifying the fault zone. Lag k=10 and Ar coefficient, $\varphi_2$ are thus selected to extract the required feature from the 3-phase currents which help distinguish the faults fed by the WF and the bulk power grid and Lag k=10 and Ar coefficient, $\varphi_7$ are selected to determine the zone. The reverse faults or the grid-based faults or faults fed by the grid have a higher value of $\varphi_2$ than the DFIG-based faults. The faults belong to the DFIG category if $\varphi_{2}$ of the 3-phases follow the condition:
 
\footnotesize
\begin{equation}\label{ch4ar2} 
\varphi_{2}(ph a) \leq  - 0.7 \ \&  \ \varphi_{2}.(ph b) \leq - 0.7 \ \&  \ \varphi_{2}(ph c) \leq - 0.7
\end{equation}
\normalsize

\noindent Furthermore, the forward faults lie in zone 2 if $\varphi_{7}$s follow:
\footnotesize
\begin{equation}\label{ch4ar3} 
\varphi_{7}(ph a) \leq  - 0.1 \ \& \ \varphi_{7}.(ph b) \leq - 0.1 \ \& \ \varphi_{7}(ph c) \leq - 0.1
\end{equation}
\normalsize

\vspace{2mm}
\section{Results and Discussions}
Table \ref{ch4ar table1} illustrates the Ar coefficients ($\varphi_{2}$) values of the 3-phases for balanced faults at locations 2 and 4 with a wind speed of 11 m/s for fault inception time of 4.0s. AR coefficients($\varphi_{2}$) of only two locations, one fed by the DFIG and the other fed by the grid are shown all through the chapter to save space. It is evident from the table that the proposed $\varphi_{2}$ based scheme and threshold chosen perform well for the 4-bus test data. The Table \ref{ch4ar table2} illustrates the Ar coefficients ($\varphi_{7}$) values of the 3-phases for locations 2 and 4 with a wind speed of 11 m/s and fault inception time of 4.0s for zone identification. 

\begin{table}[ht]
\centering
\captionsetup{justification=centering}
\caption{Auto-regressive coefficients ($\varphi_{2}$) of the 3-phase currents for 4-bus test system }
\setlength{\tabcolsep}{5pt}
\renewcommand{\arraystretch}{1.2}
\label{ch4ar table1}
\begin{tabular}{|c|c|c|c|c|c|}
\hline
\multirow{2}{*}{\begin{tabular}[c]{@{}c@{}}Fault\\  location\end{tabular}} & \multirow{2}{*}{\begin{tabular}[c]{@{}c@{}}Fault\\  resistance\\ (ohm)\end{tabular}} & \multirow{2}{*}{\begin{tabular}[c]{@{}c@{}}Crowbar\\ resistance\\ (ohm)\end{tabular}} & \multicolumn{3}{c|}{AR coefficient}                                                                                                                         \\ \cline{4-6} 
                                                                           &                                                                                      &                                                                                       & \begin{tabular}[c]{@{}c@{}}phase\\ a\end{tabular} & \begin{tabular}[c]{@{}c@{}}phase \\ b\end{tabular} & \begin{tabular}[c]{@{}c@{}}phase\\  c\end{tabular} \\ \hline
\multirow{9}{*}{\begin{tabular}[c]{@{}c@{}}Location\\ 2\end{tabular}}      & \multirow{3}{*}{2}                                                                & 0                                                                                  &                                                                                0.238 &	0.268 &	0.262                                            \\ \cline{3-6} 
                                                                           &                                                                                      & 0.01                                                                                  & 0.212 &	0.220 &	0.212                                           \\ \cline{3-6} 
                                                                           &                                                                                      & 0.1                                                                                     & -0.178 &	-0.209 &	-0.227                                             \\ \cline{2-6} 
                                                                           & \multirow{3}{*}{20}                                                                  & 0                                                                                  & 0.238 &	0.268 &	0.262                                             \\ \cline{3-6} 
                                                                           &                                                                                      & 0.01                                                                                  & 0.212 & 0.220&	0.212                                             \\ \cline{3-6} 
                                                                           &                                                                                      & 0.1                                                                                    & -0.178&	-0.209&	-0.227                                           \\ \cline{2-6} 
                                                                           & \multirow{3}{*}{200}                                                                 & 0                                                                                  & 0.237&	0.267&	0.261                                            \\ \cline{3-6} 
                                                                           &                                                                                      & 0.01                                                                                  & 0.212&	0.220&	0.212                                           \\ \cline{3-6} 
                                                                           &                                                                                      & 0.1                                                                                     & -0.180 &	-0.211 &	-0.229                                              \\ \hline
\multirow{9}{*}{\begin{tabular}[c]{@{}c@{}}Location\\ 4\end{tabular}}      & \multirow{3}{*}{2}                                                                & 0                                                                                                                                                              & -1.429 &	-1.258 &	-1.065                                             \\ \cline{3-6} 
                                                                            &                                                                                      & 0.01                    &                                                              -2.157&	-1.777 &	-1.417                                          \\ \cline{3-6} 
                                                                           &                                                                                      & 0.1                                                                                     & -1.022	 & -1.676  &	-1.255                                              \\ \cline{2-6} 
                                                                           & \multirow{3}{*}{20}                                                                  & 0                                                                                  & -1.439	& -1.163 &	-1.086                                              \\ \cline{3-6} 
                                                                           &                                                                                      & 0.01                                                                                   & -2.080	& -1.693& -1.382                                            \\ \cline{3-6} 
                                                                           &                                                                                      & 0.1                                                                                     & -1.001	& -1.650	& -1.198                                             \\ \cline{2-6} 
                                                                           & \multirow{3}{*}{200}                                                                 & 0                                                                                  & -1.494 &	-1.214&	-1.057                                               \\ \cline{3-6} 
                                                                           &                                                                                      & 0.01                                                                                   & -2.082 &	-1.783&	-1.475                                               \\ \cline{3-6} 
                                                                           &                                                                                      & 0.1                                                                                     & -1.040	& -1.628	 & -1.203

                                             \\ \hline
\end{tabular}
\end{table}

\begin{table}[ht]
\centering
\captionsetup{justification=centering}
\caption{Auto-regressive coefficients($\varphi_{7}$) of the 3-phase currents for the 4-bus test system for fault zone identification }
\setlength{\tabcolsep}{5pt}
\renewcommand{\arraystretch}{1.2}
\label{ch4ar table2}
\begin{tabular}{|c|c|c|c|c|c|}
\hline
\multirow{2}{*}{\begin{tabular}[c]{@{}c@{}}Fault\\  location\end{tabular}} & \multirow{2}{*}{\begin{tabular}[c]{@{}c@{}}Fault\\  resistance\\ (ohm)\end{tabular}} & \multirow{2}{*}{\begin{tabular}[c]{@{}c@{}}Crowbar\\ resistance\\ (ohm)\end{tabular}} & \multicolumn{3}{c|}{AR coefficient}                                                                                                                         \\ \cline{4-6} 
                                                                           &                                                                                      &                                                                                       & \begin{tabular}[c]{@{}c@{}}phase\\ a\end{tabular} & \begin{tabular}[c]{@{}c@{}}phase \\ b\end{tabular} & \begin{tabular}[c]{@{}c@{}}phase\\  c\end{tabular} \\ \hline
\multirow{9}{*}{\begin{tabular}[c]{@{}c@{}}Zone 1\end{tabular}}    & \multirow{3}{*}{2}                                                                & 0                                                                                                                                                              & 1.141&0.834&0.799                                             \\ \cline{3-6} 
                                                                            &                                                                                      & 0.01                    &                                                              1.036&0.843&0.814                                          \\ \cline{3-6} 
                                                                           &                                                                                      & 0.1                                                                                     & 1.066&0.925&0.828                                            \\ \cline{2-6} 
                                                                           & \multirow{3}{*}{20}                                                                  & 0                                                                                  & 0.880&0.953&1.002                                         \\ \cline{3-6} 
                                                                           &                                                                                      & 0.01                                                                                   & 0.775&0.951&0.901                                            \\ \cline{3-6} 
                                                                           &                                                                                      & 0.1                                                                                     & 0.813&1.090&0.976                                            \\ \cline{2-6} 
                                                                           & \multirow{3}{*}{200}                                                                 & 0                                                                                  & 1.042&1.152&0.818                                              \\ \cline{3-6} 
                                                                           &                                                                                      & 0.01                                                                                   & 1.005&1.154&0.764                                              \\ \cline{3-6} 
                                                                           &                                                                                      & 0.1                                                                                     & 1.074&1.165&0.863                                           \\ \hline
\multirow{9}{*}{\begin{tabular}[c]{@{}c@{}}Zone 2\end{tabular}}      & \multirow{3}{*}{2}                                                                & 0                                                                                                                                                              & -0.400&-0.997&-0.926                                              \\ \cline{3-6} 
                                                                            &                                                                                      & 0.01                    &                                                             -0.417&-0.947&-0.903                                         \\ \cline{3-6} 
                                                                           &                                                                                      & 0.1                                                                                     & -0.485&-0.945&-0.902                                             \\ \cline{2-6} 
                                                                           & \multirow{3}{*}{20}                                                                  & 0                                                                                  & 0.123&-0.906&-0.828                                             \\ \cline{3-6} 
                                                                           &                                                                                      & 0.01                                                                                   &-0.366&-1.008&-0.967                                            \\ \cline{3-6} 
                                                                           &                                                                                      & 0.1                                                                                     & -0.382&-1.004&-0.961                                             \\ \cline{2-6} 
                                                                           & \multirow{3}{*}{50}                                                                 & 0                                                                                  & -0.417&-0.947&-0.903                                             \\ \cline{3-6} 
                                                                           &                                                                                      & 0.01                                                                                   & 0.093&-0.916&-0.834                                          \\ \cline{3-6} 
                                                                           &                                                                                      & 0.1                                                                                     &0.159&-0.917&-0.839

                                             \\ \hline
\end{tabular}
\end{table}

\begin{figure}[h]
\vspace{-0.1cm}
\centering
\includegraphics[width=2.6 in, height= 2.3 in]{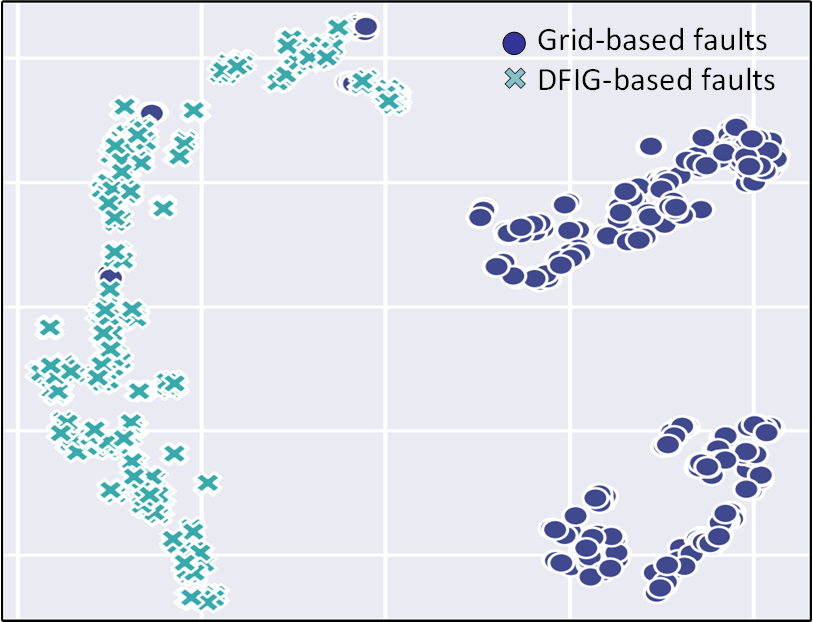}
\vspace{-0.1cm}
\caption{Scatter plot of AR coefficients($\varphi_{2}$) of 3-phase currents for DFIG and Grid-based faults}
\label{ch4tsne}
\vspace{-0.1cm}
\end{figure}

To comprehend how the AR coefficients($\varphi_{2}$) distinguish the faults better, the clusters formed by the different balanced faults are visualized. Fig.\ref{ch4tsne} shows the scatter plot for the AR coefficients($\varphi_{2}$) of the 1200 scenarios of the binary class problem at hand. T-distributed Stochastic Neighbour Embedding (t-SNE) \cite{chtsne} is used to map the 3-phase AR coefficients($\varphi_{2}$) on the 2-D plane. The two fault clusters are separable with a couple of exceptions. The grid-based faults visible in the DFIG-based faults are separable in the original domain when they are not projected on the 2-D since the AR threshold obtained distinguishes the faults accurately for all cases.

\subsection{Validation on IEEE 9-bus and IEEE-39 bus}
\begin{figure}[ht!]
\vspace{-0.1cm}
\centering
\includegraphics[width=3.9 in, height=2.0 in]{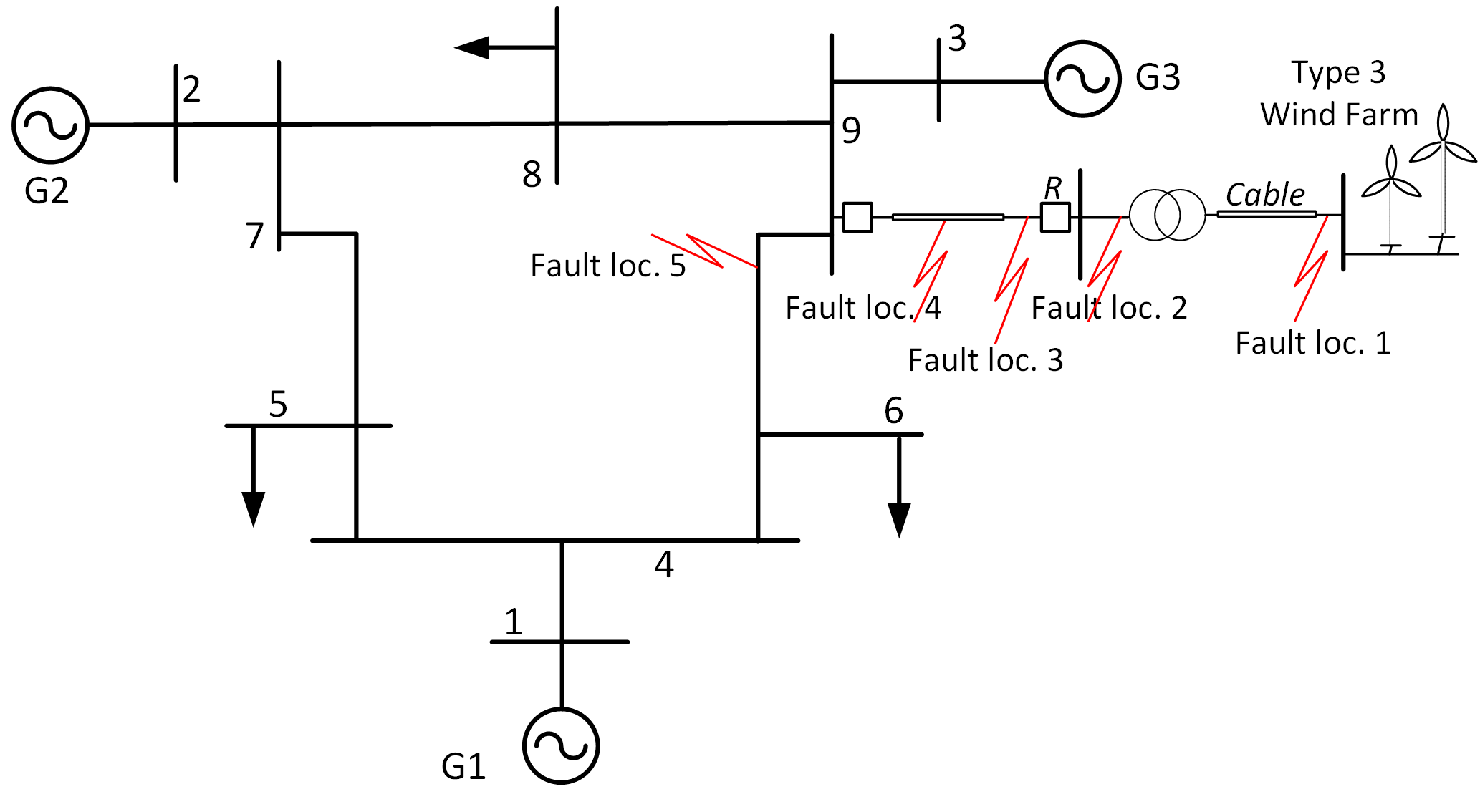}
\vspace{-0.1cm}
\caption{IEEE 9-bus with Type-3 WF at bus 9}
\label{ch49bus}
\vspace{-0.1cm}
\end{figure}

\begin{figure}[ht!]
\vspace{-0.1cm}
\centering
\includegraphics[width=4 in, height=3.6 in]{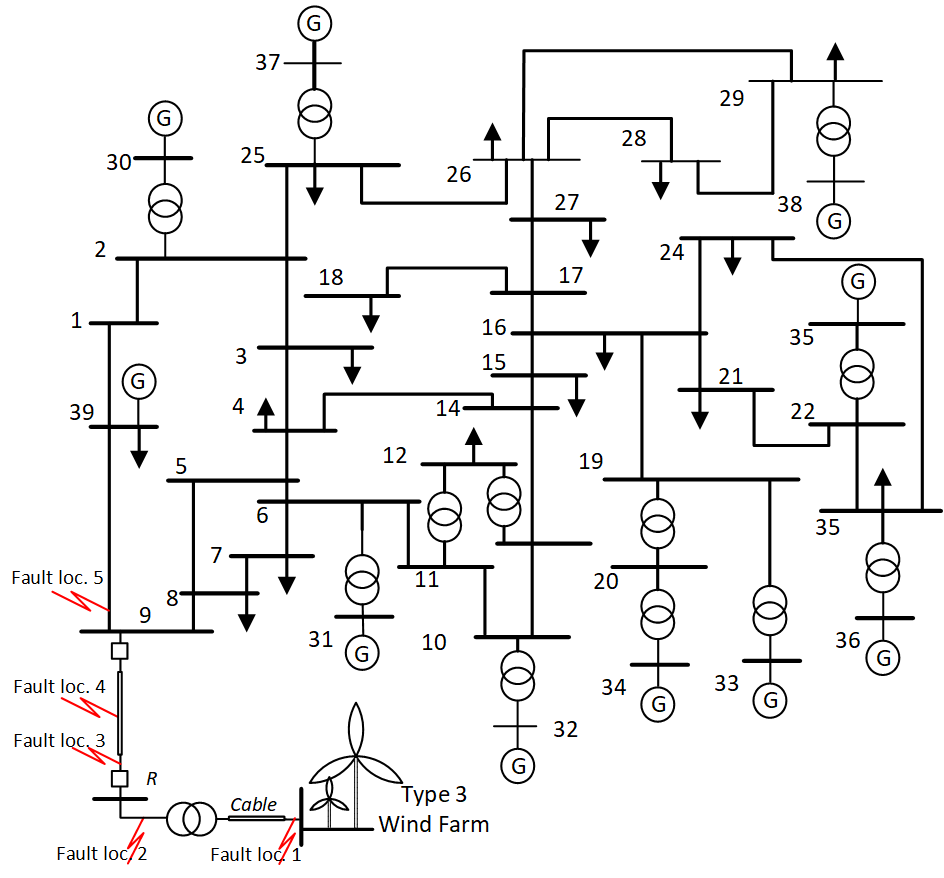}
\vspace{-0.1cm}
\caption{IEEE 39-bus with Type-3 WF at bus 9}
\label{ch439bus}
\vspace{-0.1cm}
\end{figure}

The validity of the proposed scheme is tested on IEEE-9 bus (Fig. \ref{ch49bus}) and IEEE-39 (Fig. \ref{ch439bus}) bus systems. Balanced faults are simulated for different parameters and values (see table \ref{ch4parameters}). The scheme developed for the 4-bus system holds true for the 9-bus and 39-bus systems as well. The Tables \ref{ch4ar table9bus} and \ref{ch4ar table39bus} illustrate the Ar coefficients($\varphi_{2}$) values of the 3-phases for balanced faults at location 2 and 4 with a wind speed of 11 m/s for fault inception time of 4.0s for the IEEE 9-bus and 39-bus systems.

\begin{table}[ht]
\centering
\captionsetup{justification=centering}
\caption{Auto-regressive coefficients ($\varphi_{2}$) of the 3-phase currents for IEEE 9-bus system}
\setlength{\tabcolsep}{5pt}
\renewcommand{\arraystretch}{1.2}
\label{ch4ar table9bus}
\begin{tabular}{|c|c|c|c|c|c|}
\hline
\multirow{2}{*}{\begin{tabular}[c]{@{}c@{}}Fault\\  location\end{tabular}} & \multirow{2}{*}{\begin{tabular}[c]{@{}c@{}}Fault\\  resistance\\ (ohm)\end{tabular}} & \multirow{2}{*}{\begin{tabular}[c]{@{}c@{}}Crowbar\\ resistance\\ (ohm)\end{tabular}} & \multicolumn{3}{c|}{AR coefficient}                                                                                                                         \\ \cline{4-6} 
                                                                           &                                                                                      &                                                                                       & \begin{tabular}[c]{@{}c@{}}phase\\ a\end{tabular} & \begin{tabular}[c]{@{}c@{}}phase \\ b\end{tabular} & \begin{tabular}[c]{@{}c@{}}phase\\  c\end{tabular} \\ \hline
\multirow{9}{*}{\begin{tabular}[c]{@{}c@{}}Location\\ 2\end{tabular}}      & \multirow{3}{*}{2}                                                                & 0                                                                                  &                                                                                0.483&0.307&0.434                                        \\ \cline{3-6} 
                                                                           &                                                                                      & 0.01                                                                                  & 0.495&0.369&0.436                                       \\ \cline{3-6} 
                                                                           &                                                                                      & 0.1                                                                                     & 0.463&0.403&0.420                                           \\ \cline{2-6} 
                                                                           & \multirow{3}{*}{20}                                                                  & 0                                                                                  & 0.482 &0.307 & 0.435                                          \\ \cline{3-6} 
                                                                           &                                                                                      & 0.01                                                                                  & 0.492&0.368&0.436                                             \\ \cline{3-6} 
                                                                           &                                                                                      & 0.1                                                                                    & 0.451&0.399&0.421                                           \\ \cline{2-6} 
                                                                           & \multirow{3}{*}{200}                                                                 & 0                                                                                  & 0.483& 0.306 & 0.433                                            \\ \cline{3-6} 
                                                                           &                                                                                      & 0.01                                                                                  & 0.493 &0.367& 0.435                                        \\ \cline{3-6} 
                                                                           &                                                                                      & 0.1                                                                                     & 0.454 &0.398 & 0.416                                            \\ \hline
\multirow{9}{*}{\begin{tabular}[c]{@{}c@{}}Location\\ 4\end{tabular}}      & \multirow{3}{*}{2}                                                                & 0                                                                                                                                                              & -1.868&-1.659&-1.331                                          \\ \cline{3-6} 
                                                                            &                                                                                      & 0.01                    &                                                              -2.398&-2.237&-1.823                                         \\ \cline{3-6} 
                                                                           &                                                                                      & 0.1                                                                                     & -2.192&-2.451&-1.788                                         \\ \cline{2-6} 
                                                                           & \multirow{3}{*}{20}                                                                  & 0                                                                                  & -1.879&-1.645&-1.467                                             \\ \cline{3-6} 
                                                                           &                                                                                      & 0.01                                                                                   &-2.195&-2.584&-1.771                                       \\ \cline{3-6} 
                                                                           &                                                                                      & 0.1                                                                                     &-1.133&-1.583&-0.932                                             \\ \cline{2-6} 
                                                                           & \multirow{3}{*}{200}                                                                 & 0                                                                                  & -1.936&-1.621&-1.333                                              \\ \cline{3-6} 
                                                                           &                                                                                      & 0.01                                                                                   & -2.374&-2.136&-1.818                                            \\ \cline{3-6} 
                                                                           &                                                                                      & 0.1                                                                                     & -2.168&-2.530&-1.734

                                             \\ \hline
\end{tabular}
\end{table}

\begin{table}[ht]
\centering
\captionsetup{justification=centering}
\caption{Auto-regressive coefficients ($\varphi_{2}$) of the 3-phase currents for IEEE 39-bus system }
\setlength{\tabcolsep}{5pt}
\renewcommand{\arraystretch}{1.2}
\label{ch4ar table39bus}
\begin{tabular}{|c|c|c|c|c|c|}
\hline
\multirow{2}{*}{\begin{tabular}[c]{@{}c@{}}Fault\\  location\end{tabular}} & \multirow{2}{*}{\begin{tabular}[c]{@{}c@{}}Fault\\  resistance\\ (ohm)\end{tabular}} & \multirow{2}{*}{\begin{tabular}[c]{@{}c@{}}Crowbar\\ resistance\\ (ohm)\end{tabular}} & \multicolumn{3}{c|}{AR coefficient}                                                                                                                         \\ \cline{4-6} 
                                                                           &                                                                                      &                                                                                       & \begin{tabular}[c]{@{}c@{}}phase\\ a\end{tabular} & \begin{tabular}[c]{@{}c@{}}phase \\ b\end{tabular} & \begin{tabular}[c]{@{}c@{}}phase\\  c\end{tabular} \\ \hline
\multirow{9}{*}{\begin{tabular}[c]{@{}c@{}}Location\\ 2\end{tabular}}      & \multirow{3}{*}{2}                                                                & 0                                                                                  &                                                                               0.206 & 0.169 & -0.087                                       \\ \cline{3-6} 
                                                                           &                                                                                      & 0.01                                                                                & 0.250 & 0.079 & -0.090                                       \\ \cline{3-6} 
                                                                           &                                                                                      & 0.1                                                                                 &   0.032 & -0.082 & -0.176                                           \\ \cline{2-6} 
                                                                           & \multirow{3}{*}{20}                                                                  & 0                                                                                & 0.207 & 0.170 & -0.087                                      \\ \cline{3-6} 
                                                                           &                                                                                      & 0.01                                                                                  & 0.251 & 0.080 & -0.091                                             \\ \cline{3-6} 
                                                                           &                                                                                      & 0.1                                                                                & 0.035 & -0.080 & -0.179                                           \\ \cline{2-6} 
                                                                           & \multirow{3}{*}{200}                                                                 & 0                                                                                  &0.207 & 0.170 & -0.086                                          \\ \cline{3-6} 
                                                                           &                                                                                      & 0.01                                                                                  & 0.252 & 0.080 & -0.092                                          \\ \cline{3-6} 
                                                                           &                                                                                      & 0.1                                                                                     & 0.036 & -0.079 & -0.178
                                             \\ \hline
\multirow{9}{*}{\begin{tabular}[c]{@{}c@{}}Location\\ 4\end{tabular}}      & \multirow{3}{*}{2}                                                                & 0                                                                                                                                                              & -1.170 & -0.667 & -1.441                                       \\ \cline{3-6} 
                                                                            &                                                                                      & 0.01                    &                                                             -2.270 & -1.972 & -2.047                                       \\ \cline{3-6} 
                                                                           &                                                                                      & 0.1                                                                                     & -2.289 & -2.287 & -1.963                                            \\ \cline{2-6} 
                                                                           & \multirow{3}{*}{20}                                                                  & 0                                                                                  &-1.266 & -0.762 & -1.506                                             \\ \cline{3-6} 
                                                                           &                                                                                      & 0.01                                                                                   & -2.202 & -1.930 & -2.098                                           \\ \cline{3-6} 
                                                                           &                                                                                      & 0.1                                                                                     & -2.239 & -2.170& -1.826                                           \\ \cline{2-6} 
                                                                           & \multirow{3}{*}{200}                                                                 & 0                                                                                  & -1.202 & -0.708 & -1.534                                              \\ \cline{3-6} 
                                                                           &                                                                                      & 0.01                                                                                   & -2.244 & -2.009 & -2.072                                             \\ \cline{3-6} 
                                                                           &                                                                                      & 0.1                                                                                     & -2.195& -2.231 & -1.892

                                             \\ \hline
\end{tabular}
\end{table}

\subsection{Impact of Series capacitors and phase shifters}
Series capacitors and phase shift transformers (\acrshort{PST}) are used to make the transmission system more efficient and controllable, however, they affect the functioning of the line distance protection systems. To manage the additional power because of the WF, series capacitors and PST is used in the IEEE 9-bus system. The impact of these devices on the distance relay in the intertie zone is analyzed in this section and the efficacy of the proposed AR scheme is affirmed in such scenarios.
\begin{figure}[ht!]
\vspace{-0.1cm}
\centering
\includegraphics[width=2.7 in, height= 1.4 in]{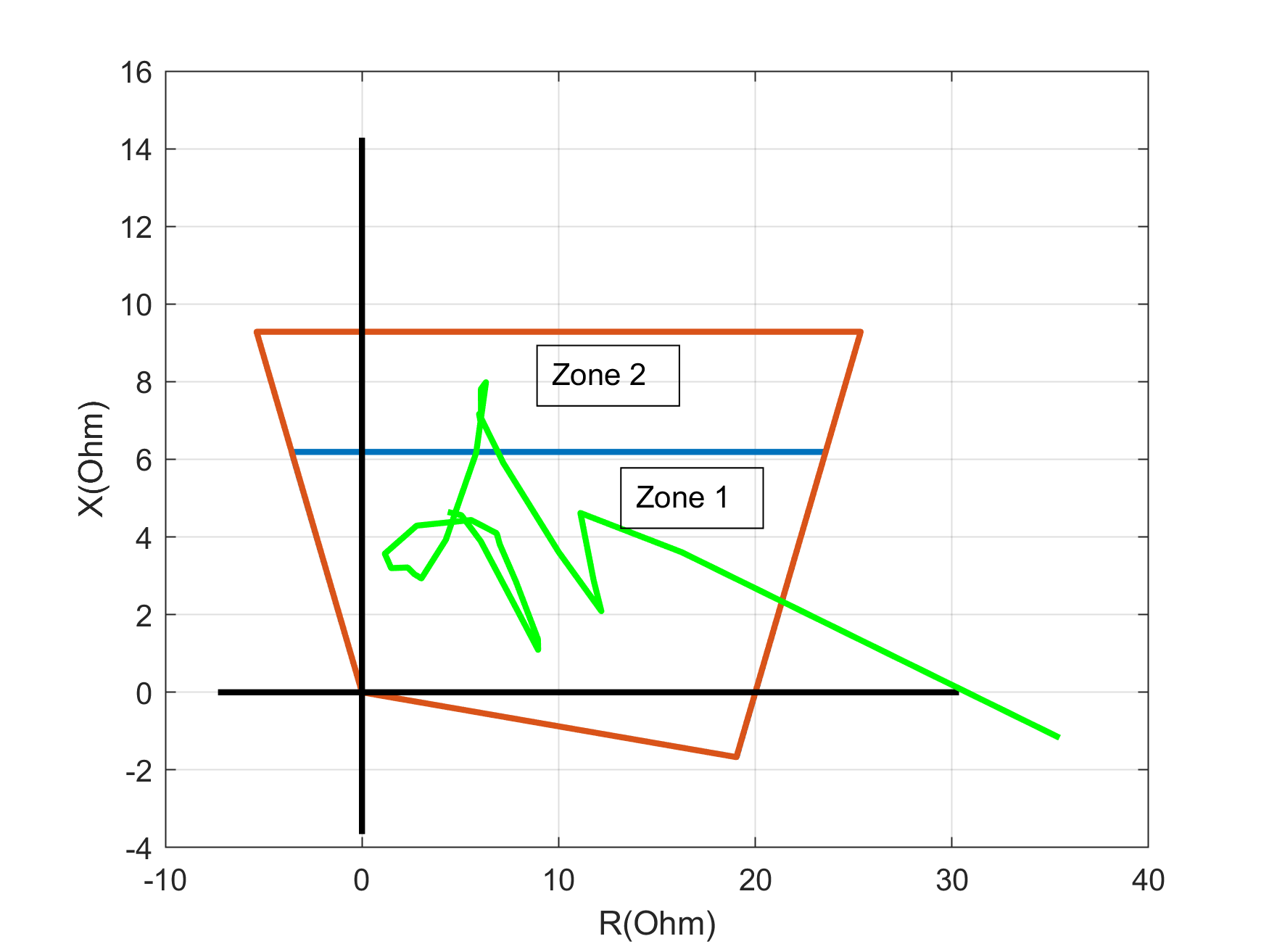}
\vspace{-0.1cm}
\caption{Operating point in impedance plane for phase A distance element for series capacitor at location 5}
\label{ch4capdia}
\vspace{-0.1cm}
\end{figure}
\subsubsection{Series capacitors} Series compensated lines require specifically designed distance relays for their protection. A comprehensive transient study is required to verify the speed, dependability, and security because the capacitor modifies the impedance of the line \cite{chstdtlines}. The series capacitors  compensate from 25 to 75\% of the inductive reactance
of the T-line. A capacitor bank with overvoltage protection which compensate 50\%  of the T-line between bus 9 and 6 is connected in series in the IEEE 9-bus system.
Fig.\ref{ch4capdia}  where  the  impedance  enters  the  zone 1  after  time  t  =  52.6  ms  of  a  balanced  fault  at  location  5, depicts the false  operation  of the distance relay  for  a  fault  in  zone  2. Table \ref{ch4ar tablecap} illustrates the Ar coefficients values of the 3-phases for balanced faults at locations 2 and 4 with a wind speed of 11 m/s for fault inception time of 4.0s for the IEEE 9-bus for the series compensated line.
\subsubsection{Phase shifting transformers (PST)} 
\begin{figure}[ht!]
\vspace{-0.1cm}
\centering
\includegraphics[width=2.7 in, height= 1.4 in]{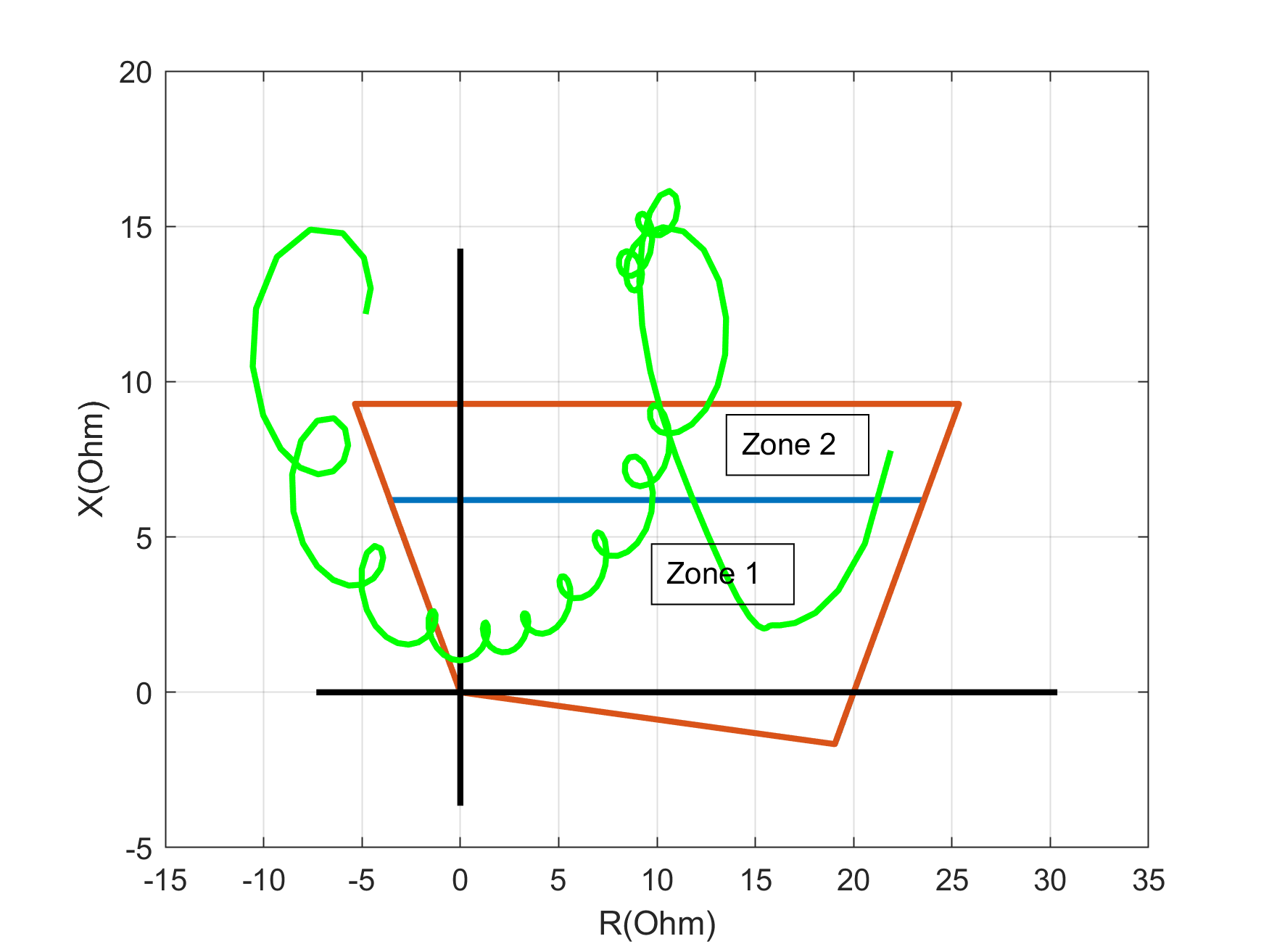}
\vspace{-0.1cm}
\caption{Operating point in impedance plane for phase A distance element for a fault at location 5 for forward shift in PST}
\label{ch4pstfdia}
\vspace{-0.1cm}
\end{figure}

\begin{figure}[ht!]
\vspace{-0.1cm}
\centering
\includegraphics[width=2.4 in, height= 1.3 in]{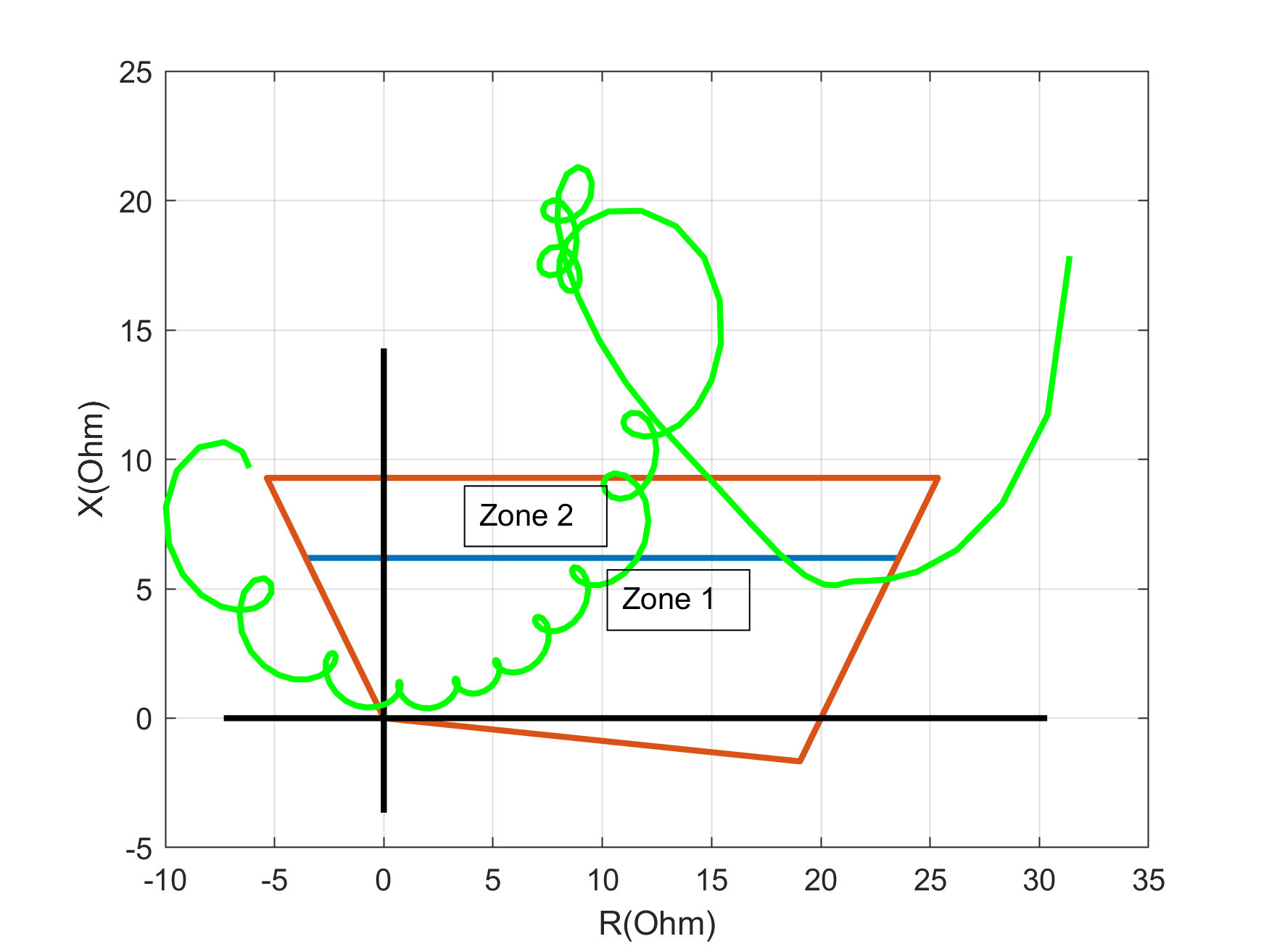}
\vspace{-0.1cm}
\caption{Operating point in impedance plane for phase A distance element for a fault at location 5 for backward shift in PST}
\label{ch4pstbdia}
\vspace{-0.1cm}
\end{figure}

Phase shifting transformers are used to control the power flow in interconnected systems. The location of CTs and VTs for the protection of lines compensated with phase shifters is critical. This study shows that the currents measured at one end can be used for the protection of the T-line. An Indirect Symmetrical PST rated at $S$=500MVA, $V$=230kV, and with maximum phase shift = $\pm25^{\circ}$ is connected between bus 9 and 6 in the IEEE 9-bus system.  Fig.\ref{ch4pstfdia} and Fig.\ref{ch4pstbdia} show the false tripping of the distance relay for a  fault  in zone 2 as  the  impedance  enters  the  zone 1  after  time  t  =  9.9  ms and time  t  = 7.8  ms of the  balanced  fault  at  location  5 for forward and backward shift respectively in the PST. Table \ref{ch4ar tablepst} illustrates the Ar coefficients ($\varphi_{2}$) values of the 3-phases  for balanced  faults with the PST in operation  at  locations  2  and  4  with  wind  speeds  of  11  m/s, crowbar resistance = 0, and half tap of the PST  for  fault  inception  time  of  4.0s.

\begin{table}[ht]
\centering
\captionsetup{justification=centering}
\caption{Auto-regressive coefficients ($\varphi_{2}$) of the 3-phase currents for IEEE 9-bus system with series compensation }
\setlength{\tabcolsep}{5pt}
\renewcommand{\arraystretch}{1.2}
\label{ch4ar tablecap}
\begin{tabular}{|c|c|c|c|c|c|}
\hline
\multirow{2}{*}{\begin{tabular}[c]{@{}c@{}}Fault\\  location\end{tabular}} & \multirow{2}{*}{\begin{tabular}[c]{@{}c@{}}Fault\\  resistance\\ (ohm)\end{tabular}} & \multirow{2}{*}{\begin{tabular}[c]{@{}c@{}}Crowbar\\ resistance\\ (ohm)\end{tabular}} & \multicolumn{3}{c|}{AR coefficient}                                                                                                                         \\ \cline{4-6} 
                                                                           &                                                                                      &                                                                                       & \begin{tabular}[c]{@{}c@{}}phase\\ a\end{tabular} & \begin{tabular}[c]{@{}c@{}}phase \\ b\end{tabular} & \begin{tabular}[c]{@{}c@{}}phase\\  c\end{tabular} \\ \hline
\multirow{6}{*}{\begin{tabular}[c]{@{}c@{}}Location\\ 2\end{tabular}}      & \multirow{2}{*}{2}                                                                & 0                                                                                  &                                                                              0.474&0.322&0.400                                        \\ \cline{3-6} 
                                                                    
                                                                           &                                                                                      & 0.1                                                                                 &  0.475&0.322&0.401                                          \\ \cline{2-6} 
                                                                           & \multirow{2}{*}{20}                                                                  & 0                                                                                & 0.484&0.328&0.425                                      \\ \cline{3-6} 
                                                               
                                                                           &                                                                                      & 0.1                                                                                & 0.487&0.328&0.425                                       \\ \cline{2-6} 
                                                                           & \multirow{2}{*}{100}                                                                 & 0                                                                                  &0.187&0.253&0.237                                        \\ \cline{3-6} 
                                                                      
                                                                           &                                                                                      & 0.1                                                                                     & 0.182&0.256&0.242
                                             \\ \hline
\multirow{6}{*}{\begin{tabular}[c]{@{}c@{}}Location\\ 4\end{tabular}}      & \multirow{2}{*}{2}                                                                & 0                                                                                                                                                              & -2.086&-1.850&-1.307                                      \\ \cline{3-6} 
                                                                            
                                                                           &                                                                                      & 0.1                                                                                     & -1.962&-1.761&-1.221                                           \\ \cline{2-6} 
                                                                           & \multirow{2}{*}{20}                                                                  & 0                                                                                  &-2.475&-2.742&-2.238                                            \\ \cline{3-6} 
                                                                                                                                                              
                                                                           &                                                                                      & 0.1                                                                                     &-2.436&-2.729&-2.261                                          \\ \cline{2-6} 
                                                                           & \multirow{2}{*}{100}                                                                 & 0                                                                                  & -1.629&-2.002&-1.391                                            \\ \cline{3-6} 
                                                                           &                                                                                      
                                                                                                                                                              & 0.1                                                                                     & -1.642&-2.010&-1.371

                                             \\ \hline
\end{tabular}
\end{table}

\begin{table}[ht]
\centering
\captionsetup{justification=centering}
\caption{Auto-regressive coefficients ($\varphi_{2}$) of the 3-phase currents for IEEE 9-bus system with phase shift transformer}
\setlength{\tabcolsep}{5pt}
\renewcommand{\arraystretch}{1.2}
\label{ch4ar tablepst}
\begin{tabular}{|c|c|c|c|c|c|}
\hline
\multirow{2}{*}{\begin{tabular}[c]{@{}c@{}}Fault\\  location\end{tabular}} & \multirow{2}{*}{\begin{tabular}[c]{@{}c@{}}Fault\\  resistance\\ (ohm)\end{tabular}} & \multirow{2}{*}{\begin{tabular}[c]{@{}c@{}}Phase \\shift\end{tabular}} & \multicolumn{3}{c|}{AR coefficient}                                                                                                                         \\ \cline{4-6} 
                                                                           &                                                                                      &                                                                                       & \begin{tabular}[c]{@{}c@{}}phase\\ a\end{tabular} & \begin{tabular}[c]{@{}c@{}}phase \\ b\end{tabular} & \begin{tabular}[c]{@{}c@{}}phase\\  c\end{tabular} \\ \hline
\multirow{6}{*}{\begin{tabular}[c]{@{}c@{}}Location\\ 2\end{tabular}}      & \multirow{2}{*}{0.1}                                                                & forward                                                                                      &                                                                              0.658&0.356&0.343                                          \\ \cline{3-6} 
                                                                    
                                                                           &                                                                                      & backward                                                                                 &  0.479&0.350&0.468                                    \\ \cline{2-6} 
                                                                           & \multirow{2}{*}{10}                                                                  & forward                                                                                  & 0.655&0.348&0.358                                      \\ \cline{3-6} 
                                                               
                                                                           &                                                                                      & backward                                                                                 &0.492&0.360&0.474                                       \\ \cline{2-6} 
                                                                           & \multirow{2}{*}{100}                                                                 & forward                                                                                  &-0.089&0.041&0.008                                    \\ \cline{3-6} 
                                                                      
                                                                           &                                                                                      & backward                                                                                      &-0.004&0.036&0.026
                                             \\ \hline
\multirow{6}{*}{\begin{tabular}[c]{@{}c@{}}Location\\ 4\end{tabular}}      & \multirow{2}{*}{0.1}                                                                & forward                                                                                                                                                             & -2.098&-1.205&-1.505                                    \\ \cline{3-6} 
                                                                            
                                                                           &                                                                                      & backward                                                                                     & -1.954&-1.272&-1.536                                           \\ \cline{2-6} 
                                                                           & \multirow{2}{*}{10}                                                                  & forward                                                                                  &-2.241&-1.831&-1.691                                            \\ \cline{3-6} 
                                                                                                                                                              
                                                                           &                                                                                      & backward                                                                                            &-2.278&-2.031&-1.662                                          \\ \cline{2-6} 
                                                                           & \multirow{2}{*}{100}                                                                 &forward                                                                                   &-1.538&-2.055&-1.799                                            \\ \cline{3-6} 
                                                                           &                                                                                      
                                                                                                                                                              & backward                                                                                            & -1.650&-2.136&-1.624

                                             \\ \hline
\end{tabular}
\end{table}

\section{Summary}
The impedance measured by a distance relay may be unpredictable in cases when the current and voltage frequencies are different. In such events, the dependability for faults in zone 1 and security for faults in adjacent zones are jeopardized. The proposed protection scheme governed by the auto-regressive coefficients of the 3-phase currents provides accurate detection and identification for the balanced faults. The balanced faults fed by the wind farm and the grid are distinguished and the correct fault zone is determined. The performance of the proposed scheme on standard IEEE systems obtained by varying the wind speeds, crowbar resistance, fault location, and fault resistance validate its effectiveness. Moreover, the protection scheme can be utilized for lines compensated with series capacitors and phase shifters.

\newpage

\chapter{Identification of Stable and Unstable Power Swings} \label{ch5}

\section{Introduction}\label{ch5intro}
Power system stability and reliability have become critical with the increase in complexity in power systems. Transmission lines are key to transfer power from generating stations to the consumers. Change in loading conditions, switching operations, and faults upset the power balance in T-lines and result in transient oscillatory power transfer among the generators known as power swings. During power swings, the 3-ph power deviates from the nominal power frequency and oscillates with slip frequency (0.3-7 Hz) \cite{chscv}. The power swings can evolve into stable or unstable swings depending on the severity of the disturbance and response of controllers. The power swings are low in magnitude and well-damped in most cases, and the apparent impedance is outside the relay operating characteristics. On the other hand, a high magnitude of power swings with lower damping is observed when multiple contingencies exist in a short duration. A system disturbance in an already stressed power system may result in such unstable power swings, causing a loss of synchronism.

The security and dependability of the T-line protection system are tested during power swings. Firstly, it should detect actual faults and trip the lines irrespective of the presence of any power swing. Secondly, it should maintain security by preventing undesired operation during stable power swings, thereby avoid worsening the situation to an unstable condition and loss of lines. Thirdly, it shall ensure the stability by isolating portions of the power system that lose synchronism during unstable power swings by blocking certain relays and unblocking pre-identified relays at the desired separation points. Various methods address these issues of power system reliability, but ensuring both security and dependability still remains a challenge. Generally, power swing blocking (\acrshort{PSB}) detects power swings and out of step tripping (\acrshort{OST}), differentiates stable and unstable swings \cite{chpsbost}.

Distance relays are predominantly used to protect high-voltage networks because of their selective and dependable tripping for line faults and simple time coordination of relays across the system. Distance relays function based on impedance (V/I ratio) measured. The relays trip with a predefined time delay when the impedance enters one of the protective zones as seen during faults. However, in power swings, the impedance trajectory may also encroach the zones and the distance relay mal-operates. During a power swing, the relay is blocked using the rate of change of impedance and then unblocked if a fault appears during the swing using zero and negative sequence components. This method would fail to detect symmetrical faults as it contains only positive sequence components. Mal-operation of distance relays is one of the primary reasons for cascaded outages \cite{chphadke}. The protection of transmission systems during power swings has always been a concern, e.g., the blackout on 14th August 2003 in the United States happened because of the operation of zone 2, and zone 3 distance relays in a system weakened by a series of outages \cite{chreport2}. Hence, reliable protection systems that differentiate actual faults from power swings are imperative.

Symmetrical and asymmetrical faults induce power swings, which can be stable or unstable. Conventionally PSB used blinders and timers to measure the rate of change of impedance to detect power swings \cite{chreport}. However, they fail to detect faults during power swings, and setting the timers appropriately is a challenge.
Many schemes have been proposed to unblock relay operation with faults during power swing. In \cite{chscv}, the rate of change of power swing center voltage is used to detect the faults during power swings, whereas the rate of change of resistance was used by \cite{chblinder}. A cross-blocking scheme that blocks the relays during power swings and quickly unblocks them when symmetrical faults occur based on change rates of active and reactive power was proposed in \cite{chlin}.
Mal-operations of distance relays during unstable power swings are discussed in \cite{chvittal} \cite{chjena}. The minimum voltage threshold evaluation method is proposed to determine the location of mis-operating relays during unstable swings in \cite{chvittal}. Reference \cite{chjena} proposed a transient potential power-based coordinate system to differentiate stable and unstable swings. In \cite{chHashemi}, the effects of asymmetrical power swings on distance protection were analyzed. 

Literature also reports the application of pattern recognition methods to differentiate power swings from faults. Wavelet Transform (WT) was used to detect faults during a power swing in \cite{chbrahma}. In \cite{chseethalekshmi}, SVMs are used to distinguish faults during power swing and voltage instability and then classify power swing and voltage instability using real power, reactive power, current, voltage, and delta and their changes as input features. Reference \cite{chDUBEY1} demonstrate the effectiveness of data-mining model (SVM, DT, Random Forest) approaches for symmetrical fault detection and out-of-step detection during power swing. Adaptive neuro-fuzzy inference system (ANFIS) with inputs: change of positive sequence impedance, positive and negative sequence currents, and power swing center voltages was used in \cite{chanfis}. Fast Fourier Transform (FFT) coefficients of the 3-ph active power were used to detect symmetrical faults during power swings in \cite{chfft}. In reference \cite{chps}, first the stable and unstable dynamic behaviors are distinguished and then the unstable contingency patterns are identified using DT, ensemble DT, and SVM.
Intelligent methods have also been used to classify faults and transients in Transformers, Phase Angle Regulators (PAR), and T-lines. The internal faults are discriminated from magnetizing inrush in a Transformer and then classified with DT, Random Forest, and Gradient Boost in \cite{chpedes}.
WT and Neural Networks were used to distinguish inrush and faults in a PAR in \cite{chtencon}.
In \cite{chsystempallav}, DT was used to discriminate internal faults and other transient disturbances in an interconnected system with PARs and Transformers. Applicability of machine learning (ML) algorithms with time and frequency features are used to distinguish and locate faults in the PAR in \cite{chietpallav}. A DT-based intelligent scheme for series-compensated T-line protection using differential phase angle of superimposed current was used in \cite{chtline}.

 \section{Contribution}  
This chapter, which was published in \cite{greentech}, explores the potential of ML algorithms to detect and classify power swings. The contributions of this work are outlined as follows:

\noindent {\raisebox{-0.4\height}{\scalebox{1.6}{\textbullet}}} It uses kNN, DT, and SVM to discriminate faults and faults during power swings from power swings and classify the power swings into stable and unstable swings. Thus, avoiding mal-operation of distance relays during faults during power swings and misoperation during unstable power swings ensuring the security and dependability of the protection system. 

\noindent {\raisebox{-0.4\height}{\scalebox{1.6}{\textbullet}}} Considering various system parameters 1320 faults, 1320 faults during swing, and 2205 power swing cases are simulated in PSCAD/EMTDC for the 9-Bus WSCC 3-machine system.  
    
\noindent {\raisebox{-0.4\height}{\scalebox{1.6}{\textbullet}}} A series of features are extracted from time, frequency, and time-frequency domains. Further, FFT coefficients of 3-ph voltages and linear trends of 3-ph currents selected using Random Forest are used to train three distinct classifiers. 

 \section{Chapter Organization}  
The remaining chapter is organized as follows. Section \ref{ch5model} describes the modeling and simulation of the power swings, faults, and faults during power swing in the 9-Bus WSCC system. Section \ref{ch5algo} consists of feature extraction and feature selection. Section \ref{ch5results} consists of the classification performance of the different classifiers. The last section summarizes the chapter.

\section{Modeling and Simulation}\label{ch5model}

\begin{figure}[htbp]
\centerline{\includegraphics[width=2.6 in, height= 1.6 in]{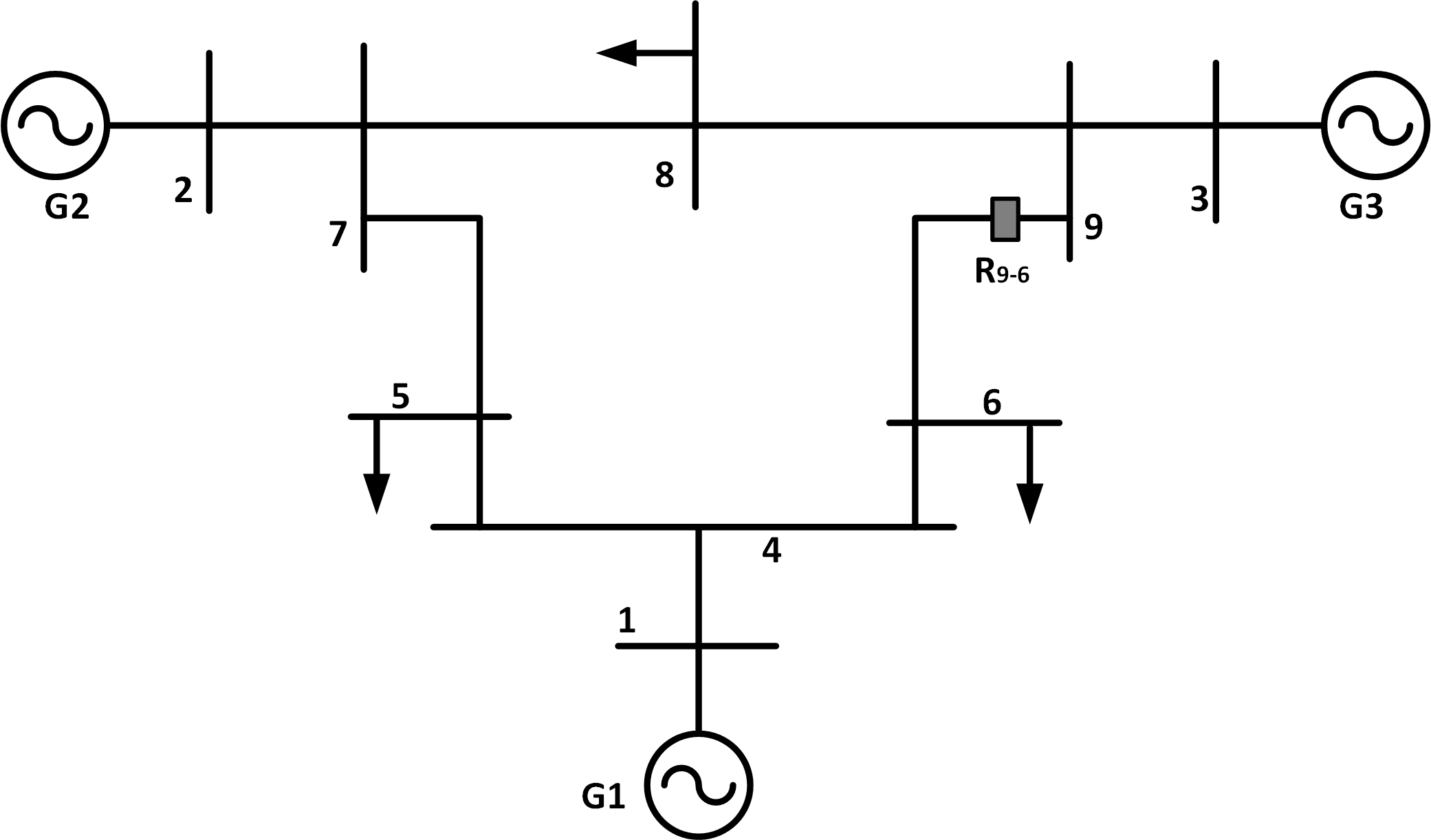}}\vspace{-1mm}
\caption{Single line diagram of 9-Bus system with generation, T-lines, and 3-ph loads}
\label{ckt}\vspace{-1mm}
\end{figure}

PSCAD/ EMTDC is used for modeling and simulation of the different power swings, faults, and faults during power swings in the 9-Bus WSCC three machine system. Fig.\ref{ckt} shows the single line diagram of the model consisting of the AC sources, T-lines, Transformers, and 3-ph loads working at 60Hz. 


Symmetrical power swings appear as 3-ph faults. Thus, it is crucial to distinguish a symmetrical swing from a 3-ph fault. Asymmetrical power swings are more likely to occur during single-pole tripping in heavy loaded long T-lines. The presence of zero and negative sequence components distinguish them from symmetrical swings. In stable power swings, the rotor angle variation is underdamped, and the system returns to a new equilibrium state. In contrast, the system experiences large fluctuations in voltage, current, power, rotor angle separation, and ultimate loss of synchronism during unstable power swings. Most asymmetrical swings are stable, but asymmetrical swings with long-dead time may develop into unstable swings.


The different scenarios for the four different power swings, faults, and faults during power swings are described in the following paragraphs.
\begin{figure}[htbp]
\vspace{-0.1cm}
\centerline{\includegraphics[width=3.7 in, height= 2.6 in]{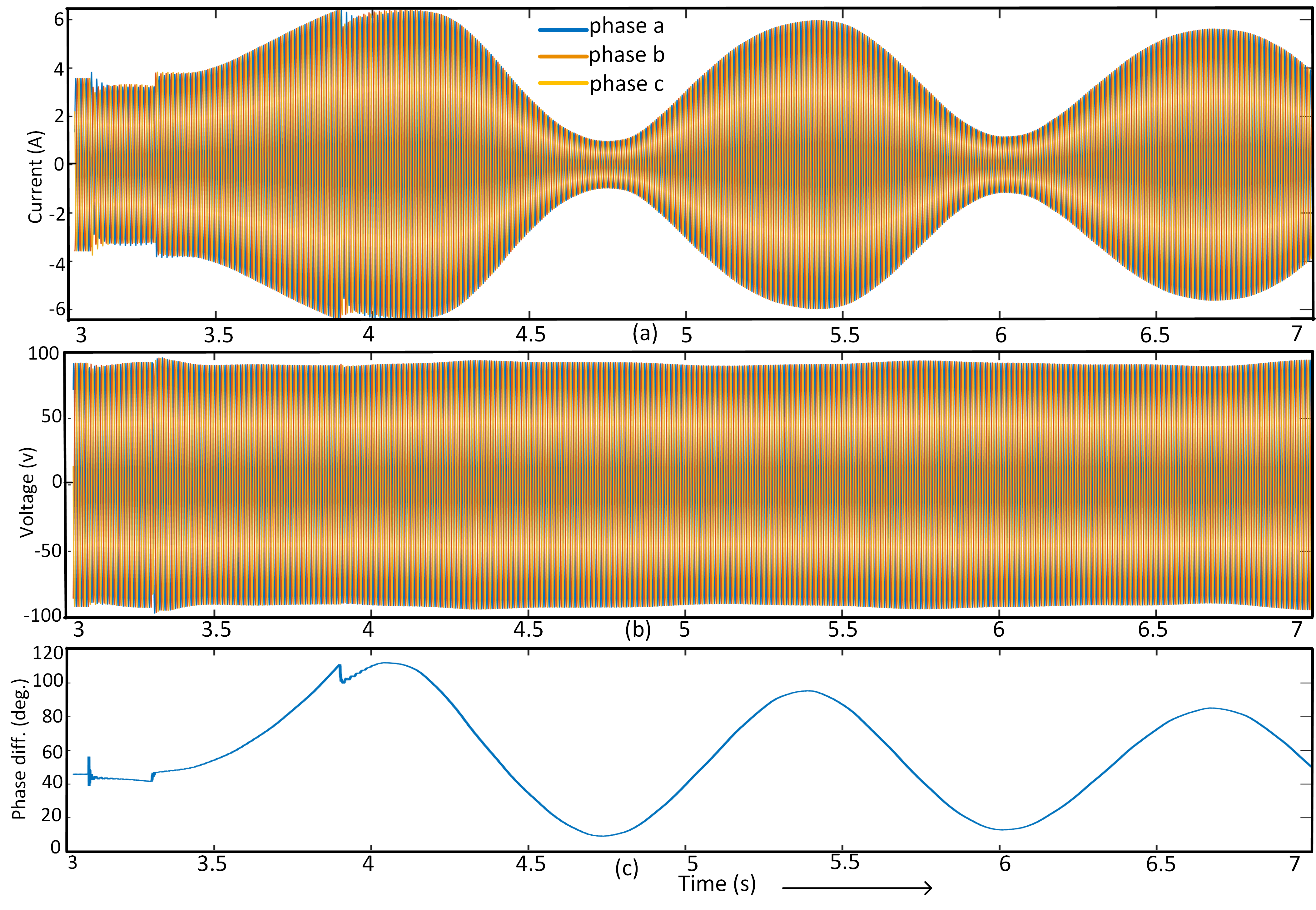}}
\vspace{0mm}\caption{(a) 3-ph currents, (b) 3-ph voltages, (c) phase difference ($\phi$) for a typical symmetrical stable power swing}
\label{ss}
\vspace{-1mm}
\end{figure}
\subsubsection{Symmetrical Stable Power Swings}
3-ph to ground (gnd) faults are assumed to occur in line 7-5 at distances of 30km, 100km, and 170 km through fault resistances of 0.1$\Omega$, 50$\Omega$, 100$\Omega$, 150$\Omega$, and 200$\Omega$ at 3.1s. Two fault clearing times are considered by opening breakers at both ends of line 7-5: 3.2s and 3.3s. The breaker closing time is varied from 3.4s to 3.9s in steps of 0.1s. The phase difference ($\phi$) between the bus-9 and bus-6 voltages are kept at 43$^{\circ}$, 33$^{\circ}$, 23$^{\circ}$, and 10$^{\circ}$. 
These result in stable power swings in the system. The swings are observed by the relay $R_{9-6}$ at bus-9. 720 cases of symmetrical stable power swings are thus obtained for training and testing the ML algorithms. Fig.\ref{ss}(a) shows the ph-A, B, and C currents at the relay location, and Fig.\ref{ss}(b) shows the voltages seen by the relay. The $\phi$ between bus-9 and bus-6 is shown in Fig.\ref{ss}(c).

\begin{figure}[htbp]
\vspace{-0.1cm}
\centerline{\includegraphics[width=3.7 in, height=  2.6 in]{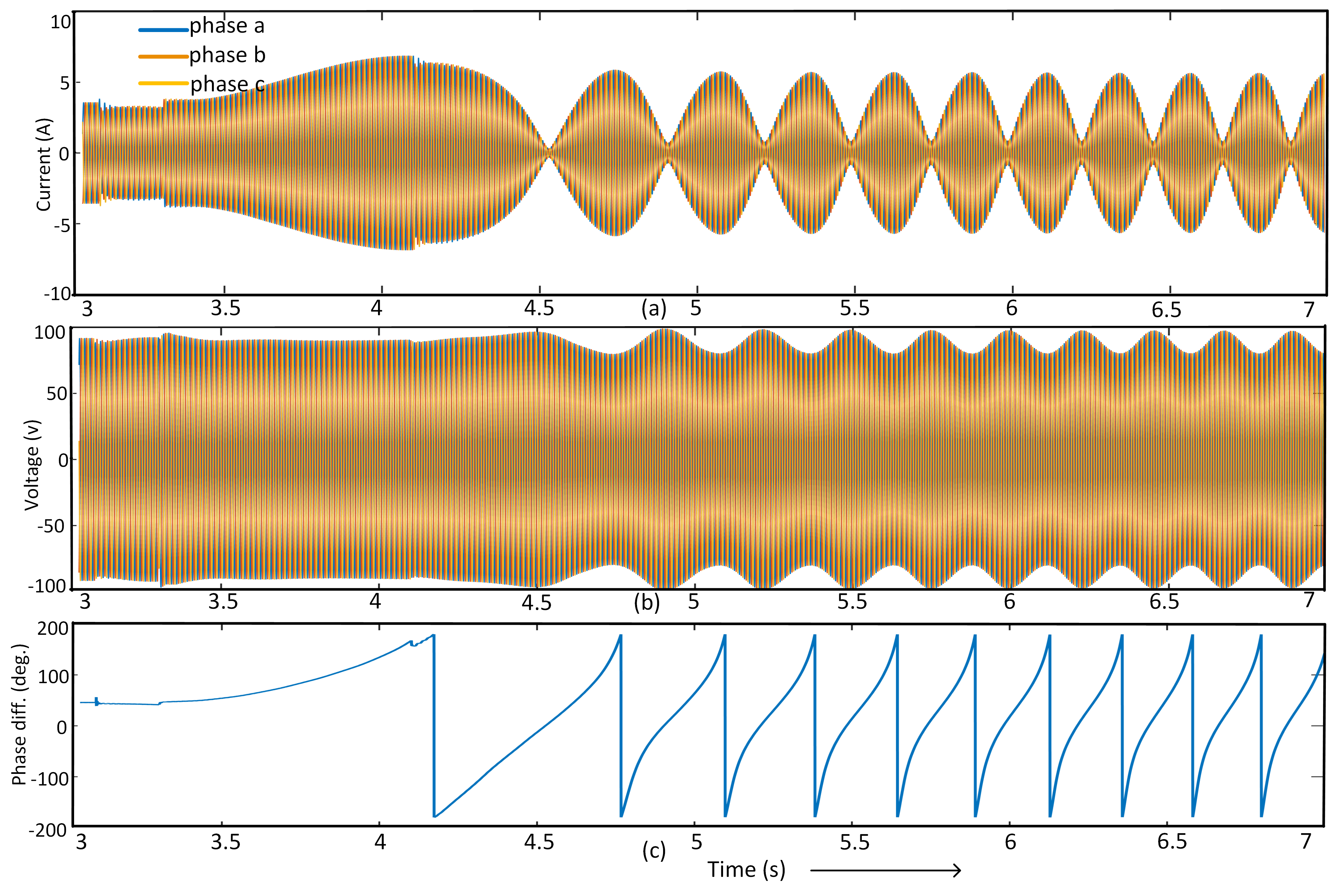}}
\vspace{0mm}\caption{(a) 3-ph currents, (b) 3-ph voltages, (c)  $\phi$ for a typical symmetrical unstable power swing}
\label{su}
\vspace{-1 em}
\end{figure}
\subsubsection{Symmetrical Unstable Power Swings}
3-ph to g faults are simulated in line 7-5 at distances of 30km, 60km, 80km, 120km, 150km, and 170 km through fault resistances of 0.1$\Omega$, 50$\Omega$, 100$\Omega$, 150$\Omega$, and 200$\Omega$ at 3.1s. The faults are cleared at 3.2s and then 3.3s by opening breakers at both ends of the line 7-5. Since a system may experience unstable swings with extended reclosing dead time, the breakers are reclosed at 4.1s, 4.2s, and then 4.3s. The $\phi$s between the bus-9 and bus-6 voltages are kept more than 43$^{\circ}$. These result in unstable power swings in the system. The swings are observed at bus-9. In total 540 cases of symmetrical unstable power swings are obtained.
 Fig.\ref{su}(a), Fig.\ref{su}(b), and Fig.\ref{su}(c) shows the ph-A, B, and C currents, voltages, and $\phi$ between bus-9 and bus-6.
\begin{figure}[htbp]
\vspace{-0.1cm}
\centerline{\includegraphics[width=3.7 in, height=  2.6 in]{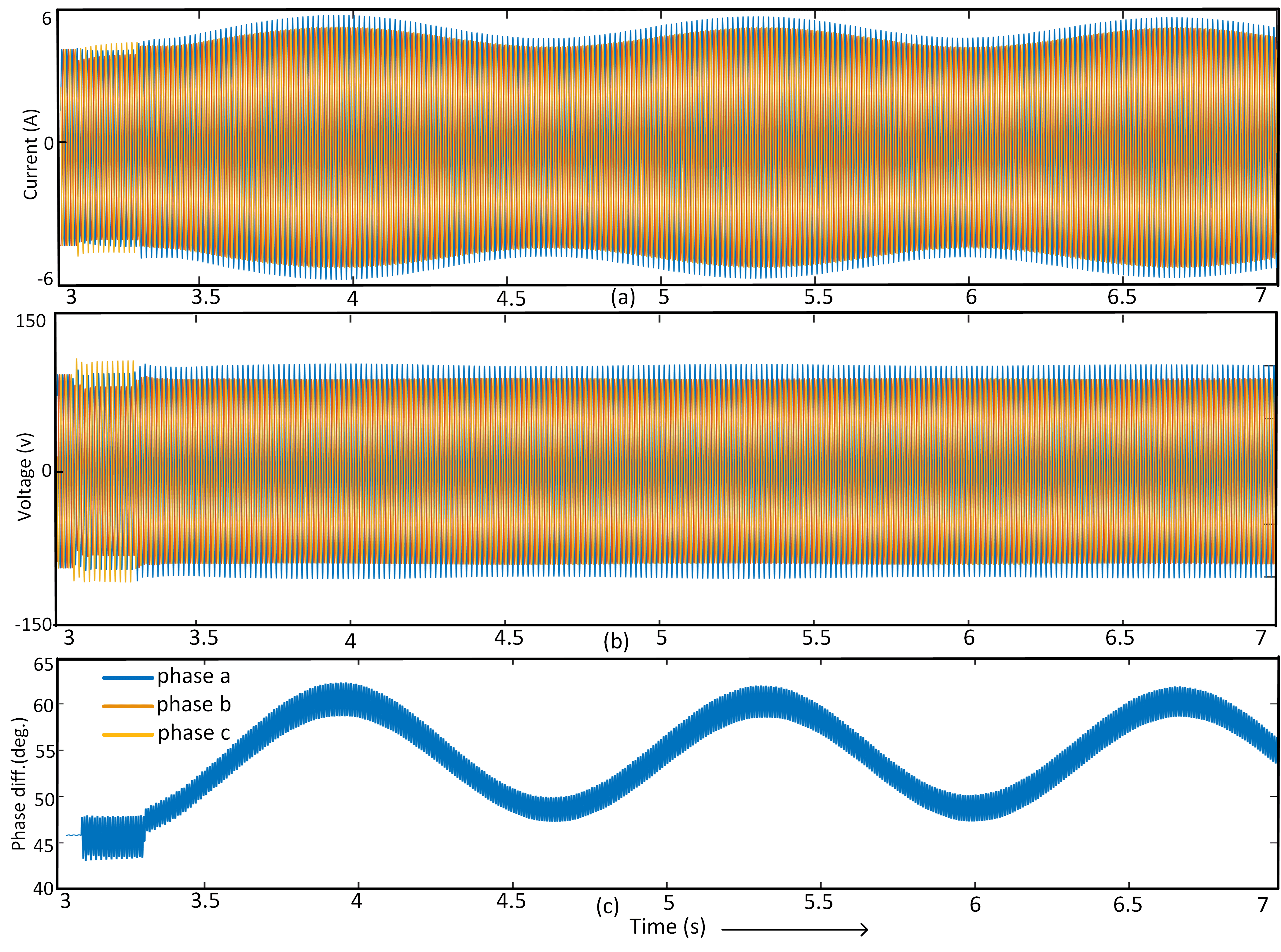}}
\vspace{-1mm}\caption{(a) 3-ph currents, (b) 3-ph voltages, (c)  $\phi$ for a typical asymmetrical stable power swing}
\label{as}
\vspace{-1em}
\end{figure}
\subsubsection{Asymmetrical Stable Power Swings}
1-ph to g faults in phs A, B, and C are created in line 7-5 at distances of 30km, 60km, 80km, 120km, and 150km through fault resistances of 0.1$\Omega$, 50$\Omega$, 100$\Omega$, 150$\Omega$, and 200$\Omega$ at 3.1s.
The faults are cleared at 3.2s and then 3.3s by opening breakers at both ends of line 7-5.
The $\phi$s between the bus-9 and bus-6 voltages are kept at  23$^{\circ}$, 33$^{\circ}$, and 43$^{\circ}$. These result in stable power swings in the system. The swings are observed by the relay $R_{9-6}$. In total, 450 cases of asymmetrical stable power swings are obtained. Fig.\ref{as}(a) shows the ph-A, B, and C currents at the relay location, and Fig.\ref{as}(b) shows the voltages seen by the relay. The $\phi$ between bus-9 and bus-6 is shown in Fig.\ref{as}(c).

\subsubsection{Asymmetrical Unstable Power Swings}
\begin{figure}[htbp]
\vspace{-0.1cm}
\centerline{\includegraphics[width=3.7 in, height=  2.6 in]{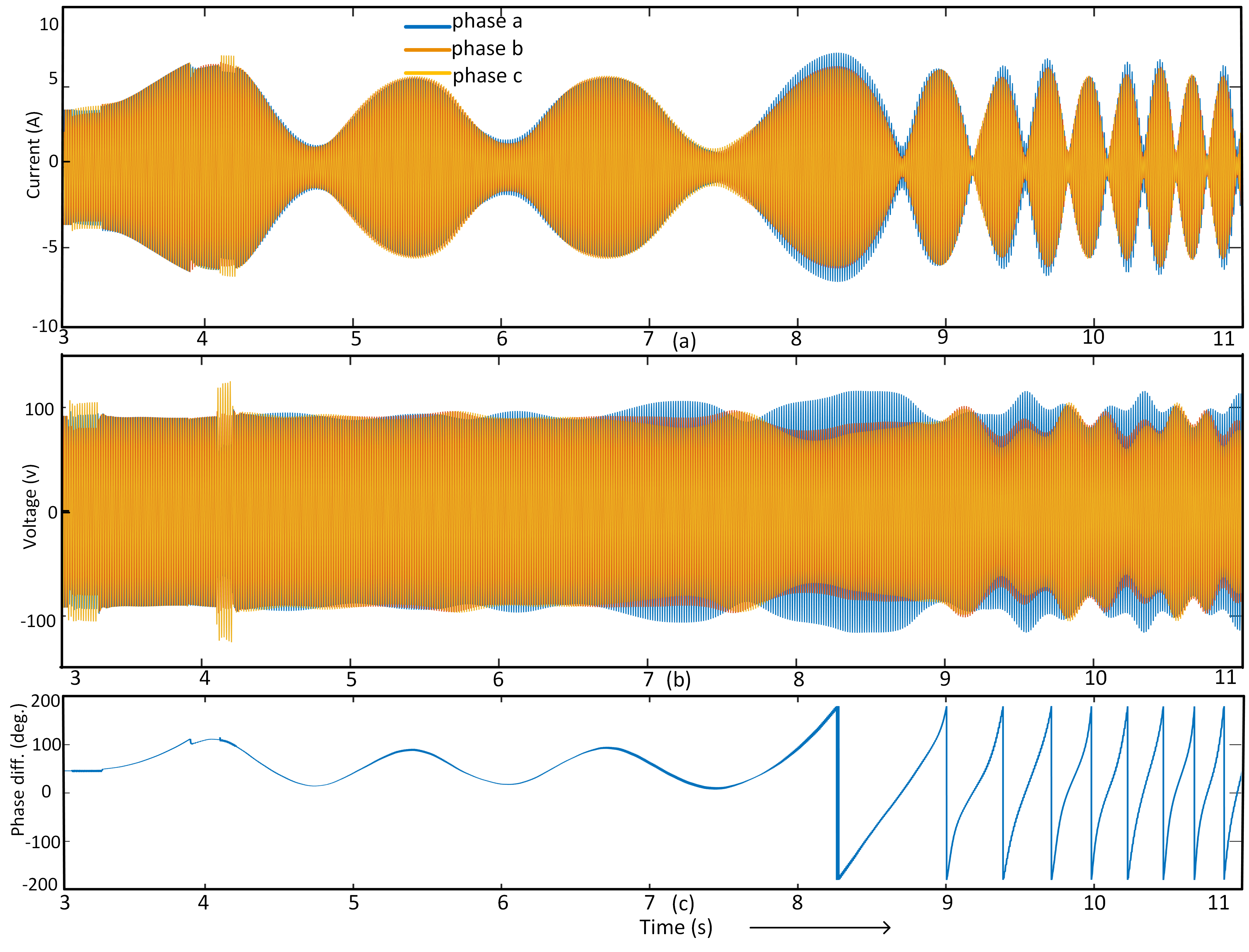}}
\vspace{-1mm}\caption{(a) 3-ph currents, (b) 3-ph voltages, (c)  $\phi$ for a typical asymmetrical unstable power swing}
\label{au}
\vspace{-1 em}
\end{figure}

A 3-ph to g fault is simulated in line 7-5 at 3.1s. The breakers are opened at 3.3s and reclosed again at 3.9s.
Then 1-ph to g faults in ph A, B, and C are created in line 7-8 at distances of 30km, 50km, and 70km through fault resistances of 0.1$\Omega$, 20$\Omega$, 40$\Omega$, 60$\Omega$, 80$\Omega$, 100$\Omega$, 120$\Omega$, 140$\Omega$, 160$\Omega$, 180$\Omega$, and 200$\Omega$.
The fault inception and clearing time pairs are at: (4.1,4.3s), (4.2s,4.4s), (4.3,4.5s), (4.4,4.6s), and (4.5,4.7s). The fault is cleared by opening the breakers at both ends of the line 7-8. The $\phi$ between the bus-9 and bus-6 voltages is kept more than 43$^{\circ}$ during this. These result in unstable power swings in the system. The swings are observed by the relay $R_{9-6}$ at bus-9. In total, 495 cases of asymmetrical unstable power swings are obtained. Fig.\ref{au}(a) shows the ph-A, B, and C currents at the relay location, and Fig.\ref{au}(b) shows the voltages seen by the relay. The $\phi$ between bus-9 and bus-6 is shown in Fig.\ref{au}(c).

\subsubsection{T-line faults}
The ph and gnd T-lines faults ($l$g, $ll$g, $ll$, $lll$ \& $lll$g) are simulated in line 9-6 at 30km and 170km through fault resistances of 0.1$\Omega$, 50$\Omega$, 100$\Omega$, 150$\Omega$, and 200$\Omega$ at 3.1s. The fault inception time is varied from 3.1 to 3.152778 in steps of 0.0038s. Thus, 1320 cases of faults are obtained.

\subsubsection{Fault during Symmetrical Stable Swings}
3-ph to g faults are created in line 7-5 at 100km distance at 3.1s. The breakers at both ends are opened at 3.3s and then closed at 3.8s, which results in symmetrical stable swings. 11 different ph and gnd faults are then simulated on line 9-6 through fault resistances 0.1$\Omega$, 50$\Omega$, 100$\Omega$, 150$\Omega$, and 200$\Omega$ at fault inception times of 5.5s to 5.5152778 in steps of 0.0038s. Thus, 660 faults during symmetrical swings are obtained.
\begin{figure}[htbp]
\vspace{-0.1cm}
\centerline{\includegraphics[width=3.7in, height=  2.6 in]{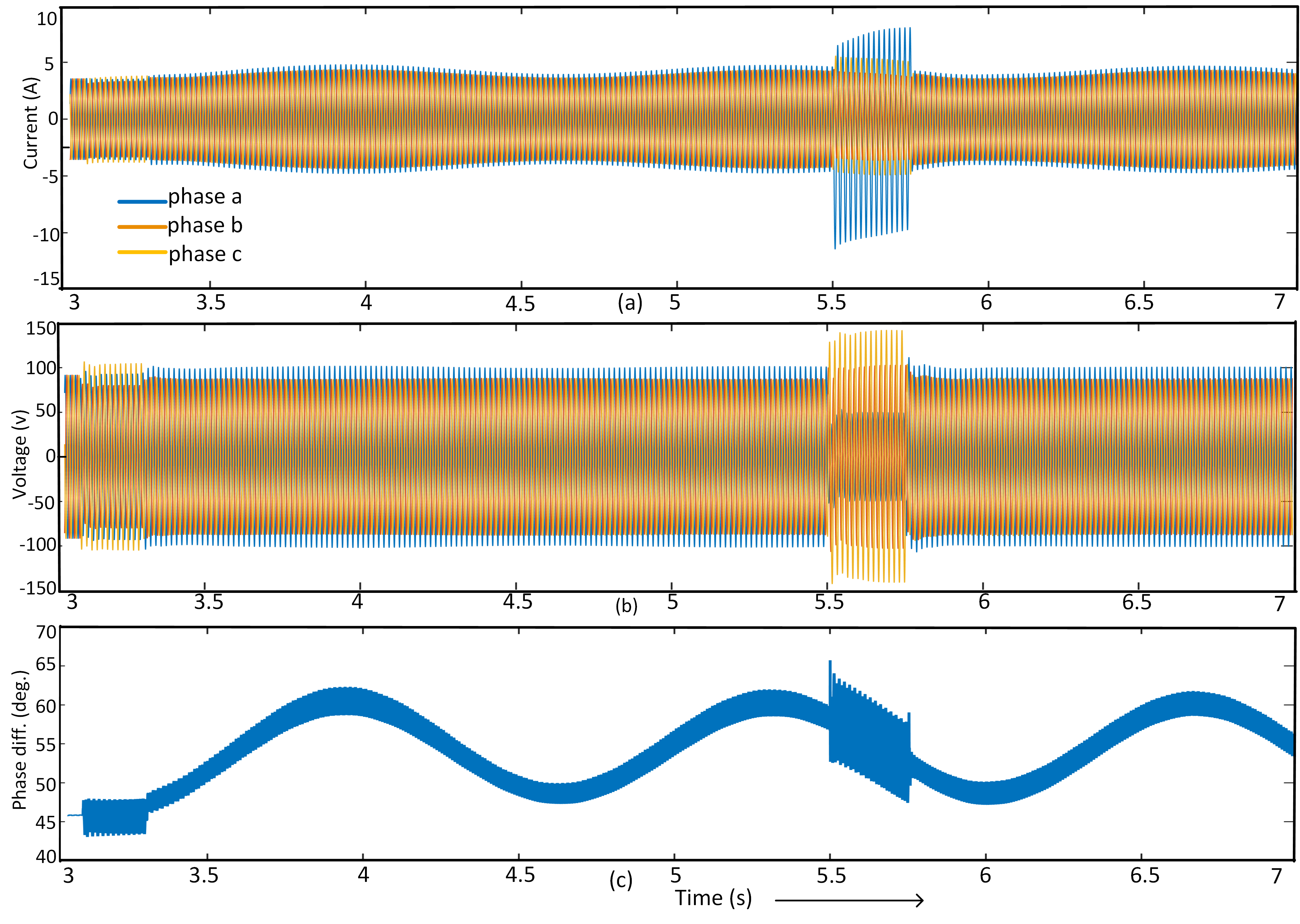}}
\vspace{-1mm}\caption{(a) 3-ph currents, (b) 3-ph voltages, (c)  $\phi$ for a typical fault during asymmetrical stable power swing}
\label{asf}
\vspace{-1em}
\end{figure}
\subsubsection{Fault during Asymmetrical Stable Swings}
1-ph to g faults are created in line 7-5 at a distance of 100km at 3.1s. The breakers at both ends are opened at 3.3s, which results in asymmetrical stable swings. The 11 different ph and gnd faults are then simulated on line 9-6 through fault resistances of 0.1$\Omega$, 50$\Omega$, 100$\Omega$, 150$\Omega$, and 200$\Omega$ at fault inception times of 5.5s to 5.5152778 in steps of 0.0038s. Thus, 660 faults during asymmetrical swings are obtained.
 Fig.\ref{asf}(a), Fig.\ref{asf}(b), and Fig.\ref{asf}(c) shows the ph-A, B, and C currents, the voltages seen by the relay, and $\phi$ between bus-9 and bus-6.

\section{Classification Framework}\label{ch5algo}
\begin{figure}[ht!]
\vspace{-0.1cm}
\centering
\includegraphics[width=4.0 in, height= 1.7 in]{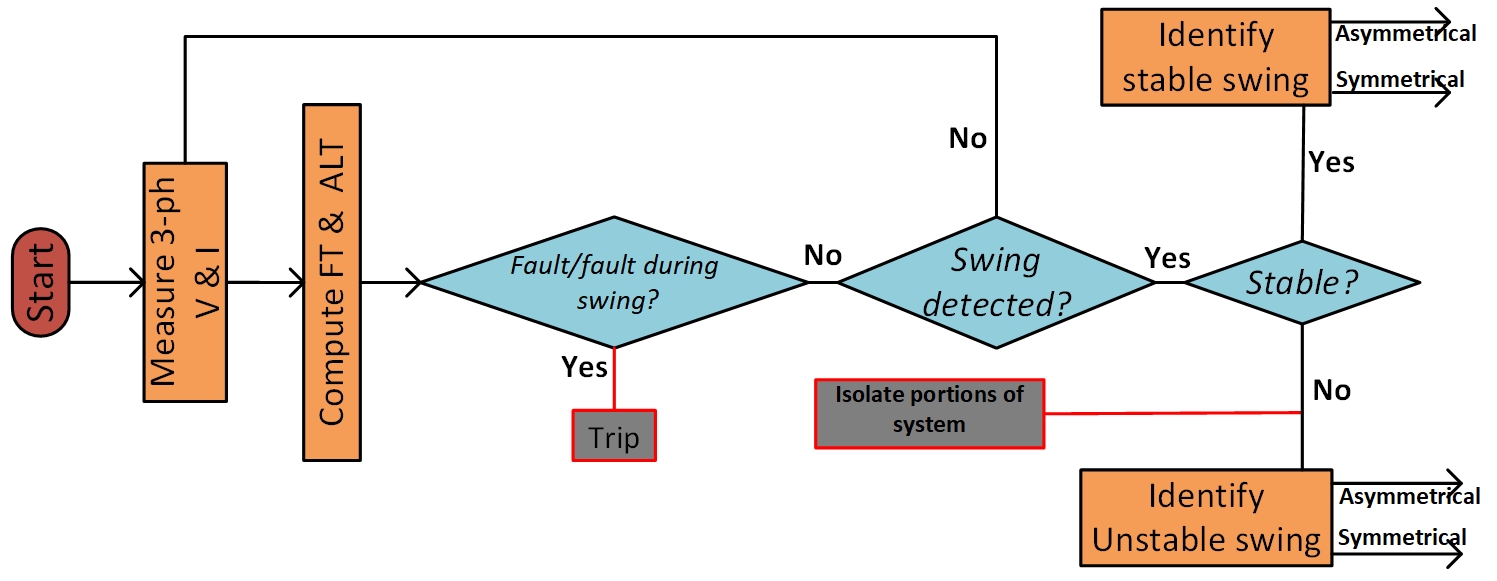}
\vspace{-0.1cm}
\caption{Detection and classification of swings}
\label{flowchart}
\vspace{-0.1cm}
\end{figure}

Fig.\ref{flowchart} shows the classification framework that is used to detect and classify power swings. The 3-ph relay voltage and current samples are used to compute the features (FFT coefficient and linear trend). 83 samples (one-cycle) are used to detect faults and faults during power swings and send the trip signal. If a power swing is detected, it is classified into stable and unstable swings using samples from 10 cycles. A trip/block signal is sent to the pre-identified relays to avoid unintentional islanding if an unstable swing is detected. The unstable and stable swings are then classified into symmetrical and asymmetrical swings.

\subsection{Feature Extraction and Selection}
The 3-ph currents and voltages at bus-9 are used to derive non-redundant, informative, and interpretable values. 
Random Forest is used to obtain the most relevant voltage and current features. 
The FFT coefficients of ph-A, B, and C voltages and aggregate linear trends of ph-A, B, and C  currents are used to detect power swings, distinguish swings into stable and unstable  swings, and then into symmetrical and asymmetrical swings. 



\begin{figure}[ht!]
\vspace{-0.1cm}
\centering
\includegraphics[width=4.1 in, height= 4 in]{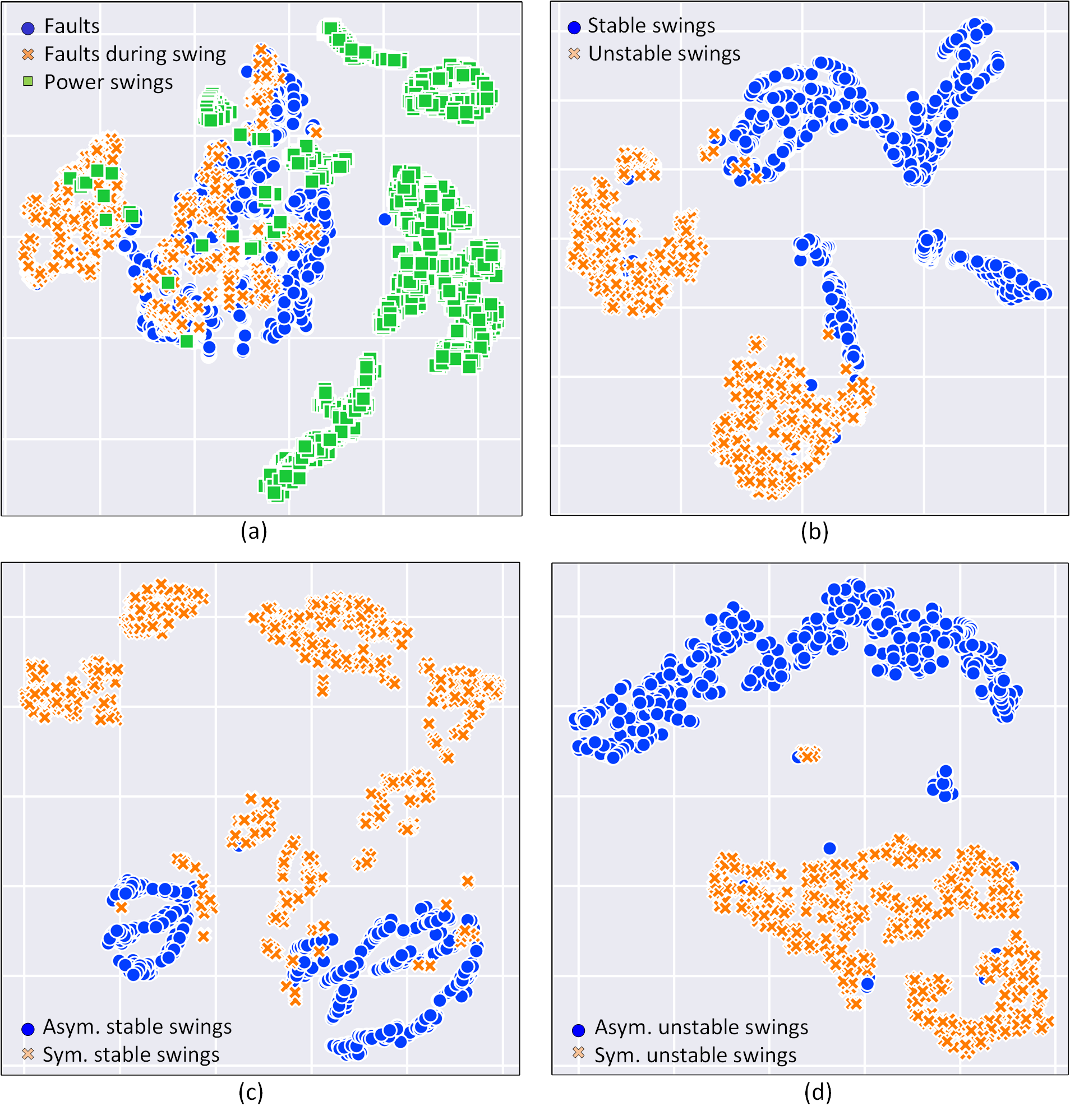}
\vspace{-0.1cm}
\caption{Scatter plot of input data for (a) faults and power swings, (b) power swings, (c) stable power swings, and (d) unstable power swings }
\label{tsne}
\vspace{-0.1cm}
\end{figure}

In order to visualize, the six features (3-ph FFT coefficients and 3-ph linear trends) are mapped non-linearly on a two-dimensional plane by preserving both the important local and global structures in high dimensional space using t-distributed Stochastic Neighbour Embedding (t-SNE)\cite{chtsne}. The scatter plots of the six identified features to differentiate: faults and swings; stable and unstable swings; asymmetrical and symmetrical stable swings; and  asymmetrical and symmetrical unstable swings are illustrated in Fig.\ref{tsne}.

\section{Results and Discussion}\label{ch5results}
First, features obtained from one cycle of post transient 3-ph currents and voltages are used to distinguish faults and power swings. 
Then 10 cycles of 3-ph currents and voltages are used to identify first, stable and unstable swings and second, symmetrical and asymmetrical swings. At a sampling rate of 5 kHz, 83 data samples are obtained per cycle. Some factors that influence an ML algorithm's classification accuracy are cross-validation, grid search, data split, and performance metrics. Ten-fold stratified cross-validation is applied to the training data to avoid overfitting and underfitting of the classifier on the test set. Grid search is used to optimize the parameters over a hyperparameter space. The data is split randomly into training and test set in a  4:1  ratio. The selected hyperparameters' performance is then tested on the unseen test data that is not used during the training process using balanced accuracy. 
Accuracy is used as the typical metrics to measure a classifier's performance, which gives biased results if the data is imbalanced. Hence, balanced accuracy which is the average accuracy obtained on all classes and computed as $\bar{\eta}$ = $ \frac{1}{2}\cdot[\frac{TP}{(TP+FN)} + \frac{TN}{(TN+ FP)}]$ for a two-class problem has been used to compute the performance where, T=true, F=false, N=negative, and P=positive.

\subsubsection{Differentiate faults and Power Swings}
Faults and faults during power swing are differentiated from power swings using one cycle of post-event samples. The classifiers are trained and tested on FFT coefficients of the 3-ph voltages and linear trends of 3-ph currents obtained from 1320 faults, 1320 faults during the swing, and 2205 power swing cases simulated in section II.
Table \ref{tabpsflt2} shows the performance of kNN for the three class classification and Table \ref{tabscom} compares the $\bar{\eta}$s of kNN, DT, and SVM. The best parameters for SVM, kNN, and DT obtained from grid searching are [$\gamma$:1, C:100, kernel:rbf], [leaf size:1, n neighbors:1, p:1], and [criterion:entropy] respectively.

\begin{table}[ht]
\centering
\caption{Differentiate Faults and Power Swings \label{tabpsflt}}
\setlength{\tabcolsep}{3pt}
\begin{subtable}[b]{.65\linewidth}
\centering
\caption{Performance with kNN \label{tabpsflt2}}
{\begin{tabular}{lllll}\toprule
  Transient event  &   Total &   TP  &   FN &   FP\\
\midrule
 Fault&  251 &   248 & 3 &  4\\
 Fault during swing & 273 & 272 &  1 &  3\\
 Power swing & 443 & 442 & 3&  0\\
\hline
\end{tabular}}{}
\end{subtable}{}
\begin{subtable}[b]{.26\linewidth}
\captionsetup{justification=centering}
\caption{ Comparison\label{tabscom}}
{\begin{tabular}{lc}\toprule
Classifier & $\bar{\eta}$ \\
\midrule
kNN &  {99.25\%}\\
DT  &97.82\%\\
SVM & 98.40\%\\

\hline
\end{tabular}}
\end{subtable}{}
\vspace{-0em}
\end{table}

\subsubsection{Differentiate Stable \& Unstable Power Swings}
Once a power swing is detected, 10 cycles of 3-ph voltage and current samples are used to identify the swing as stable or unstable. The classifiers are trained on the six features obtained from 1764 cases and then tested on 441 cases.
Table \ref{tabps} compares  the  $\bar{\eta}$s  of  kNN,  DT, and  SVM. The best parameters for SVM, kNN, and DT are [$\gamma$:1, C:10, kernel:rbf], [leaf size:1, n neighbors:1, p:1], and [criterion:entropy] respectively.

\subsubsection{Differentiate asymmetrical \& symmetrical stable swings}
If the swing is stable, it is further classified into asymmetrical and symmetrical. The classifiers are again trained on the same features obtained from 936 cases and then tested on 234 cases.
The Table \ref{tabs} compares the $\bar{\eta}$s of kNN, DT, and SVM. The hyperparameters obtained from grid search for SVM, kNN, and DT are [$\gamma$:1, C:1000, kernel:rbf], [leaf size:1, n neighbors:1, p:2], and [criterion:entropy] respectively.

\subsubsection{Differentiate asymmetrical \& symmetrical unstable swings}
On the contrary, if the swing is unstable, it is classified into asymmetrical and symmetrical unstable swings. The classifiers are trained on the same six features obtained from 824 cases and then tested on 207 cases.
The Table \ref{tabu} compares the $\bar{\eta}$s of  kNN, DT, and SVM classifiers. The best parameters for SVM, kNN, and DT are [$\gamma$:1, C:1, kernel:rbf], [leaf size:1, n neighbors:1, p:2], and [criterion:gini] respectively.
\begin{table}[ht]
\centering
\captionsetup{justification=centering}
\caption{Classification of Power Swings}
\setlength{\tabcolsep}{2pt}
\begin{subtable}[b]{.26\linewidth}
\centering
\caption{Stable \& Unstable Power Swings \label{tabps}}
{\begin{tabular}{lc}\toprule
Classifier & $\bar{\eta}$ \\
\midrule
kNN & {99.77\%}\\
DT  & 99.32\%\\
SVM & 99.78\%\\

\hline
\end{tabular}}{}
\end{subtable}{}\hfill
\begin{subtable}[b]{.32\linewidth}
\centering

{\caption{Asymmetrical \& Symmetrical Stable Swing\label{tabs}}}
{\begin{tabular}{lc}\toprule
Classifier & $\bar{\eta}$ \\
\midrule
kNN &  {100.0\%}\\
DT  & 99.01\%\\
SVM & 100.0\%\\

\hline
\end{tabular}}
\end{subtable}{}\hfill
\begin{subtable}[b]{.34\linewidth}
\centering
\caption{Asymmetrical \& Symmetrical Unstable Swing\label{tabu}}
{\begin{tabular}{lc}\toprule
Classifier & $\bar{\eta}$ \\
\midrule
kNN & {99.04\%}\\
DT  & 100.0\%\\
SVM & 100.0\%\\

\hline
\end{tabular}}{}
\end{subtable}{}
\vspace{-1.0 em}
\end{table}

\section{Summary}\label{ch5summary}
Faults during symmetrical power swings may cause mal-operation of distance relay. Unwanted operation also happens during unstable power swings causing unintentional and uncontrolled islanding. Faster and accurate detection of the faults and faults during power swings and the classification of power swings can help the protection system make reliable decisions during power swings. This chapter discriminates faults from power swings with one-cycle and identifies different swings that may occur in a power system with 10 cycles of post transient 3-ph voltage and current samples. The different faults and swings are simulated by varying the line length, fault resistance, fault type, phase difference, breaker opening time, loading, etc. SVM, DT, and kNN trained on six features distinguish the faults from power swing with an accuracy of 99.3\% and discriminate an unstable swing from a stable one with 99.8\% accuracy. The results obtained ensure the security/dependability and indicate the effectiveness of the proposed scheme in distinguishing power swings from faults.

\newpage

\chapter{Conclusions and Future Work} \vspace{-4mm}
In this chapter, the main contributions and conclusions of this dissertation are presented. Directions and ideas for future are also suggested. 

\vspace{-3mm}

\section{Summary}
In this dissertation, the use of data-driven techniques for power system protection was investigated. To accurately analyze the faults and other transients, it was important to correctly model the power transformers and phase angle regulators in Chapters 2 and 3. Additionally, it was important to simulate enough transient cases for accurate performance evaluation of the classifiers. Appropriate time and frequency domain features were selected using different selection algorithms to train the classifiers. The following outcomes summarize this dissertation and imply that ML approaches may be used in the near future.

In Chapter 2, we looked at whether time and time-frequency feature-based estimators could be used to discriminate internal faults from other transient conditions for ISPARs.
Models of two- and three-winding transformers were built to simulate faults.
The defective core unit was discovered, and the transients were recognized using features extracted from a single cycle of post-transient 3-phase differential currents filtered by an event detector.
The appropriate wavelet energy, time-domain, and wavelet coefficient features were selected using Maximum Relevance Minimum Redundancy, Random Forest, and exhaustive search with Decision Trees approaches, respectively.
The fault detection scheme with the XGBoost classifier and hyperparameters from Bayesian Optimization performed the best.
The suggested scheme's reliability was tested at several tap positions, noise levels, and ratings, as well as under diverse conditions such as CT saturation, fault during magnetizing inrush, series core saturation, low current faults, and wind energy integration.

Next, Chapter 3,
solved the problem of fault detection and transient classification in a 5-bus interconnected system for Phase Angle Regulators and Power Transformers. Six transients other than faults which include magnetizing inrush, sympathetic inrush, external faults with CT saturation, capacitor switching, non-linear load switching, and ferroresonance were simulated.
 It was found that the gradient boosting classifier outperformed the others for detection of internal faults, localization of faulty unit, identification of fault type, and transient classification. Five most relevant frequency and time domain features obtained using Information Gain were used for training and testing. The reliability of the scheme was verified for different ratings and connections of the transformers involved, CT saturation, and noise level in the signals.

Next,  Chapter 4, proposed a waveshape property-based protection of the intertie zone between WF and grid during 3-phase faults. Autoregressive coefficients of the 3-phase currents obtained from the CT at one end were used to distinguish the faults fed by the type-3 WFs and the bulk grid. The validity of the technique was tested on 3 systems. The findings suggested that feature-based algorithms could be used to improve the power system distance relaying system. 

Finally, in Chapter 5, mal-operation of distance relay during faults with symmetrical power swings and during unstable power swings was considered.  The faults, faults during power swing were distinguished from power swings in one cycle. The different power swings were identified in the 9-bus WSCC system.  Six features obtained from 3-phase relay voltage and current were used to test the validity of the detection and classification scheme. 

\section{Conclusion}\vspace{-2mm}

The research provides a comprehensive study of the different power systems transients which includes the faults, magnetizing inrush, sympathetic inrush, overexcitation, external faults with CT saturation, ferroresonance, load switching, capacitor switching, CT saturation, faults during inrush, series core saturation, and low current faults considering an exhaustive list of 3-phase current features for differential protection.

Ensemble methods presented here which combine several machine learning algorithms improved the performance compared to single algorithms, such as Neural Networks, Bayesian Methods, and  other supervised classification methods. Particularly Gradient Boosting Classifier and XGBoost algorithms which construct a set of hypotheses generated by several base learners worked well.

Time and time-frequency domain features performed well in detecting and distinguishing faults, and other transients in transformers, as well as detecting faults, and power swings and determining the fault zone in transmission lines connected to wind farms.

Although it is observed that Ensemble methods performed satisfactorily across the board, the choice of features – wavelet coefficients, time-domain, or wavelet energy – is application dependent. Time-domain features had the better performance in terms of accuracy; however, their execution time is more than wavelet-coefficients. In an actual implementation scenario, the particular selection of features must take into account the computational speed of the microcontroller while maximizing the overall accuracy.


While effects of noise, change in rating, and transformer connections were analyzed, real-world data may differ in other aspects from simulated data used in this dissertation and may result in performance degradation not studied here. Therefore, it’s important to test the algorithms with actual systems prior to their deployment. 

\section{Future Work}\vspace{-2mm}

Some promising directions for future work related to non-conventional ML algorithms and transmission line protection are listed below:

\begin{itemize}

\item Data processing and analysis for intelligent decision-making in large-scale complex multi-energy systems with lightweight machine learning-based solutions.

    \item Investigate the use and effectiveness of Tiny ML which deploys ML on ARM-based microcontrollers which are cheap and draw low power.
    
    \item Investigate attributes of real-world data such as data loss and communication delays, and address coordination issues with ML-based methods for protection at higher and lower voltages.
    
\item Applications of trending machine learning methods like deep learning, reinforcement learning, unsupervised learning and so forth to derive network protection algorithms from vast data.
\item Analyze the major issues that cause distance relays to malfunction, such as the effect of the DC component, close-in faults, high resistance faults, fault inception angle and power flow angle, load encroachment and uncoordinated zone 3 relay settings, transient faults and auto-reclosing schemes, power swings, and series compensation in transmission lines, to name a few. In addition, the impact of various wind turbine generators (type-3 and type-4) will be investigated. 

    
    
    
  
    
    
\end{itemize}


\newpage
 \addcontentsline{toc}{chapter}{\protect\numberline{}{\hspace{-0.75cm} Appendix}}
\appendix 
\chapter{Script for 2-winding and 3-winding Transformer fault models}
\label{appendix2}
\begin{table}[ht]
\renewcommand{\arraystretch}{1}
\setlength{\tabcolsep}{10pt}
\centering
\caption{\textbf{Fortran script for single-phase 2-winding transformer fault model\label{ch2a1}}}

 \begin{tabular}{ll}\toprule
  1. nw = 4 &         18. L2l = Lk1/2*$fb$\\
  2. $I_m2$ = $I_m1$ = $I_m$ &           19. L3l = Lk2/2*$fc$\\
  3. $fa=fault1*0.01 $ &       20. L4l = Lk2/2*$fd$\\
  4. $fb=1.0-fa$ &       21. L1m = (v1/(w*$I_m1$*i1))*$fa*fa$\\
  5. $fc=fault2*0.01$ &       22. L2m = (v1/(w*$I_m1$*i1))*$fb*fb$\\
  6. $fd=1.0-fc$ &          23. L3m = (v2/(w*$I_m2$*i2))*$fc*fc$\\
  7. i1 = MVA/v1 &        24. L4m = (v2/(w*$I_m2$*i2))*$fd*fd$\\
  8. i2 = MVA/v2 &        25. L1 = L1l + L1m\\
  9. z1 = v1/i1 &           26. L2 = L2l + L2m\\
  10. z2 = v2/i2  &          27. L3 = L3l + L3m\\
 11. w = 2*pi*f   &             28. L4 = L4l + L4m\\
  12. l1 = v1/(w*$I_m1$*i1) &           29. M12 = sqrt(L1m*L2m)\\
  13. l2 = v2/(w*$I_m2$*i2) &            30. M13 = sqrt(L1m*L3m)\\
  14. Lk1 = Xl*z1/w &          31. M14 = sqrt(L1m*L4m)\\
   15. Lk2 = Xl*z2/w  &         32. M23 = sqrt(L2m*L3m)\\
   16. tr = v1/v2 & 33. M24 = sqrt(L2m*L4m)\\
  17. L1l = Lk1/2*$fa$ & 34. M34 = sqrt(L3m*L4m)\\ \hline
\end{tabular}
\end{table}

\begin{table}[ht]
\renewcommand{\arraystretch}{1}
\setlength{\tabcolsep}{10pt}
\centering
\caption{\textbf{Fortran script for single-phase 3-winding transformer fault model}}
 \begin{tabular}{ll}\toprule
  1. nw = 6, w = 2*pi*f         &         24. L3m = v2/(w*$I_m2$*i2)*$fc^2$\\
  2. $I_m3$ = $I_m2$ =$I_m1$    &           25. L4m = v2/(w*$I_m2$*i2)*$fd^2$ \\
  3. fa = fault1*0.01        &       26.  L5m =v3/(w*$I_m3$*i3)*$fe^2$ \\
  4. fb = 1.0-fa        &       27. L6m =v3/(w*$I_m3$*i3)*$ff^2$ \\
  5. fc = fault2*0.01    &       28. L1 = L1l + L1m, L2 = L2l + L2m \\
  6. fd = 1.0-fc          &        29. L3 = L3l + L3m, L4 = L4l + L4m  \\
  7. fe = fault1*0.01     &  30. L5 = L5l + L5m, L6 = L6l + L6m  \\
  8. ff = 1.0-fe           &   31. M12 = sqrt(L1m*L2m)  \\
  9. z1=v1/i1, z2=v2/i2, z3=v3/i3 &        32. M13 = sqrt(L1m*L3m)\\
  10. l1 = v1/(w*$I_m1$*i1)              &       33. M14 = sqrt(L1m*L4m)\\
  11. l2 = v2/(w*$I_m2$*i2)              &        34.  M15 = sqrt(L1m*L5m)\\
  12. l3 = v3/(w*$I_m3$*i3)          &   35. M16 = sqrt(L1m*L6m)\\
  13. X1 = (x13-x23+x12)/2    & 36. M23 = sqrt(L2m*L3m)\\
  14. X2 = (x23-x13+x12)/2   & 37. M24 = sqrt(L2m*L4m)\\
  15. X3 = (x13-x12+x23)/2    & 38. M25 = sqrt(L2m*L5m)\\
  16. Lk1 = X1*z1/w &             39.  M26 = sqrt(L2m*L6m)\\
  17. Lk2 = X2*z2/w &           40. M34 = sqrt(L3m*L4m)\\
  18. Lk3 = X3*z3/w &            41. M35 = sqrt(L3m*L5m)\\
  19. L1l= Lk1*$fa$, L2l=Lk1*$fb$   &          42. M36 = sqrt(L3m*L6m)\\
  20. L3l= Lk2*$fc$, L4l=Lk2*$fd$   &         43. M45 = sqrt(L4m*L5m)\\
  21. L5l= Lk3*$fe$, L6l=Lk3*$ff$   & 44. M46 = sqrt(L4m*L6m)\\
  22. L1m =v1/(w*$I_m1$*i1)*$fa^2$  & 45.  M56 = sqrt(L5m*L6m)\\ 
  23. L2m =v1/(w*$I_m1$*i1)*$fb^2$   &                         \\ \hline
\end{tabular}
\end{table}

\chapter{4-Bus Test System}
\label{appendix5}
\section{System Details for 4-bus test system}
The test system is 230 kV, 60 Hz.

\noindent \textbf{Sources:}
$Z_{S4}$=60$\angle$86.7$^{\circ}$ $\Omega$, $Z_{S5}$= 20$\angle$80$^{\circ}$$\Omega$.

\noindent \textbf{DFIG rating:} S = 200 MW, V = 33 kV,
Rstator = 0.0054 p.u.,
Rrotor = 0.006 pu, Xm = 4.5 pu, Xstator = 0.01 pu, Xrotor = 0.11 pu

\noindent \textbf{DC-link:} Rated voltage= 1700 V, capacitor=7.5 mF.

\noindent \textbf{Turbine Transformer:} 2.5 MVA, 0.69/0.9/33 kV, ygygYg.

\noindent \textbf{Main transformer:} 250 MVA, 33/230 kV, ygYg.

\section{Impedance Measurement in case of 4-bus test system}

$Z_1$=0.189$\angle$84$^{\circ}$$\Omega$/km, $Z_0$=1.06$\angle$84.17$^{\circ}$$\Omega$/km.

\noindent \textbf{Impedance obtained after CT and PT:}

\noindent Zone 1 impedance is 80\% of the line length to be protected
$Z_{sec1}$ = R*(0.8*$Z_1$) = 4.69$\angle$86.7$^{\circ}$,

\noindent Zone 2 impedance is 120\% of the line length to be protected
$Z_{sec2}$ = R*(1.2*$Z_1$) = 5.62$\angle$86.7$^{\circ}$

\noindent where, $R=\frac{Nc}{Nv}=\frac{current\ transformer\ ratio}{CVT\ ratio}=\frac{500:1}{2021:1}=0.248$
, $Z_{sec}=R* Z_{primary}$

\noindent \textbf{Positive sequence impedance for phase A to ground fault}: $Z_A=\frac{V_A}{I_A + 3k_0I_0}$ where, $3I_0=I_A +I_B +I_C$, $k_0=\frac{Z_0 -Z_1}{3Z_1}$


\singlespacing 
\addcontentsline{toc}{chapter}{References}
\centerline{\Large{List of References}}
\bibliography{bibpkb}
\bibliographystyle{IEEEtran}


\end{document}